\documentclass[epj]{svjour}
\pdfoutput=1

\usepackage{atlasphysics}
\usepackage{subfigure}
\usepackage{amsmath}
\usepackage{amssymb}
\usepackage{rotating}
\usepackage{multirow}

\usepackage{fancyhdr}
\usepackage{mathrsfs}
\usepackage{booktabs}
\usepackage[switch]{lineno}
\usepackage{cite}
\usepackage{afterpage}
\usepackage{preprintcover}
\usepackage{hyperref}

\def\Zmm{{Z\to\mu\mu}}
\def\Jpsimm{{J/\psi\to\mu\mu}}
\def\Upsilonmm{{\Upsilon\to\mu\mu}}

\PreprintCoverPaperTitle{Measurement of the muon reconstruction performance of the ATLAS
  detector using 2011 and 2012 LHC proton--proton collision data}

\PreprintIdNumber{CERN-PH-EP-2014-151}
\PreprintCoverAbstract{This paper presents the performance of the ATLAS muon reconstruction
during the LHC run with $pp$ collisions at $\sqrt{s}=7-8$~TeV in 2011-2012, focusing mainly on data collected in 2012.
Measurements of the reconstruction efficiency and of the momentum
scale and resolution, based on large reference samples of $\Jpsimm$,
$\Zmm$ and $\Upsilonmm$ decays, are presented and compared to Monte Carlo
simulations. Corrections to the simulation, to be used
in physics analysis, are provided. Over most of the covered phase
space (muon $|\eta|<2.7$
and $5 \lesssim \pt \lesssim 100$~GeV) the efficiency is above $99\%$ and is
measured with per-mille precision.  The momentum resolution ranges
from $1.7\%$ at central rapidity and for transverse momentum $\pt \simeq 10$~GeV,
to $4\%$ at large rapidity and $\pt \simeq 100$~GeV. The momentum scale is
known with an uncertainty of $0.05\%$ to $0.2\%$  depending on rapidity.
A method for the recovery of final state radiation from the muons is also presented.}

\PreprintJournalName{EPJC}

\begin{document}


\title{Measurement of the muon reconstruction performance of the ATLAS
  detector using 2011 and 2012 LHC proton--proton collision data}

\titlerunning{Measurement of the muon reconstruction performance of the ATLAS detector}

\institute{CERN, CH-1211 Geneva 23. Switzerland}


\author{The ATLAS collaboration}

\abstract{
This paper presents the performance of the ATLAS muon reconstruction
during the LHC run with $pp$ collisions at $\sqrt{s}=7-8$~TeV in 2011-2012, focusing mainly on data collected in 2012.
Measurements of the reconstruction efficiency and of the momentum
scale and resolution, based on large reference samples of $\Jpsimm$,
$\Zmm$ and $\Upsilonmm$ decays, are presented and compared to Monte Carlo
simulations. Corrections to the simulation, to be used
in physics analysis, are provided. Over most of the covered phase
space (muon $|\eta|<2.7$
and $5 \lesssim \pt \lesssim 100$~GeV) the efficiency is above $99\%$ and is
measured with per-mille precision.  The momentum resolution ranges
from $1.7\%$ at central rapidity and for transverse momentum $\pt \simeq 10$~GeV,
to $4\%$ at large rapidity and $\pt \simeq 100$~GeV. The momentum scale is
known with an uncertainty of $0.05\%$ to $0.2\%$  depending on rapidity.
A method for the recovery of final state radiation from the muons is also presented.
}
\maketitle

\section{Introduction}

The efficient identification of muons and the accurate measurement of their momenta are two of the main features of the ATLAS detector~\cite{atlas-det} at the LHC.
These characteristics are often crucial in physics analysis, as for example in  precise measurements of Standard Model processes~\cite{Aad:2011dm,Aad:2014xca,LowMassDY}, in the discovery of the Higgs boson, in the determination of its mass~\cite{Aad:2012tfa,Higgs-mass}, and in searches for physics beyond the Standard Model~\cite{Aad:2012bsa, Aad:2013lna}.
This publication presents the  performance of the ATLAS muon reconstruction during the LHC run at $\sqrt{s}=7-8$~TeV, focusing mainly on data collected in 2012. 
 The performance of the ATLAS muon reconstruction has already been presented in a recent publication~\cite{perf2010} based on 2010 data. The results presented here are based on an integrated luminosity $\approx500$ times larger, which allows a large reduction of the uncertainties.
The measurements of the efficiency, of the momentum scale and resolution are discussed with a particular emphasis on the comparison between data and Monte Carlo (MC) simulation, on the corrections used in the physics analyses and on the associated systematic uncertainties.
 Muons with very large transverse momentum\footnote{ATLAS uses a right-handed coordinate system with its origin at the nominal interaction point (IP) in the centre of the detector and the $z$-axis along the beam pipe. The $x$-axis points from the IP to the centre of the LHC ring, and the $y$-axis points upward. Cylindrical coordinates $(r,\phi)$ are used in the transverse plane, $\phi$ being the azimuthal angle around the beam pipe. The pseudorapidity and the transverse momentum are defined in terms of the polar angle $\theta$ as $\eta=-\ln\tan(\theta/2)$ and $\pt= p \sin \theta$, respectively. The $\eta-\phi$ distance between two particles is defined as  \mbox{$\Delta R=\sqrt{\Delta\eta^2 +\Delta\phi^2}$}.}, $\pt> 120$~GeV, are not treated here as they will be the subject of a forthcoming publication on the alignment of the ATLAS muon spectrometer and its high-$\pt$ performance.

This publication is structured as follows: Sect.~\ref{Sec:MuonReco} gives a short description of muon detection in ATLAS and Sect.~\ref{Sec:Datasets} describes the real and simulated data samples used in the performance analysis. The measurement of the reconstruction efficiency is described in Sect.~\ref{Sec:Efficiency} while Sect.~\ref{Sec:ScaleResolution} reports the momentum scale and resolution. A method for including photons from final-state radiation in the reconstruction of the muon kinematics, is described in Sect.~\ref{Sec:FSR}.  Conclusions are given in Sect.~\ref{sec:conclusions}.

\section{Muon identification and reconstruction}\label{Sec:MuonReco}

A detailed description of the ATLAS detector can be found elsewhere~\cite{atlas-det}.  The ATLAS experiment uses the information from the muon spectrometer (MS) and from the inner detector (ID) and, to a lesser extent, from the calorimeter, to identify and precisely reconstruct muons produced in the $pp$ collisions.

The MS is the outermost of the ATLAS sub-de\-tectors: it is designed to detect charged particles in the pseudorapidity  region up to $|\eta| = 2.7$, and to provide momentum measurement with a relative resolution better than 3\% over a wide $\pt$ range and up to 10\% at $\pt \approx 1$~TeV. The MS consists of one barrel part (for $|\eta| < 1.05$) and two end-cap sections. A system of three large superconducting air-core toroid magnets provides a
 magnetic field with a bending integral of about $2.5$~Tm in the barrel and up to $6$~Tm in the end-caps. 
Triggering and $\eta$, $\phi$ position measurements, with typical spatial resolution of $5-10$~mm, are provided by the Resistive Plate Chambers (RPC, three doublet layers for $|\eta| < 1.05$) and by the Thin Gap Chambers (TGC, three triplet and doublet layers for $1.0<|\eta| < 2.4$). Precise muon momentum measurement is possible up to $|\eta|=2.7$ and it is provided by three layers of Monitored Drift Tube Chambers (MDT), each chamber providing six to eight $\eta$ measurements along the muon track. For $|\eta|>2$ the inner layer is instrumented with a quadruplet of Cathode Strip Chambers (CSC) instead of MDTs.  The single hit resolution in the bending plane for the MDT and the CSC is about $80$~$\mu$m and $60$~$\mu$m, respectively. Tracks in the MS are reconstructed in two steps: first local track segments are sought within each layer of chambers and then local track segments from different layers are combined into full MS tracks.

The ID provides an independent measurement of the muon track close to the interaction point. It consists of three sub-detectors: the Silicon Pixels and the Semi-Con\-ductor Tracker (SCT) detectors for $|\eta| < 2.5$ and the Transition Radiation Tracker (TRT) covering $|\eta| < 2.0$. They provide high-resolution coordinate measurements for track reconstruction inside an axial magnetic field of $2$~T. A track in the barrel region has typically 3 Pixel hits, 8 SCT hits, and approximately 30 TRT hits.

The material between the interaction point and  the MS  ranges approximately from 100 to 190 radiation lengths, depending on $\eta$, and consists mostly of  calorimeters. 
The sampling liquid-argon (LAr) electromagnetic calorimeter covers $|\eta|<3.2$ and is surrounded by hadronic calorimeters based on iron and scintillator tiles for $|\eta|\lesssim 1.5$ and on LAr for larger values of $|\eta|$.
 
Muon identification is performed according to several reconstruction criteria (leading to different muon ``types''), according to the available information from the ID, the MS, and the calorimeter sub-detector systems. The different types are:

\begin{itemize}
\item Stand-Alone (SA) muons: the muon trajectory is reconstructed only in the MS. The parameters of the muon track at the interaction point are determined by extrapolating the track back to the point of closest approach to the beam line, taking into account  the estimated energy loss of the muon in the calorimeters. In general the muon has to traverse at least two layers of MS chambers to provide a track measurement. SA muons are mainly used to extend the acceptance to the range $2.5<|\eta|<2.7$ which is not covered by the ID;

\item Combined (CB) muon: track reconstruction is performed independently in the ID and MS, and a combined track is formed from the successful combination of a MS track with an ID track. This is the main type of reconstructed muons; 

\item Segment-tagged (ST) muons: a track in the ID is classified as a muon if, once extrapolated to the MS, it is associated with at least one local track segment in the MDT or CSC chambers. ST muons can be used to increase the acceptance in cases in which the muon crossed only one layer of MS chambers, either because of its low $\pt$ or because it falls in regions with reduced MS acceptance;

\item Calorimeter-tagged (CaloTag) muons: a track in the ID is identified as a muon if it could be associated to an energy deposit in the calorimeter compatible with a minimum ionizing particle. This type has the lowest purity of all the muon types but it recovers acceptance in the uninstrumented regions of the MS. The identification criteria of this muon type are optimized for a region of $|\eta|<0.1$ and a momentum range of $25\lesssim \pt \lesssim 100$~GeV.
\end{itemize}
CB candidates have the highest muon purity. The reconstruction of tracks in the spectrometer, and as a consequence the SA and CB muons, is affected by acceptance losses mainly in two regions:
 at $\eta\approx 0$, where the MS is only partially equipped with muon chambers in order to provide space for the services for the ID and the calorimeters, and 
 in the region ($1.1 < \eta< 1.3$) between the barrel and the positive $\eta$ end-cap, where there are regions in $\phi$ with only one layer of chambers traversed by muons in the MS, due to the fact that some of the chambers of that region were not yet installed\footnote{The installation of all the muon chambers in this region has been completed during the 2013-2014 LHC shutdown.}.

The reconstruction of the SA, CB and ST muons  (all using the MS information) has been performed using two independent reconstruction software packages, implementing different strategies~\cite{cscbook} (named ``Chains'')  both for the reconstruction of muons in the MS and for the ID-MS combination. For the ID-MS combination,  the first chain (``Chain 1'')  performs a statistical combination of the track parameters of the SA and ID muon tracks using the corresponding covariance matrices. The second (``Chain 2'') performs a global refit of the muon track using the hits from both the ID and MS sub-detectors. The use of two independent codes provided redundancy and robustness in the ATLAS commissioning phase. A unified reconstruction programme  (``Chain 3'')  has been developed to incorporate the best features of the two chains and has been used, in parallel to the other two, for the reconstruction of 2012 data. It is planned to use only Chain~3 for future data taking. So far, the first two chains were used in all ATLAS publications. As the three chains have similar performance, only results for ``Chain 1''  are shown in the present publication. A summary of the results for the other two chains is reported in Appendix~\ref{appendix:A}.

The following  quality requirements are applied to the ID tracks used for CB, ST or CaloTag muons:
\begin{itemize}
\item at least 1 Pixel hit;
\item at least 5 SCT hits;
\item at most 2 active Pixel or SCT sensors traversed by the track but without hits;
\item in the region of full TRT acceptance, $0.1<|\eta|<1.9$,  at least 9 TRT hits.
\end{itemize}
The number of hits required in the first two points is reduced by one if the track traverses a sensor known to be inefficient according to a time-dependent database. 
The above requirements are dropped in the region  $|\eta|>2.5$,  where short ID track segments can be matched to SA muons to form a CB muon.


\section{Data and Monte Carlo Samples}\label{Sec:Datasets}

\subsection{Data Samples}\label{sec:data_samples}
The results presented in this article are mostly obtained from the analysis of $\sqrt{s}=8$~TeV $pp$ collision events corresponding to an integrated luminosity of $20.3$~fb$^{-1}$ collected by the ATLAS detector in 2012. 
Results from $pp$ collisions at $\sqrt{s}=7$~TeV, collected in 2011, are presented in Appendix~\ref{appendix:B}.
Events are accepted only if the ID, the MS and the calorimeter detectors were operational and both solenoid and toroid magnet systems were on. 

The online event selection was performed by a three-level trigger system described in Ref.~\cite{Aad:2012xs}. The performance of the ATLAS muon trigger during the 2012 data taking period is reported in Ref.~\cite{Muontrigger2012}.
The $Z\to\mu\mu$ candidates have been selected online by requiring at least one muon candidate with \mbox{$\pt>24$}~GeV, isolated from other activity in the ID.
The $\Jpsimm$ and the $\Upsilonmm$  samples used for momentum scale and resolution studies have been selected online with two dedicated dimuon triggers that require two opposite-charge muons compatible with the same vertex, with transverse momentum $\pt>6$~GeV, and the dimuon invariant mass in the range 2.5-4.5 GeV for the $J/\psi$ and  8-11 GeV for the $\Upsilon$ trigger. The $\Jpsimm$ sample used for the efficiency measurement was instead selected using a mix of single-muon triggers and a dedicated trigger requiring a muon with  $\pt>6$~GeV and an  ID track with $\pt > 3.5$~GeV, such that the invariant mass of the muon+track pair, under a muon mass hypothesis, is in the window $2.7-3.5$~GeV. This dedicated trigger operated during the whole data taking period with a prescaled rate of $\approx 1$~Hz.

\subsection{Monte Carlo Samples}\label{sec:mc_samples}

Monte Carlo samples for the process $pp \rightarrow (Z/\gamma^*) X \rightarrow \mu^+\mu^- X$, called $\Zmm$ in the following, were generated using POWHEG~\cite{Alioli:2010xd} interfaced to \linebreak PYTHIA8 \cite{Pythia8}.  The CT10~\cite{Lai:2010vv} parton density functions (PDFs) have been used. The PHOTOS~\cite{Golonka:2005pn} package has been used to
simulate final state photon radiation (FSR), using the exponentiated mode that leads to multi-photon emission taking
into account $\gamma^*$ interference in $Z$ decays. 
To improve the description of the dimuon invariant mass distribution, the generated lineshape was reweighted using 
an improved Born approximation with a  running-width definition of the $Z$ lineshape parameters.  The ALPGEN~\cite{alpgen} generator, interfaced with PYTHIA6~\cite{Sjostrand:2006za}, was also used to generate alternative $\Zmm$ samples.

Samples of prompt $\Jpsimm$ and of $\Upsilonmm$ were generated using PYTHIA8, complemented with  PHOTOS to simulate the effects of final state radiation. The samples were generated requiring each muon to have $\pt>6.5$($6$) GeV for $J/\psi$ ($\Upsilon$). The $J/\psi$ distribution in rapidity and transverse momentum has been reweighted in the simulated samples to match the distribution observed in the data.
The samples used for the simulation of the backgrounds to $\Zmm$ are described in detail in~\cite{Aad:2012en
}, they include $Z\to\tau\tau$, $W\to\mu\nu$ and $W\to\tau\nu$, generated with POWHEG, $WW$, $ZZ$ and $WZ$ generated with SHERPA~\cite{sherpa},  $t\bar{t}$ samples generated with MC@NLO~\cite{Frixione:2002ik}  and  $b\bar{b}$ as well as $c\bar{c}$ samples generated with PYTHIA6.

All the generated samples were passed through the simulation of the ATLAS detector based on GEANT4~\cite{simuAtlas,GEANT4} and were reconstructed with the same programs used for the data. 
The ID and the MS were simulated with an ideal geometry without any misalignment. To emulate the effect of the misalignments of the MS chambers in real data, the reconstruction of the muon tracks in the simulated samples was performed  using a random set of MS alignment constants.  The amount of random smearing applied to these alignment constants was derived from an early assessment of the precision of the alignment, performed with special runs in which the toroidal magnetic field was off.  The knowledge of the alignment constants improved with time. In particular the alignment constants used for the reconstruction of the data were more precise than those used to define the random smearing applied in the simulation, resulting in some cases in a worse MS resolution in MC than in data.

\section{Efficiency}
\label{Sec:Efficiency}

The availability of two independent detectors to reconstruct the muons
(the ID and the MS) enables a precise determination of the muon reconstruction efficiency in the region $|\eta|<2.5$. This is obtained with the
so called tag-and-probe method described in the next section. A different methodology, described in Sect.~\ref{sec::muon_reco_forward}, is used in the region  $2.5<|\eta|<2.7$ in which only one detector (the MS) is available.

\subsection{Muon reconstruction efficiency in the region $|\eta|<2.5$ }
\label{sec::muon_reco_central}

The tag-and-probe method  is employed to measure the reconstruction efficiencies of all muon types within the acceptance of the ID ($|\eta|<2.5$).
The conditional probability that a muon reconstructed by the ID is also reconstructed using the MS as a particular muon type,
$P (\textrm{Type} | \textrm{ID})$, with $\textrm{Type} = (\textrm{CB, ST})$, can be measured using ID probes. Conversely, the conditional probability  that a muon reconstructed by the MS is also reconstructed in the ID, $P (\textrm{ID} | \textrm{MS})$, is measured using MS tracks as probes.

For each muon type, the total reconstruction efficiency is given by:
\begin{equation} \label{eq:TagProbeEffiDef}
  \varepsilon(\textrm{Type}) = \varepsilon(\textrm{Type} | \textrm{ID}) \cdot \varepsilon(\textrm{ID})\;\; ,
\end{equation}
where $\varepsilon(\textrm{ID})$ is the probability that a muon is reconstructed as an ID track.
The quantity $\varepsilon(\textrm{ID})$ cannot be measured directly and is replaced by  $\varepsilon(\textrm{ID} | \textrm{MS})$ to give the tag-and-probe approximation:
\begin{equation} \label{eq:TagProbeEffiDef2}
  \varepsilon(\textrm{Type}) \simeq \varepsilon(\textrm{Type} | \textrm{ID}) \cdot \varepsilon(\textrm{ID}|\textrm{MS})\;\; .
\end{equation}
The level of agreement of the measured efficiency, \linebreak $\varepsilon^{{\textrm{Data}}}(\textrm{Type}),$ with the efficiency measured with the same method in MC,  $\varepsilon^{{\textrm{MC}}}(\textrm{Type})$, is expressed as the ratio between these two numbers, called ``efficiency scale factor'' or SF:
\begin{equation}
  SF=\frac{\varepsilon^{\textrm{Data}}(\textrm{Type})}{\varepsilon^{\textrm{MC}}(\textrm{Type})} \textrm{.}
\end{equation}
Possible biases introduced by the tag-and-probe approximation and other systematic effects on the efficiency measurement, which appear both in data and in MC, cancel in the SF. 
The SF is therefore used to correct the simulation in physics analysis.

\subsubsection{The tag-and-probe method with $\Zmm$ events}
\label{sec::T&P}
For $Z\to\mu\mu$ decays, events are selected by requiring two oppositely charged isolated muons\footnote{Here a muon is considered to be isolated when the sum of the momenta of the other tracks with $\pt>1$~GeV   in a cone of $\Delta R=0.4$ around the muon track is less than $0.15$ times the muon momentum itself. Different cone sizes and cuts on the momentum fraction are used in other parts of this paper.} with transverse momenta of at least $\pt> 25 $ and $10$~GeV respectively and a dimuon invariant mass within $10$~GeV of the $Z$-boson mass. The muons are required to be back to back in the transverse plane ($\Delta \phi > 2$). One of the muons is required to be a CB muon, and to have triggered the readout of the event. This muon is called the ``tag''.
The other muon, the so-called ``probe'', is required to be a MS track (i.e. a SA or a CB muon) when $\varepsilon(\textrm{ID} | \textrm{MS})$ is to be measured. 
The probe is required to be a CaloTag muon for the measurement of  $\varepsilon (\textrm{Type} | \textrm{ID})$. The use of CaloTag muons as the ID probes reduces the background in the $\Zmm$ sample by an order of magnitude without biasing the efficiency measurement.
The MS probes are also used to measure the  efficiency of CaloTag muons.
After selecting all tag-probe pairs, an attempt is made to match the probe to a reconstructed muon: a match is successful when the muon and the probe are close in the $\eta-\phi$ plane ($\Delta R<0.01$ for CaloTag probes to be matched with CB or ST muons and  $\Delta R<0.05$ for MS probes to be matched to ID or CaloTag muons).

\subsubsection{Background treatment in $\Zmm$ events}
\label{sec::T&Pbg}
\begin{figure}[hbt]
\begin{center}
 \includegraphics[width=0.9\linewidth]{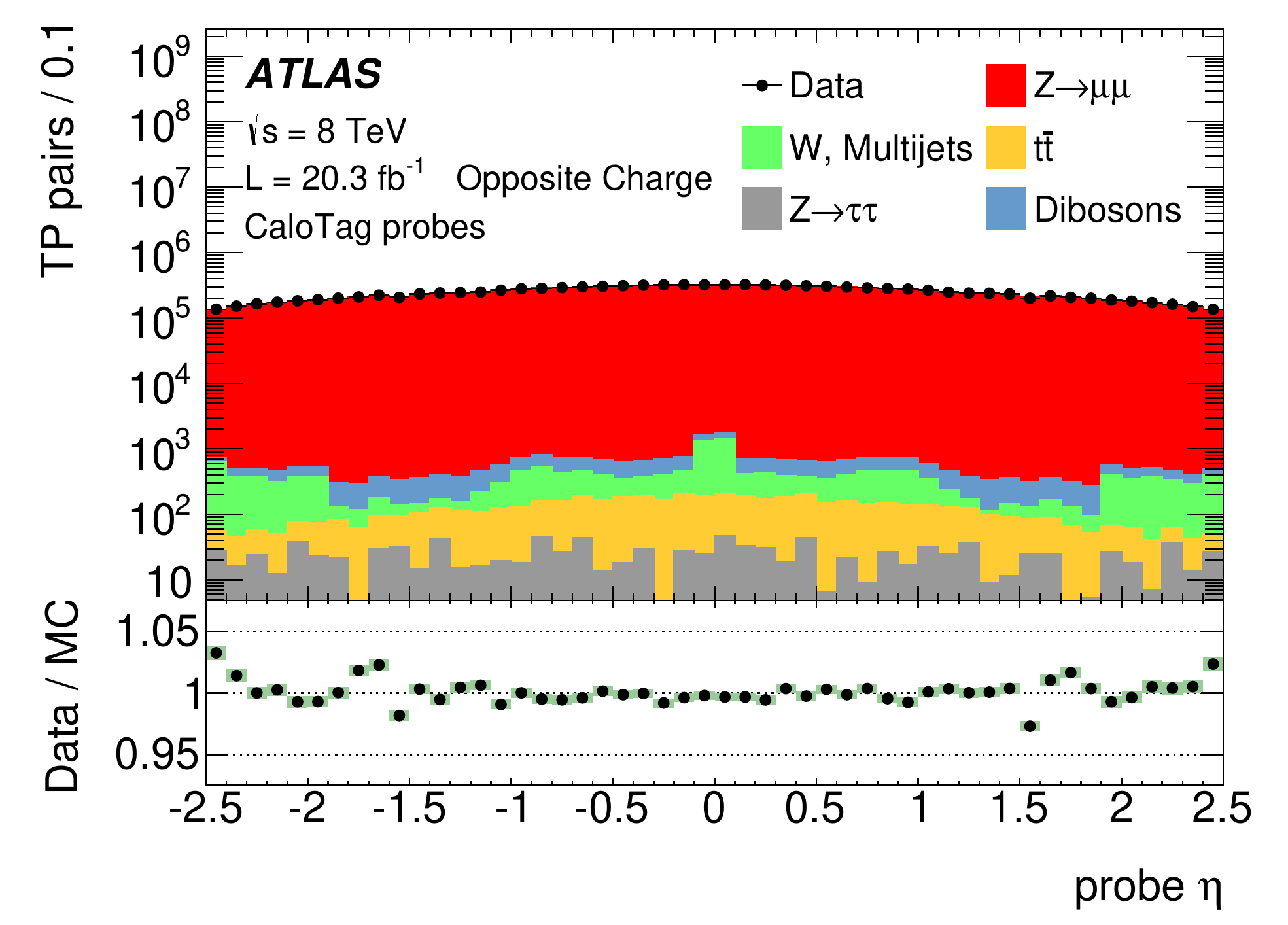}
  \includegraphics[width=0.9\linewidth]{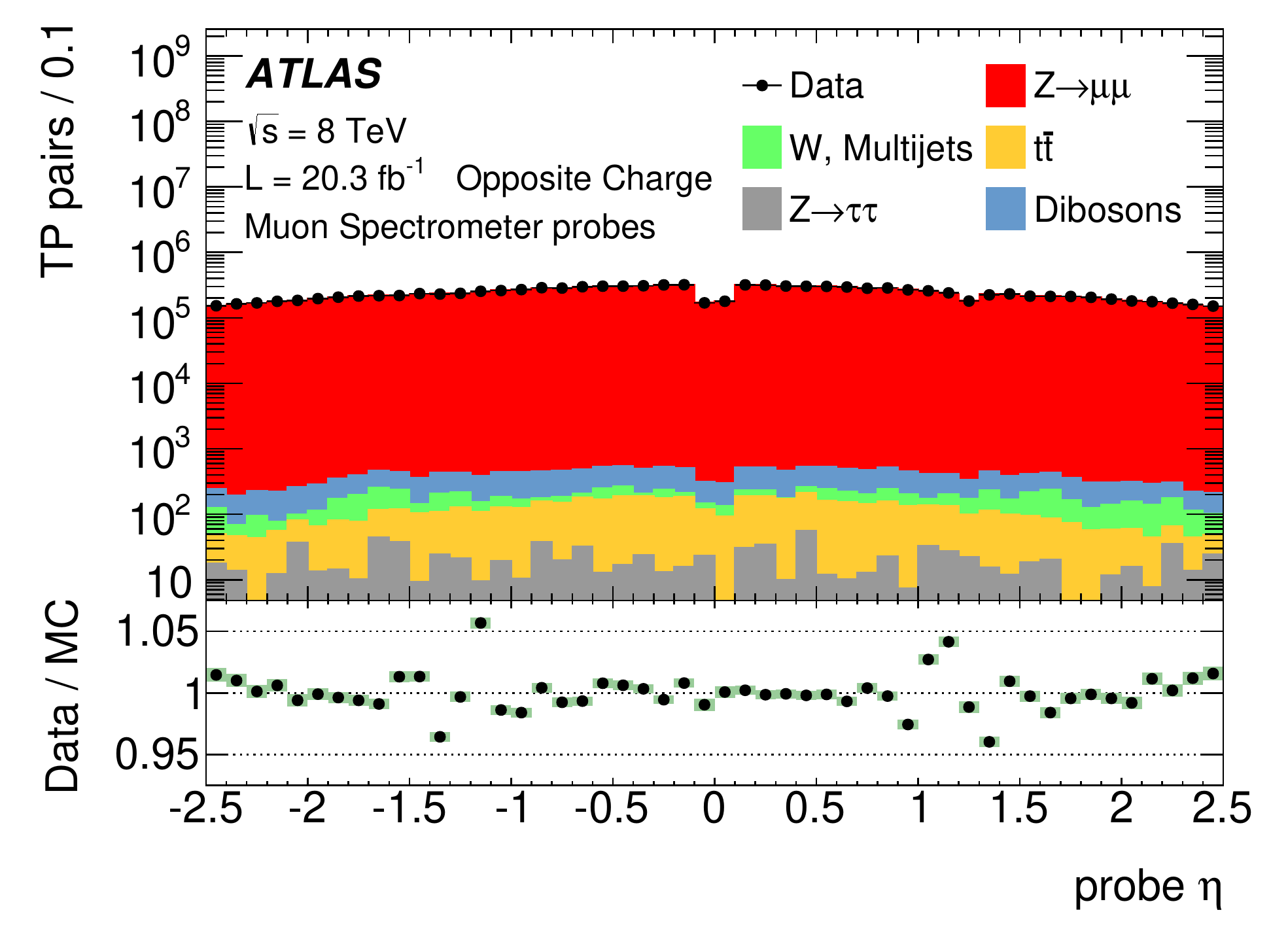}
    \caption{Pseudorapidity distribution of the CaloTag (top) or MS (bottom) probes used in the tag-and-probe analysis.
    The bottom panel shows the ratio between observed and expected counts. The sum of the MC samples is normalized to the number of events in the data. The green band represents the statistical uncertainty.}\label{Fig:ProbeEtaTagProbe}
\end{center}
\end{figure}

Apart from $Z\to\mu\mu$ events, a small fraction of the selected tag-probe pairs may come from other sources.
For a precise efficiency measurement, these backgrounds have to be estimated and subtracted.
Contributions from $Z\to\tau\tau$ and $t\bar{t}$ decays are estimated using MC simulation.
Additionally, QCD multijet events and $W\to\mu\nu$ decays in association with jet activity ($W+$jets) can yield tag-probe pairs through secondary muons from
heavy- or light-hadron decays. As these backgrounds are approximately charge-symmetric, they are estimated from the data using same-charge (SC) tag-probe pairs.
This leads to the following estimate of the \linebreak opposite-charge (OC) background for each region of the kinematic phase-space: 
\begin{equation}
N\left(\text{Bkg}\right) = N_{\text OC}^{Z,t\bar{t}\text{ MC}} + T \cdot \left( N_{\text{SC}}^{\text{Data}} - N_{\text{SC}}^{Z,t\bar{t}\text{ MC}} \right)
\end{equation}
where $N_{\text OC}^{Z,t\bar{t}\text{ MC}}$ is the contribution from  $Z\to\tau\tau$ and $t\bar{t}$ decays, $N_{\text{SC}}^{\text{Data}}$ is the number of SC pairs measured in data and $N_{\text{SC}}^{Z,t\bar{t}\text{ MC}}$ is the estimated contribution of the  $Z\to\mu\mu$, $Z\to\tau\tau$ and $t\bar{t}$ processes to the SC sample.
 $T$ is a global transfer factor that takes into account the residual charge asymmetry of the QCD multijet and W+jets samples, estimated using the simulation:
\begin{equation}
T = 1 + \theta;\phantom{aaaa} \theta = \frac{N_{\text{OC}}^{\text{QCD+W MC}} - N_{\text{SC}}^{\text{QCD+W MC}}}{N_{\text{SC}}^{\text{Data}}} .
\label{eq:transfer_factor}
\end{equation}
For the kinematic region covered by the measurement, the transfer factor is  $T=1.15$ for CaloTag probes. For the MS probes the misidentification rate is low
and the residual  QCD multijet background has a large contribution from oppositely charged muon pairs in $b\bar{b}$ decays, leading to $T=2.6$.
The efficiency for finding a muon of type $\text{A}$ given a probe of type $\text{B}$, corrected for the effect of background, can then be computed as:
\begin{equation}
\varepsilon(\text{A}|\text{B}) = \frac{N^{\rm Match}_{\rm Probes}\left(\text{Data}\right) - N^{\rm Match}_{\rm Probes}\left(\text{Bkg}\right)} {N^{\rm All}_{\rm Probes}\left(\text{Data}\right) - N^{\rm All}_{\rm Probes}\left(\text{Bkg}\right)}\textrm{,}
\end{equation}
where $N^{\rm All}_{\rm Probes}$ stands for the total number of probes considered and $N^{\rm Match}_{\rm Probes}$ is the number of probes successfully matched to a reconstructed muon of type  $\text{A}$.
According to the background estimate reported above, the sample of selected CaloTag probes is more than $99.5\%$ pure in $\Zmm$ decays, as shown in Fig.~\ref{Fig:ProbeEtaTagProbe}. The $\Zmm$ purity is maximal for muon $\pt \simeq 40$~GeV and decreases to $98.5\%$ ($97\%$) for $\pt=10$ ($100$)~GeV.  The $\Zmm$ purity  has a weak dependence on the average number of  inelastic $pp$ interactions per bunch crossing, $\langle \mu \rangle$, decreasing from $99.8\%$ at  $\langle \mu \rangle = 10$ to $99.5\%$ at $\langle \mu \rangle=34$. A purity above $99.8\%$ is obtained in the selection of MS probes, with weaker dependence on $\pt$ and $\langle \mu \rangle$.

\subsubsection{Low $\pt$ efficiencies from $\Jpsimm$ decays}\label{Sec:LowPtEff}

The efficiencies extracted from $Z\to\mu\mu$ decays are complemented at low $\pt$ with results derived from a sample of $J/\psi\to\mu\mu$ events.
In 2012 ATLAS collected approximately 2M $J/\psi\to\mu\mu$ decays which were not biased by dimuon triggers requirements, using a combination of single muon triggers (isolated and non-isolated) and the dedicated ``muon + track'' trigger described in Sect.~\ref{sec:data_samples}.

The analysis proceeds in a similar manner to the \linebreak $\Zmm$ with some modifications due to the different kinematics of the $J/\psi$. Tags are required to be CB muons with $\pt>4$~GeV and $|\eta|<2.5$. As with the $Z$, the tag must have triggered the read-out of the event. Probes are sought from amongst the ID tracks and must have $\pt > 2.5$~GeV and $|\eta| < 2.5$, opposite charge to the tag muon, and must form with the tag an invariant mass in the window $2.7-3.5$~GeV.
Finally the tag-probe pairs must fit to a common vertex with a very loose quality cut of $\chi^2 < 200$ for one degree of freedom, which removes tracks from different vertices, without any significant efficiency loss.
Muon reconstruction efficiencies are then derived by binning in small cells of $\pt$ and $\eta$ of the probe tracks. Invariant mass distributions are built in each cell for two samples:  (a) all tag-probe pairs and (b) tag-probe pairs in which the probe failed to be reconstructed in the MS. The invariant mass distributions are fitted with a signal plus background model  to obtain the number of $J/\psi$ signal events in the two samples, called  $N_a(\pt,\eta)$ and  $N_b(\pt,\eta)$, respectively. The fit model is a Gaussian plus a second order polynomial for the background. The two samples are fitted simultaneously using the same mean and width to describe the signal. The MS reconstruction efficiency in a given $(\pt,\eta)$ cell is then defined as:
\begin{equation}
\varepsilon_{\pt, \eta}(\mathrm{Type}|\mathrm{ID}) = 1 - \frac{N_b(\pt,\eta)}{N_a(\pt,\eta)}\textrm{.}
\end{equation}
The largest contribution to the systematic uncertainty originates from the model used in the fit. This uncertainty was estimated by changing the background model to a first or a third order polynomial and by relaxing the constraint that the mass and the width of the $J/\psi$ signal are the same between the two samples. The resulting variations in the efficiency are added in quadrature to the statistical uncertainty to give the total uncertainty on the efficiency.
The efficiency integrated over the full $\eta$ region is obtained as an average of the efficiencies of the different $\eta$ cells. This method ensures a reduced dependency on local variations of background and resolution, and on the kinematic distribution of the probes.

\subsubsection{Systematic Uncertainties}
\label{sec::T&Psys}
The main contributions to the systematic uncertainty on the measurement of the efficiency SFs are shown in Fig.~\ref{Fig:sfsyste_staco_st}, as a function of $\eta$ and $\pt$, and are discussed below (the labels in parenthesis refer to the legend of Fig.~\ref{Fig:sfsyste_staco_st}):
\begin{figure}[!htb]
\begin{center}
  \includegraphics[width=0.9\linewidth]{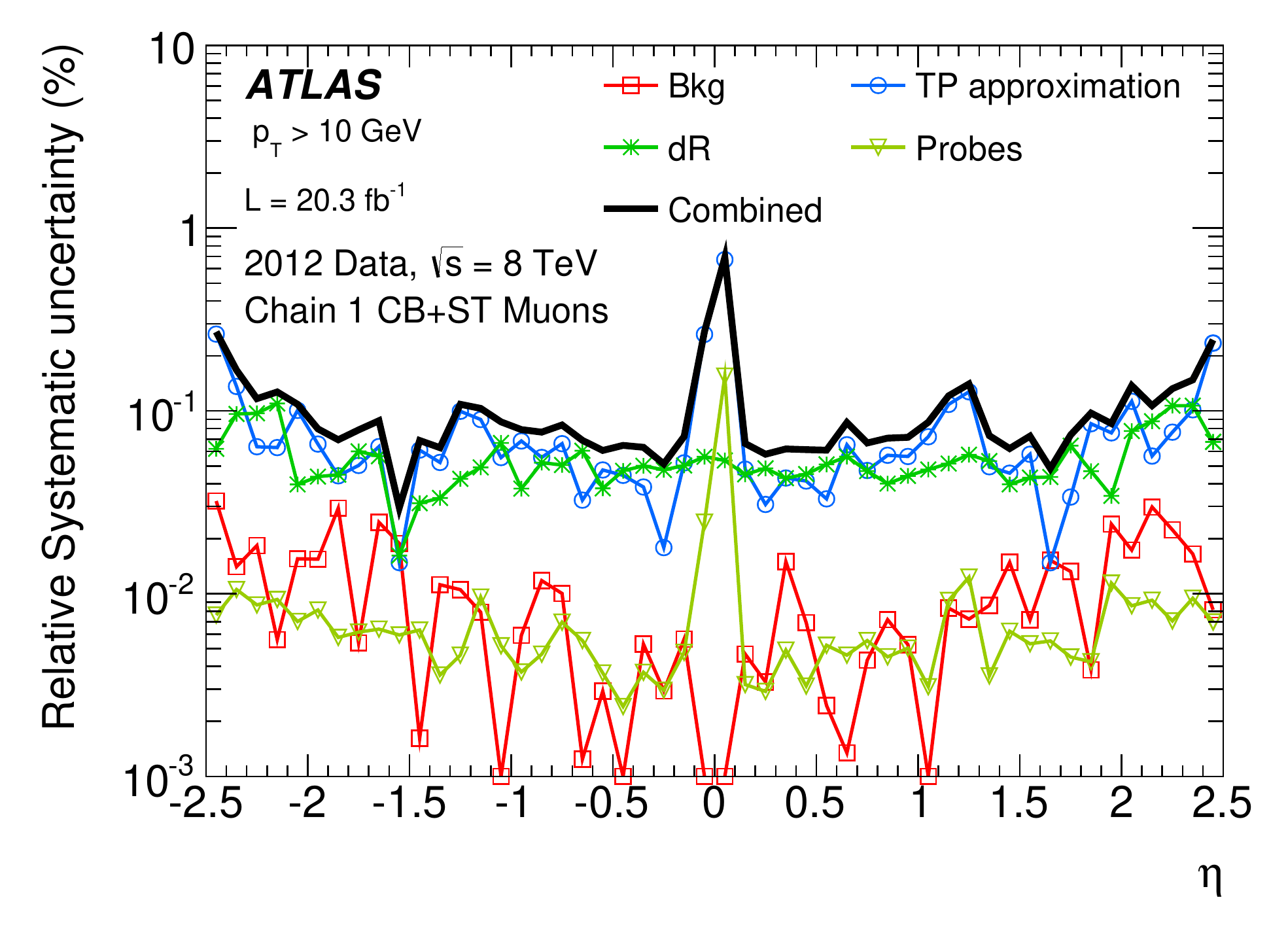}
  \includegraphics[width=0.9\linewidth]{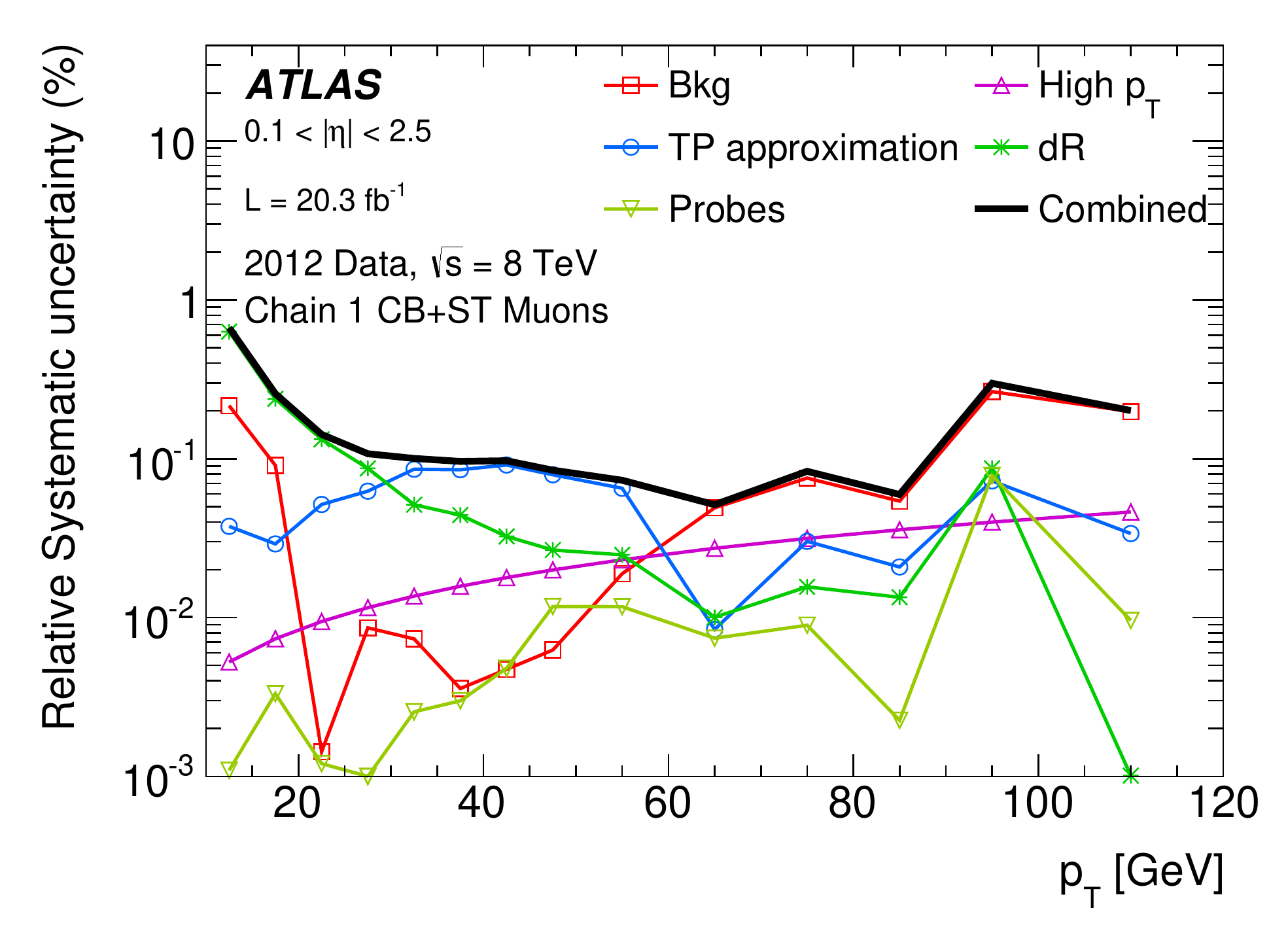}
  \caption{Systematic uncertainty on the efficiency scale factor for  CB+ST muons, obtained from $\Zmm$ data, as a function of $\eta$ (top) and $\pt$ (bottom) for muons with $\pt>10$~GeV.  The background systematic uncertainty in the last two bins of the bottom plot is affected  by  a large statistical uncertainty. The combined systematic uncertainty is the sum in quadrature of the individual contributions.}\label{Fig:sfsyste_staco_st}
\end{center}
\end{figure}
\begin{itemize}
\item{(Bkg)} the uncertainty on the data-driven background estimate is evaluated by varying the charge-asymmetry parameter $\theta$ of Eq.~\ref{eq:transfer_factor}
by $\pm 100\%$. This results in an uncertainty of the efficiency measurement below $0.1\%$  in a large momentum range, reaching up to $0.2\%$ for low muon momenta where the contribution of the background is most significant. 
\item{(dR)} the choice of the cone size used for matching reconstructed muons to probe objects has been optimized to minimize the amount of matches with wrong tracks while keeping the maximum match efficiency for correct tracks. A systematic uncertainty is evaluated by varying the cone size by $\pm50\%$. This yields an uncertainty of $\approx 0.1\%$.
\item{(TP approximation)}
possible biases in  the tag-and-probe method, for example due to different distributions between MS probes and ``true'' muons or due to correlation between ID and MS efficiencies, are investigated. The simulation is used to compare the efficiency measured with the tag-and-probe method with the ``true'' MC efficiency calculated  as the fraction of generator-level muons that are successfully reconstructed.  Agreement within less than $0.1\%$ is observed, with the exception of the region $|\eta|<0.1$.  In the extraction of the data/MC scale factors, the difference between the measured and the ``true'' efficiency cancels  to first order.
To take into account possible imperfection of the simulation, half the observed difference is used as an additional systematic uncertainty  on the SF.
\item{(Probes)} the scale factor maps may be sensitive to disagreements between data and simulation in the kinematic distributions of the probes. The corresponding systematic uncertainty is estimated by reweighting the  distribution of the probes in the simulation to bring it into agreement with the data. The resulting effect on the efficiency is below $0.1\%$ over most of the phase space.
\item{(Low $\pt$)}
for $4<\pt<10$~GeV the systematic uncertainties are obtained from the analysis performed with the  $\Jpsimm$ sample, as discussed in Sec.\ref{Sec:LowPtEff} (not shown in  Fig.~\ref{Fig:sfsyste_staco_st}).
The resulting uncertainty on the low-$\pt$ SFs ranges between 0.5\% and 2\%, depending on $\pt$ and $\eta$ and is dominated by the uncertainty on the background model.
\item{(High $\pt$)}
no significant dependence of  the measured SFs with $\pt$ was observed  in the momentum range considered. An upper limit on the SF variation for large muon momenta has been extracted by using a MC simulation with built-in imperfections, including a realistic residual misalignment of the detector components or a 10\% variation of the muon energy loss. On the basis of this, a systematic uncertainty of $\pm 0.42\% \times (\pt / 1$~TeV~$)$ is obtained.
\end{itemize}

\subsubsection{Results}
\label{sec::T&Pres}

\begin{figure}[bh]
\begin{center}
  {\includegraphics[width=0.99\linewidth]{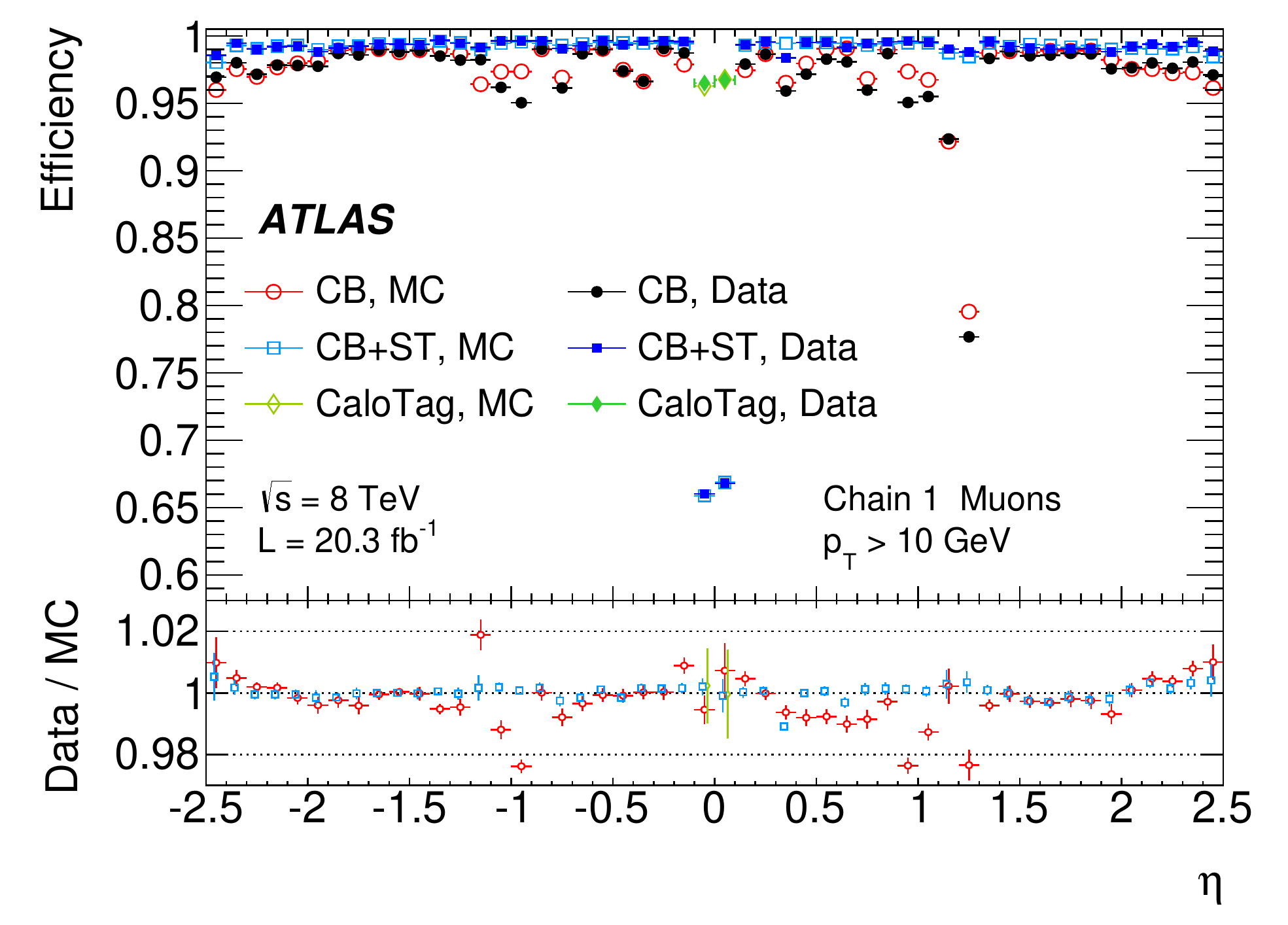}}
    \caption{Muon reconstruction efficiency as a function of $\eta$ measured in $\Zmm$ events for muons with $\pt>10$~GeV and different muon reconstruction types. 
    CaloTag muons are only shown in the region $|\eta|<0.1$, where they are used in physics analyses. The error bars on the efficiencies indicate the statistical uncertainty. The panel at the bottom shows the ratio between the measured and predicted efficiencies. The error bars on the ratios are the combination of statistical and systematic uncertainties.}\label{Fig:AllTypeEff_staco}
\end{center}
\end{figure}

Figure~\ref{Fig:AllTypeEff_staco} shows the muon reconstruction efficiency $\varepsilon(\textrm{Type})$ as a function of $\eta$ as measured from $\Zmm$ events. The combination of all the muon reconstruction types (for CB, ST, and CaloTag muons) gives a uniform muon reconstruction efficiency of about $99\%$ over most the detector regions. The use of ST muons allows the recovery of efficiency especially in the region $1.1<\eta < 1.3$ (from $85\%$ to $99\%$) in which part of the MS chambers were not installed, as discussed in Sect.~\ref{Sec:MuonReco}. The remaining inefficiency of the combination of CB or ST muons (CB+ST) at $|\eta|<0.1$ ($66\%$) is almost fully recovered by the use of CaloTag muons ($97\%$).

The efficiencies measured in experimental and simulated data are in good agreement, in general well within 1\%. The largest differences are observed in the CB muons. To reconstruct an MS track, the Chain~1 reconstruction requires track segments in at least two layers of precision chambers (MDT or CSC) and at least one measurement of the $\phi$ coordinate from trigger chambers (RPC or TGC). These requirements introduce some dependency on detector conditions and on the details of the simulation in the regions in which only two layers of precision chambers or only one layer of trigger chambers are crossed by the muons. This results in a reduction of efficiency in data with respect to MC of approximately 1\% in the region of $\eta \sim 0.5$ due  the RPC detector conditions and to local deviations up to about 2\% at $0.9<|\eta|<1.3$ related to imperfections in the simulation of the barrel-endcap transition region.  For the CB+ST muons the agreement between data and MC is very good, with the only  exception of a 
low-efficiency region in data at $\eta = 0.3-0.4$ related to an inactive portion of an MDT chamber (not included in MC) in a region with reduced coverage due to the supporting structure of the ATLAS detector\footnote{This effect is also visible in Fig.~\ref{Fig:2deff_stacost_data} at $\phi \simeq -1$.}.

\begin{figure}[!tbh]
\begin{center}
    \includegraphics[width=0.9\linewidth]{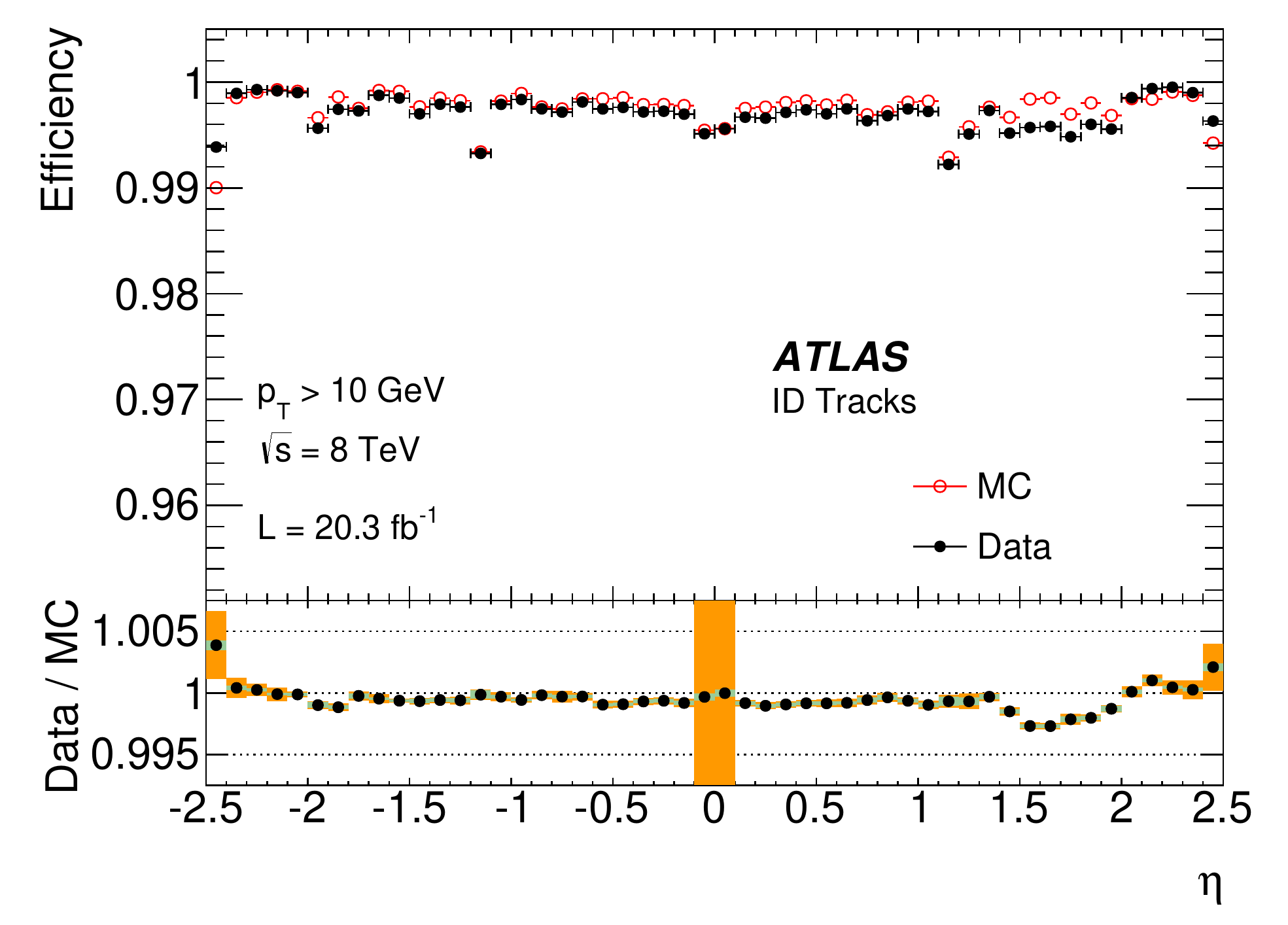}
    \includegraphics[width=0.9\linewidth]{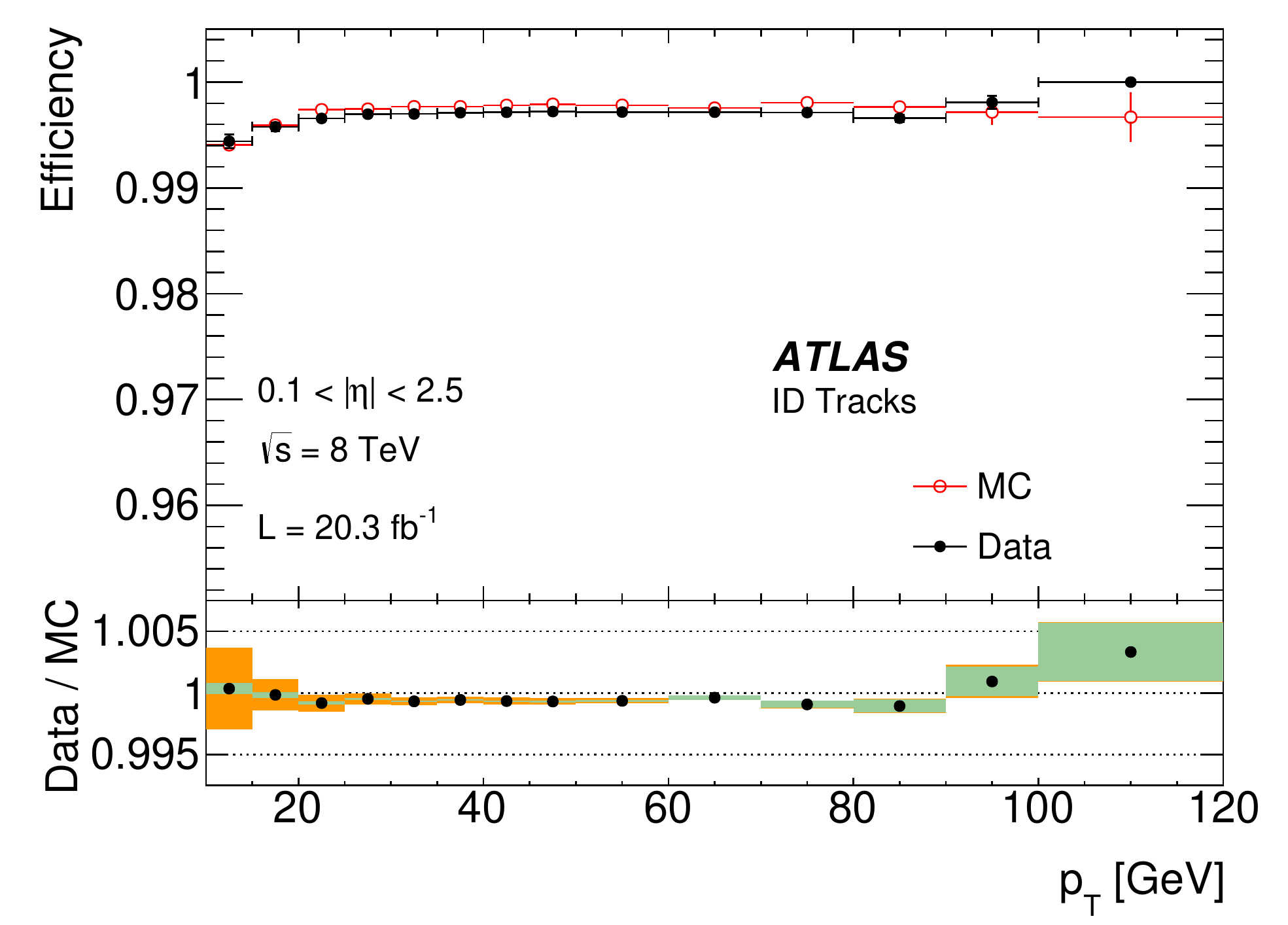}
    \caption{ID muon reconstruction efficiency as a function of $\eta$ (top) and $\pt$ (bottom) measured in $\Zmm$ events for muons with $\pt>10$~GeV. The error bars on the efficiencies indicate the statistical uncertainty. The panel at the bottom shows the ratio between the measured and predicted efficiencies. The green areas depict the pure statistical uncertainty, while the orange areas also include systematic uncertainties.}\label{Fig:IDeff_b_layer}
\end{center}
\end{figure}

The ID muon reconstruction efficiency,  $\varepsilon(\textrm{ID}|\textrm{MS})$, for $\pt>10$~GeV as a function of $\eta$ and $p_{T}$ is shown in Fig.~\ref{Fig:IDeff_b_layer}. The efficiency is greater than $0.99$ and there is very good agreement between data and MC. The small efficiency reduction in the region $1.5<\eta<2$ is related to temporary hardware problems in the silicon detectors. The larger uncertainty at $|\eta|<0.1$ is related to the limited MS coverage in that region.

\begin{figure}[!h]
\begin{center}
\includegraphics[width=0.9\linewidth]{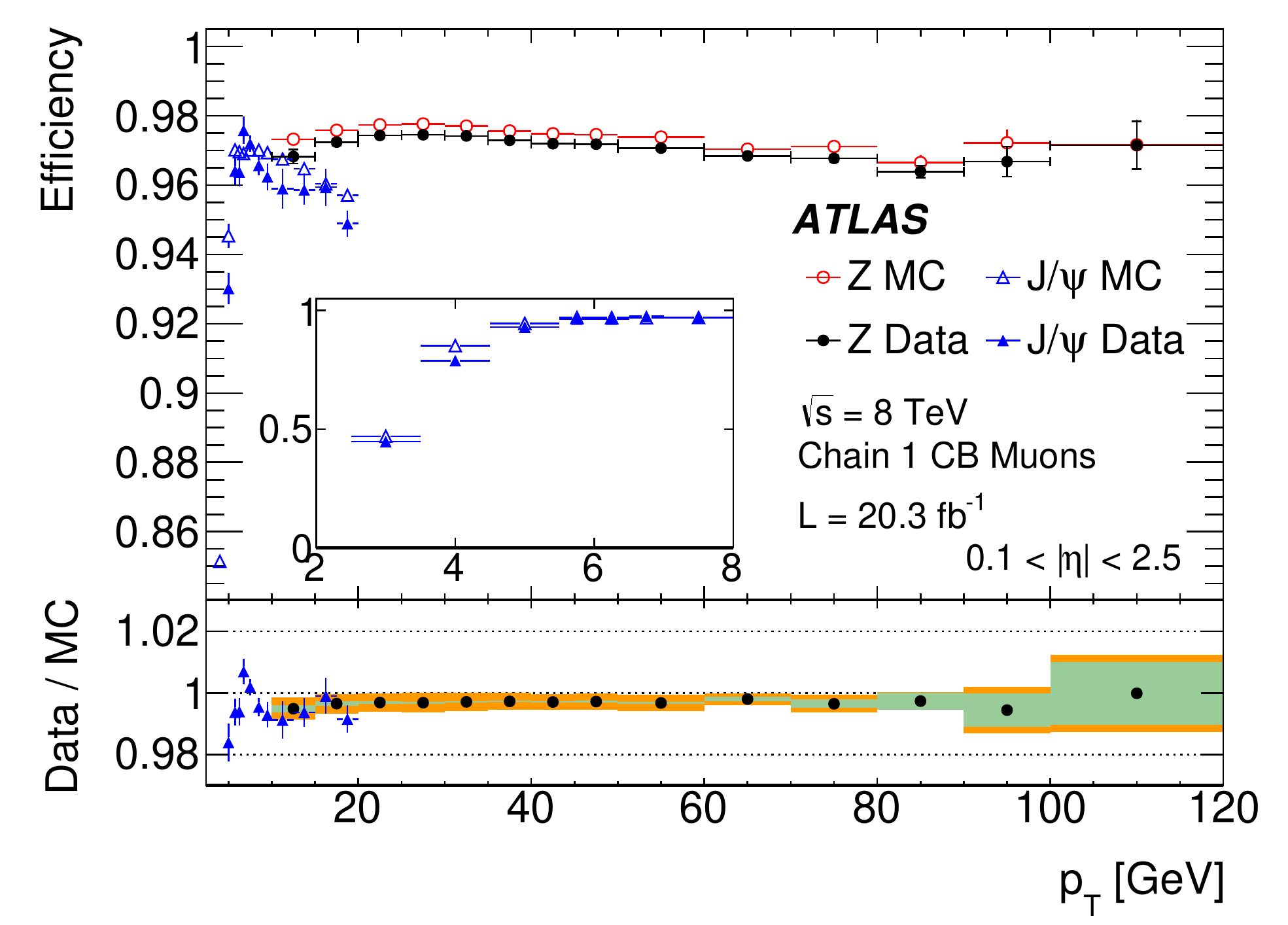}
\includegraphics[width=0.9\linewidth]{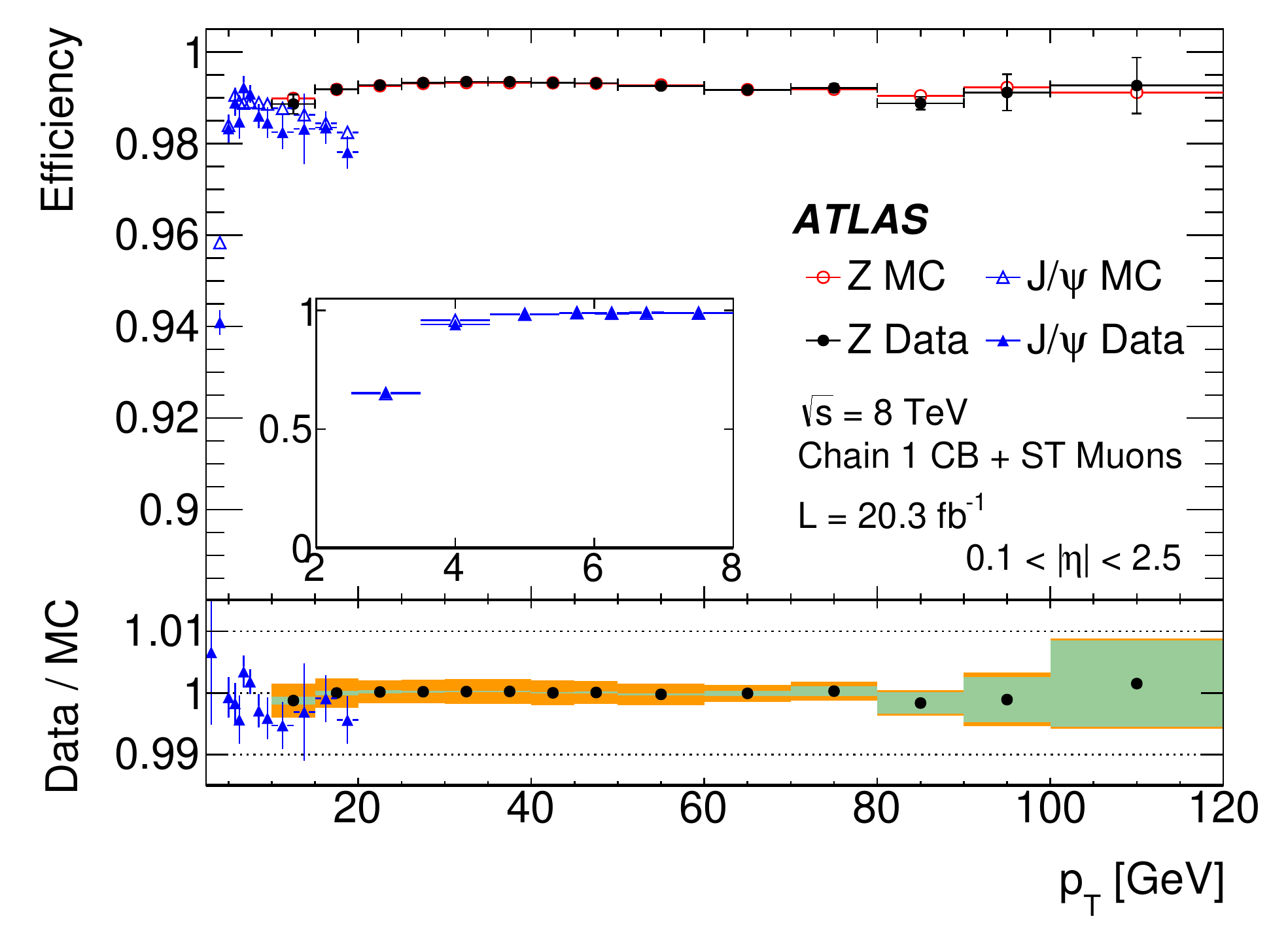}
  \includegraphics[width=0.9\linewidth]{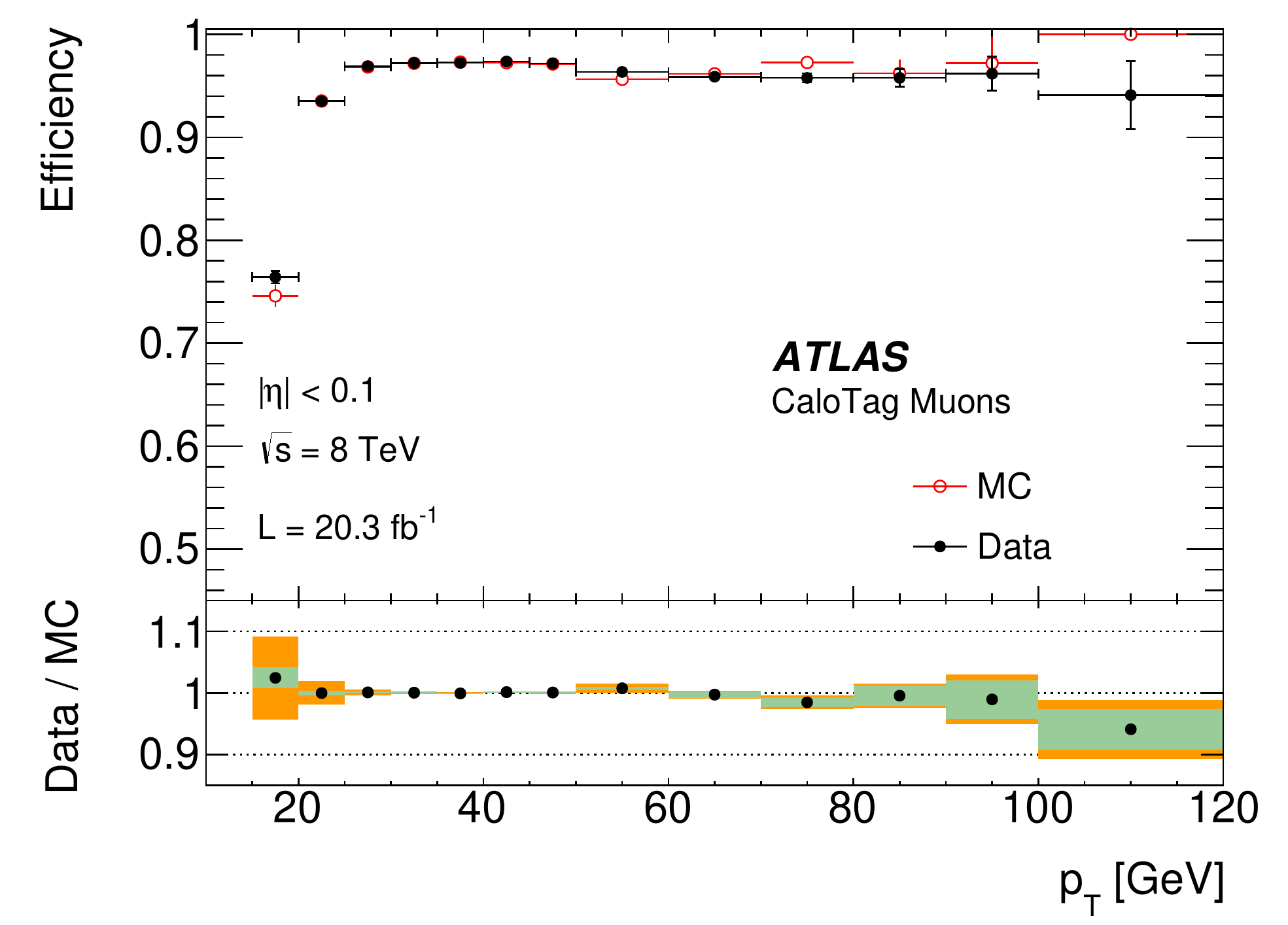}
  \caption{Reconstruction efficiency for CB (top), CB+ST (middle) and CaloTag (bottom) muons as a function
 of the $\pt$ of the muon, for muons with  $0.1 <|\eta|< 2.5$  for CB and CB+ST muons and for $|\eta| < 0.1$ 
for CaloTag muons. The upper two plots also show the result obtained with $\Zmm$ and $\Jpsimm$ events. The inserts on the upper plots show the detail of the efficiency as a function of $\pt$ in the low $\pt$ region. The CaloTag muon efficiency (bottom) is only measured with $\Zmm$ events. The error bars on the efficiencies indicate the statistical uncertainty for $\Zmm$ and include also the fit model uncertainty for $\Jpsimm$. The panel at the bottom shows the ratio between the measured and predicted efficiencies. The green areas show the pure statistical uncertainty, while the orange areas also include systematic uncertainties.}\label{Fig:CBSTEff_pt}
\end{center}
\end{figure}

Figure~\ref{Fig:CBSTEff_pt} shows the reconstruction efficiencies for CB and for CB+ST muons as a function of the transverse momentum, including results from $\Zmm$ and $\Jpsimm$. A steep increase of the efficiency is observed at low $\pt$, in particular
for the CB reconstruction, since a minimum momentum of approximately 3~GeV is required for a muon to traverse the calorimeter material and cross at least two layers of MS stations before being bent back by the magnetic field. Above $\pt \approx 20$~GeV, the
reconstruction efficiency for both CB and CB+ST muons is expected to be independent of the transverse momentum. This is confirmed within $0.5\%$ by the  $\Zmm$  data. The drop in efficiency observed in the $J/\psi$ data at $\pt>15$~GeV is due to the
inefficiency of the MS reconstruction for muon pairs with small angular separation as in the case of highly boosted $J/\psi$. This effect is  well reproduced by MC and the SF of the $\Jpsimm$ analysis are in good agreement with those from $\Zmm$ in the
overlap region.  The CaloTag muon efficiency reaches a plateau of approximately $0.97$ above $\pt\gtrsim 30$~GeV, where it is well predicted by the MC.

\begin{figure}[!h]
\begin{center}
    \includegraphics[width=0.9\linewidth]{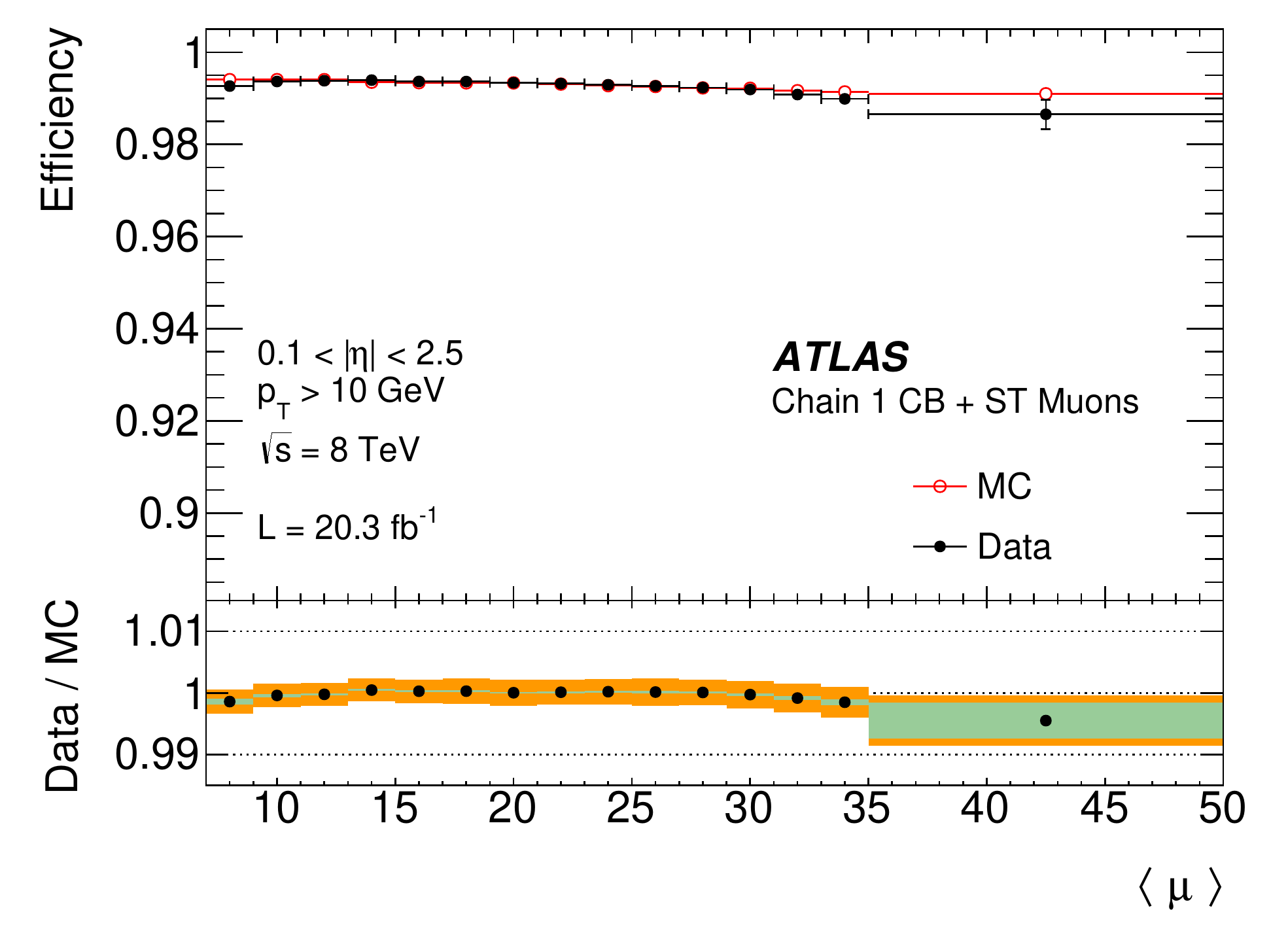}
    \caption{Measured CB+ST  muon reconstruction efficiency for muons with $\pt>10$~GeV  as a function of the average number of inelastic $pp$ collisions per bunch crossing $\langle \mu \rangle$. The error bars on the efficiencies indicate the statistical uncertainty. The panel at the bottom shows the ratio between the measured and predicted efficiencies. The green areas depict the pure statistical uncertainty, while the orange areas also include systematic uncertainties.}\label{Fig:mu_staco}
\end{center}
\end{figure}

Figure~\ref{Fig:mu_staco} shows the reconstruction efficiency for CB+ST muons as a function of $\langle \mu \rangle$, showing a high value (on average above $0.99$) and remarkable stability. A small efficiency drop of about $1\%$ is only observed for $\langle \mu \rangle>\gtrsim 35$. This is mainly caused by  limitations of the MDT readout electronics in the high-rate regions close to the beam lines. These limitations are being addressed in view of the next LHC run.

\subsection{Muon reconstruction efficiency for $|\eta|> 2.5$}
\label{sec::muon_reco_forward}
As described in the previous sections, the CB muon reconstruction is limited by the ID acceptance which covers the pseudo-rapidity region $|\eta|<2.5$. Above $|\eta|=2.5$, SA muons are the only muon type that provides large efficiency. A measurement of the efficiency SF for muons in the range $2.5<|\eta|<2.7$, hereafter called high-\eta, is needed for the physics analyses that  exploit the full MS acceptance.

A comparison with the Standard Model calculations for $\Zmm$ events is used to measure the reconstruction efficiency SF in the high-$\eta$ region.
To reduce the theoretical and experimental uncertainties, the efficiency SF is calculated from the double ratio 
\begin{equation}
\label{eq:doubleRatio}
 {\rm SF} =  
\frac{ \frac{N^{\rm Data}(2.5<|\eta_{\rm fwd}|<2.7)}  {N^{\rm MC}(2.5<|\eta_{\rm fwd}|<2.7)}}
        {  \frac{N^{\rm Data}(2.2<|\eta_{\rm fwd}|<2.5)}  {N^{\rm MC}(2.2<|\eta_{\rm fwd}|<2.5)}}\textrm{,}
\end{equation}
where the numerator is the ratio of the number of $\Zmm$ candidates in data and in MC for which one of the muons, called the \emph{forward} muon, is required to be in the high-\eta\ region  $2.5<|\eta_{\rm fwd}|<2.7$  while the other muon from the \Zboson\ decay, called the \emph{central} muon,  is required to have $|\eta|<2.5$. The denominator is the ratio of $\Zmm$ candidates in data over MC with the forward muon lying in the control region $2.2<|\eta_{\rm fwd}|<2.5$ and the central muon in the region $|\eta| < 2.2$. In both the numerator and denominator the central muon is required to be a CB muon while the forward muon can either be a CB or SA muon.  The simulation of muons with $|\eta|<2.5$ is corrected using the standard SF described in the previous section.

The selection of the central muon is similar to that of the tag muon in the tag-and-probe method. It is required to have triggered the event readout, to be isolated and to have transverse momentum $\pt>25$~\GeV. 
The requirements for the forward muon include calorimeter-based isolation, requiring  the transverse energy $E_T$ measured in the calorimeter in a cone of $\Delta R = 0.2$ (excluding the energy lost by the muon itself) around the muon track, to be less than $10\%$ of the muon \pt. The central and forward muons are required
to have opposite charge, a dimuon invariant mass within 10~GeV of the $Z$ mass, and a separation in $(\eta,\phi)$ space of $\Delta R>0.2$.

Different sources of systematic uncertainties have been considered:
a first group  is obtained by varying the $\pt$ and isolation cuts on the central muons and  the dimuon mass window. These variations produce effects of less than 0.3\% in the efficiency SF for the \pt\ range 20-60~\GeV.
The effect of the calorimetric isolation on the efficiency SF yields an uncertainty of less than 1\%, which is estimated by comparing the nominal SF values with the ones extracted when no calorimetric isolation is applied on the forward muons and by studying the dependence of this cut on the number of $pp$ interactions.
The contribution from the background processes, mainly dimuons from $b$ and $\bar{b}$ decays, has been studied using MC background  samples and found to be negligible.

\begin{figure}[!hbt]
\centering
\includegraphics[width=0.8\linewidth]{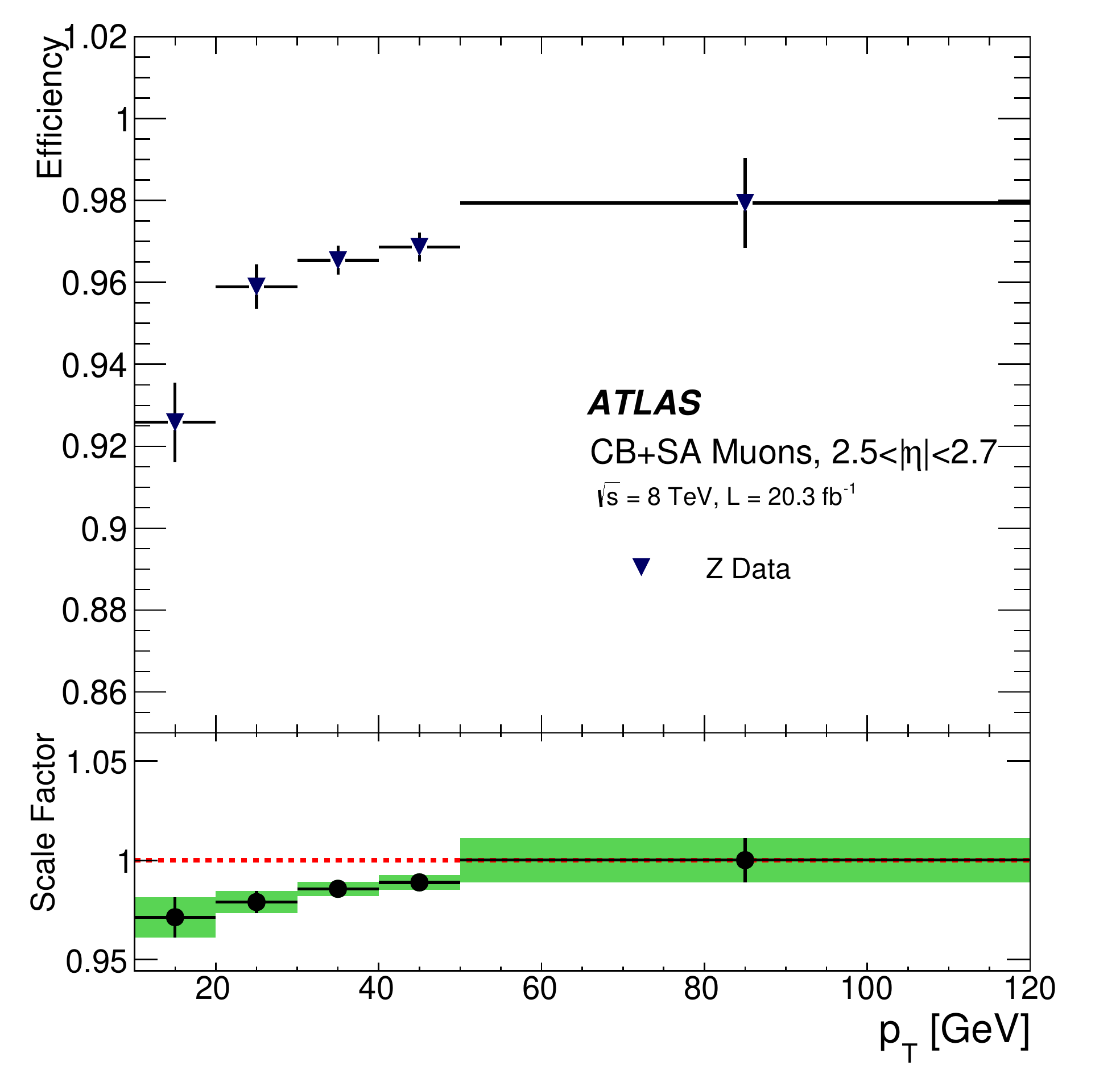}
\caption{Reconstruction efficiency for  muons within $2.5<|\eta|<2.7$ from $\Zmm$ events. The upper plot shows the efficiency obtained as the product of scale factor (Eq. ~\ref{eq:doubleRatio}) and the MC efficiency. The lower plot shows the scale factor. The error bars correspond to the statistical uncertainty while the green shaded band corresponds to the statistical and systematic uncertainty added in quadrature.}
 \label{fig:SF_staco}
\end{figure}

The theoretical uncertainty from higher-order corrections is estimated by varying the renormalization and factorization scales in the POWHEG NLO calculation at the generator level and is found to produce a negligible effect on the ratio of Eq.~\ref{eq:doubleRatio}.
The uncertainty from the knowledge of the parton densities is estimated by reweighting the PDFs used in the MC samples from CT10 to \linebreak MSTW2008NLO~\cite{Martin:2009iq} and by studying, at the generator level, the effect of the uncertainty associated to the \linebreak MSTW2008 PDF set on the double ratio of Eq.~\ref{eq:doubleRatio}, obtaining an overall theoretical uncertainty of less than $0.55\%$.

The efficiency in this region is obtained as the product of the SF and the ``true'' MC efficiency, calculated  as the fraction of generator-level muons that are successfully reconstructed.   The reconstruction efficiency and the  SF for muons in the high-$\eta$ region is shown in Fig.~\ref{fig:SF_staco} as a function of the muon $\pt$.


\subsection{Scale factor maps}

The standard approach used in ATLAS for physics analysis is to correct the muon reconstruction efficiency in the simulation using efficiency scale factors (SFs). The SFs are obtained with the tag-and-probe method using $\Zmm$ events, as described above, and are provided to the analyses in the form of  $\eta - \phi$ maps. Since no significant
$\pt$ dependence of the SF has been observed, no $\pt$ binning is used in the SF maps. Different maps are produced for different data taking sub-periods  with homogeneous detector conditions. The whole 2012  dataset is divided into 10 sub-periods. For each analysis, the final map is obtained as an average of the maps for all sub-periods, weighted by the periods' contribution to the integrated luminosity under study.

\begin{figure}[!h]
\begin{center}
  \includegraphics[width=0.9\linewidth]{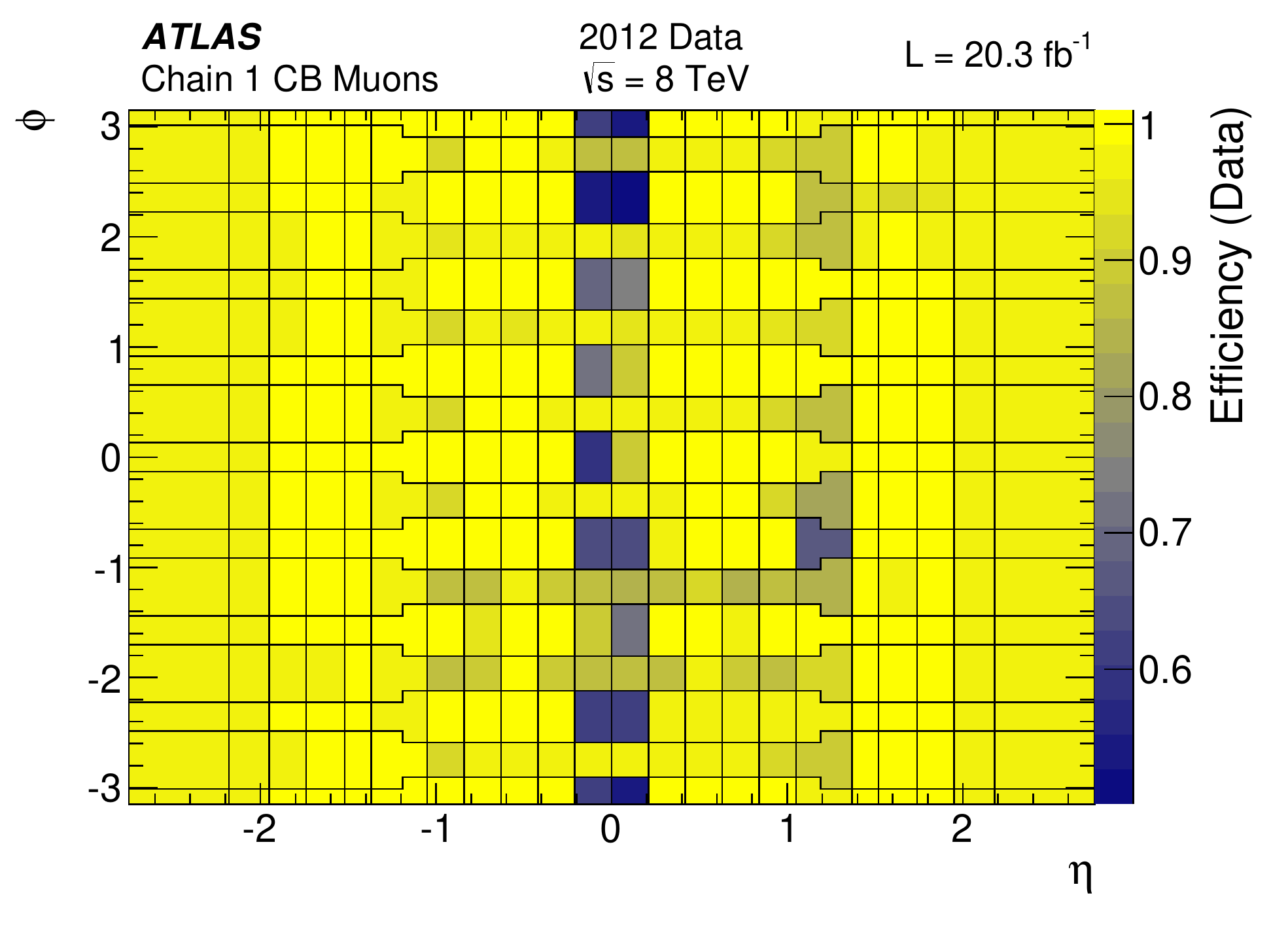}
  \includegraphics[width=0.9\linewidth]{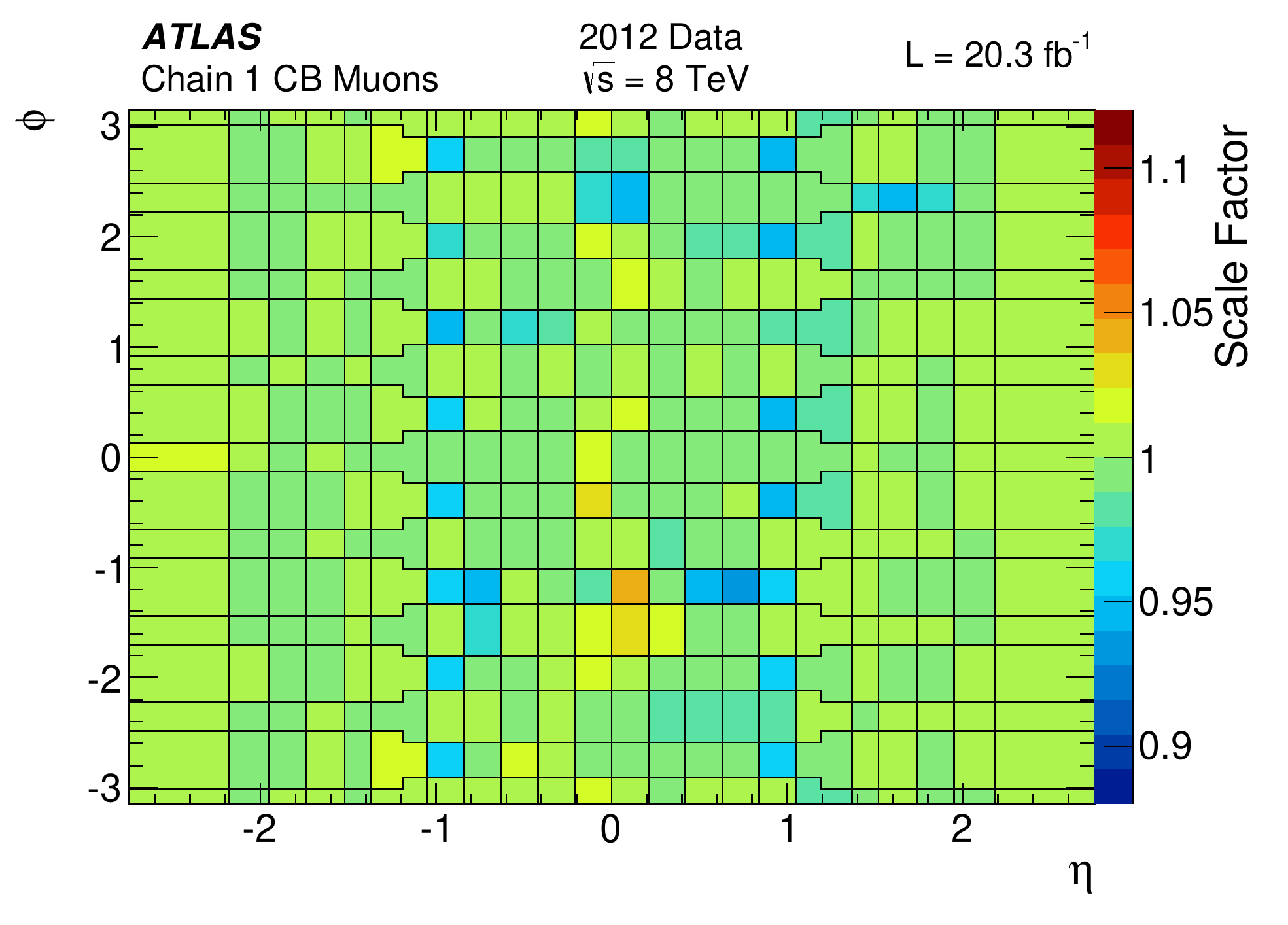}
  \caption{Reconstruction efficiency measured in the experimental data (top),
and the data/MC efficiency scale factor (bottom) for  CB muons as a function of $\eta$ and $\phi$ for muons with $\pt>10$~GeV. }\label{Fig:2deff_stacocb_data}
\end{center}
\end{figure}

\begin{figure}[!h]
\begin{center}
  \includegraphics[width=0.9\linewidth]{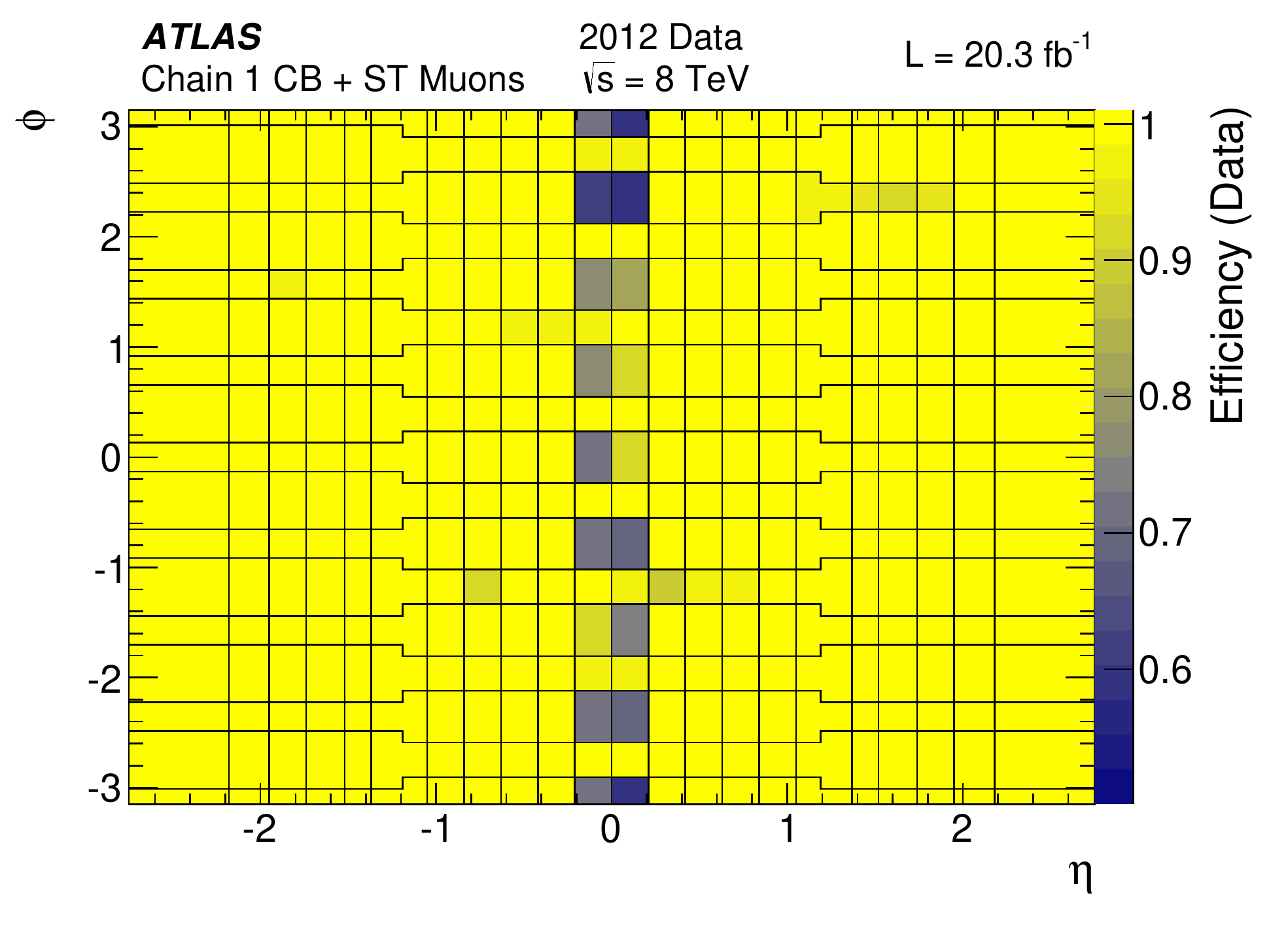}
  \includegraphics[width=0.9\linewidth]{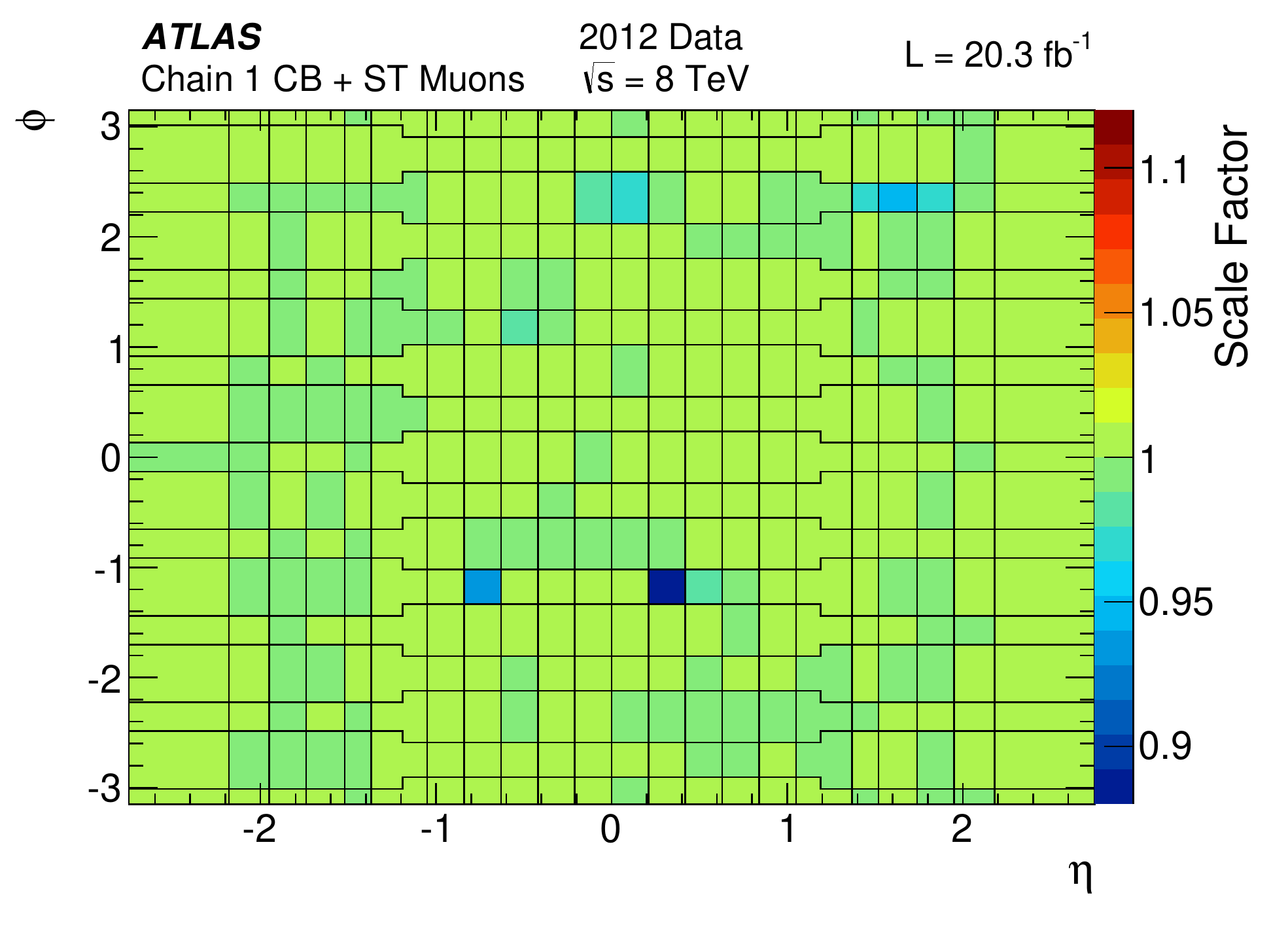}
  \caption{Reconstruction efficiency measured in the experimental data (top) and the data/MC  efficiency scale factor (bottom) for  CB+ST muons as a function of $\eta$ and $\phi$ for muons with $\pt>10$~GeV. 
 }\label{Fig:2deff_stacost_data}
\end{center}
\end{figure}

Figure~\ref{Fig:2deff_stacocb_data} and~\ref{Fig:2deff_stacost_data}
 show the maps of the efficiencies measured using the data in the $\eta$-$\phi$ plane and the corresponding Scale Factors. The large data sample allows for a precise resolution of localized efficiency losses, for example in the muon spectrometer for $|\eta| \sim 0$ due to limited coverage.  The SF maps
show local differences between data and MC related to detector conditions as discussed in Sect.~\ref{sec::T&Pres}.

\section{Momentum Scale and Resolution}
\label{Sec:ScaleResolution}

The large samples of  $\Jpsimm$, $\Upsilonmm$ and $\Zmm$ decays collected by ATLAS are used to study in detail the muon momentum scale and resolution.
The ATLAS simulation includes the best knowledge of the detector geometry, material distribution, and physics model of the muon interaction at the time of the MC events were generated. Additional corrections are needed to reproduce the muon momentum resolution and scale of experimental data at the level of precision that can be obtained using high-statistics samples of dimuon resonances.  Section~\ref{Sec:CorrectionExtraction} describes the methodology used to extract the corrections to be applied to the MC simulation. In Sect.~\ref{sec::validation}, the muon momentum scale and resolution is studied in the data and in MC samples with and without corrections.

\subsection{Corrections to the Muon Momentum in MC}\label{Sec:CorrectionExtraction}

Similarly to Ref.~\cite{perf2010}, the simulated muon transverse momenta reconstructed in the ID and in the MS sub-de\-tec\-tors,  $\pt^{\rm MC, Det}$, where Det$ = {\rm ID, MS}$, are corrected using the following equation:
\begin{eqnarray}\label{eq:pt_cor}
  \pt^{\rm Cor,Det} &=&  \frac{\pt^{\rm MC,Det} + \sum\limits^{1}_{n=0} s^{\rm Det}_{n}(\eta,\phi)\left(\pt^{\rm MC,Det}\right)^{n}}{1+\sum\limits^{2}_{m=0} \Delta r^{\rm Det}_{m}(\eta,\phi)\left(\pt^{\rm MC,Det}\right)^{m-1}g_{m}} \\\nonumber 
&&\textrm{ (with }s^{\rm ID}_{0}=0\textrm{ and } \Delta r^{\rm ID}_{0} =0\textrm{),} 
\end{eqnarray}
where $g_{m}$ are normally distributed random variables with mean 0 and width 1 and the terms  $\Delta r_{m}^{\rm Det}(\eta,\phi)$ and $s_n^{\rm Det}(\eta, \phi)$  describe, respectively, the momentum resolution smearing and the scale corrections applied in a specific $\eta$, $\phi$ detector region. The motivations for Eq.~\ref{eq:pt_cor} are the following:
\begin{itemize}
\item corrections are defined in $\eta - \phi$ detector regions such that in each region the variation of momentum resolution and scale, and therefore of their possible corrections, are expected to be small. In particular the nominal muon identification acceptance region (up to $|\eta|=2.7$) is divided in 18 $\eta$ sectors of size $\Delta \eta$ between 0.2 and 0.4, for both the MS and the ID. In addition, the MS is divided into two types of $\phi$ sectors of approximate size of $\pi/8$, exploiting the octagonal symmetry of the magnetic system
: the sectors that include the magnet coils (called ``small sectors'') and the sectors between two coils (called ``large sectors'').
\item The $\Delta r_{m}^{\rm Det}(\eta,\phi)$ correction terms introduce a $\pt$ dependent momentum smearing that effectively increases the relative momentum resolution, $\frac{ \sigma(\pt)}{\pt}$, when under-estimated by the simulation. The $\Delta r_{m}^{\rm Det}(\eta,\phi)$  terms can be related to different sources of experimental resolution by comparing the coefficient of the $\pt$ powers in the denominator of Eq.~\ref{eq:pt_cor} to the following empirical parametrization of the muon momentum resolution (see for example~\cite{detectors_grupen}):
\begin{equation}\label{eq:qoverpt_reso}
 \frac{ \sigma(\pt)}{\pt} = r_0/\pt \oplus r_1 \oplus r_2\cdot \pt\textrm{,}
\end{equation}
where $\oplus$ denotes a sum in quadrature. The first term (proportional to $1/\pt$) accounts for fluctuations of the energy loss in the traversed material. Multiple scattering, local magnetic field inhomogeneities and local radial displacements are responsible for
 the second term (constant in $\pt$). The third term (proportional to $\pt$) describes intrinsic resolution effects caused by the spatial resolution of the hit measurements and by residual misalignment. Energy loss fluctuations are relevant for muons traversing the calorimeter in front of the MS but they are negligible in the ID measurement. For this reason $ \Delta r^{\rm ID}_{0}$ is set to zero in Eq.~\ref{eq:pt_cor}.
\item  Imperfect knowledge of the magnetic field integral and of the radial dimension of the detector are reflected in the multiplicative momentum scale difference $s_1^{\rm Det} $ between data and simulation.  In addition, the $s^{\rm MS}_0(\eta, \phi)$ term is necessary to model the $\pt$ scale dependence observed in the MS momentum reconstruction due to differences between data and MC  in the energy loss of muons passing through the calorimeter and other materials between the interaction point and the MS. As the energy loss between the interaction point and the ID is negligible,  $s^{\rm ID}_0(\eta)$ is set to zero. 
\end{itemize}

The separate correction of ID and MS momentum reconstruction allows a direct understanding of the sources of the corrections. In a second step the corrections are propagated to the CB  momentum reconstruction, $\pt^{\rm Cor,CB}$, using a weighted average:
  \begin{equation}\label{eq:cb_correction1}
   \pt^{\rm Cor,CB} = f\cdot \pt^{\rm Cor,ID} + (1 - f)\cdot \pt^{\rm Cor, MS}\textrm{,}
 \end{equation}
with the weight $f$ derived for each muon by expressing the CB transverse momentum before corrections, $\pt^{\rm MC,CB}$, as a linear combination of $\pt^{\rm MC,ID}$ and $\pt^{\rm MC,MS}$:
   \begin{equation}\label{eq:cb_correction2}
   \pt^{\rm MC,CB} = f\cdot \pt^{\rm MC,ID} + (1 - f)\cdot \pt^{\rm MC, MS}
 \end{equation}
 and solving the corresponding linear equation.

\subsubsection{Correction extraction using a template fit to \texorpdfstring{$\Jpsimm$}{J/psi to mumu} and \texorpdfstring{$\Zmm$}{Z to mumu} events}\label{sec:corr_extract}

The MS and ID correction parameters contained in Eq.~\ref{eq:pt_cor} need to be extracted from data. For this purpose, a MC template maximum likelihood fit is used to compare the simulation to the data for $\Jpsimm$ and $\Zmm$ candidate events: this gives sensitivity to reconstructed muon momenta in the  $\pt$ range from a few GeV to $\approx 100$~GeV.
The dataset used for the correction extraction consists of $6$M $\Jpsimm$ and $9$M $\Zmm$ candidates passing the final selection. 

The $\Jpsimm$ and $\Zmm$ candidates have been selected online according to the requirements described in Sect.~\ref{sec:data_samples} and, offline, by requiring two CB muons. For the correction extraction in a specific $\eta - \phi$ Region Of Fit (ROF), the ID and MS reconstructed momenta  are considered individually. All the events with at least one of the two muons in the ROF contribute to the correction extraction fit. The angles from the CB reconstruction are used to define the ROF and to calculate the invariant mass distributions.

The ID corrections are extracted using the distribution of the ID dimuon invariant mass, $m_{\mu\mu}^{\textrm{ID}}$. Events with $m_{\mu\mu}^{\textrm{ID}}$ in the window $2.76 - 3.6$~GeV and $\pt^{\textrm{ID}}$ in the range $8 - 17$~GeV are selected as $\Jpsimm$ candidate decays; events with $m_{\mu\mu}^{\textrm{ID}}$ between $76$ and $96$~GeV and the leading (sub-leading) muons with \mbox{$26< \pt^{\textrm{ID}}< 300$~GeV} (\mbox{$15< \pt^{\textrm{ID}}< 300$~GeV})\linebreak are selected as $\Zmm$ candidate decays. To enhance the sensitivity to the $\pt$ dependent correction effects, the  $m_{\mu\mu}^{\textrm{ID}}$ is classified according to the $\pt$ of the muons: for $\Jpsimm$ candidates the $\pt^{\textrm{ID}}$ of the sub-leading muon defines three bins with lower thresholds at $\pt^{\textrm{ID}}={8, 9, 11}$~GeV, 
for $\Zmm$ candidates the $\pt^{\textrm{ID}}$ of the leading muon defines three bins with lower thresholds at $\pt^{\textrm{ID}}={26, 47, 70}$~GeV.

Similarly, the MS corrections are extracted using the distribution of the MS reconstructed dimuon invariant mass, $m_{\mu\mu}^{\textrm{MS}}$, in the same way as for the ID. However, as in the MS part of Eq.~\ref{eq:pt_cor} more correction parameters and more ROFs are present, an additional variable sensitive to the momentum scale and resolution is added to the MS fit. The variable, used only in $\Zmm$ candidate events, is defined by the following equation: 
\begin{equation}
  \label{eq:rho}
  \rho= \frac{\pt^{\textrm{MS}} - \pt^{\textrm{ID}}}{\pt^{\textrm{ID}}}\textrm{,}
\end{equation}
representing a measurement of the $\pt$ imbalance between the measurement in the ID and in the MS.
The $\rho$ variable is binned according to $\pt^{\textrm{MS}}$ of the muon in the ROF: the lower thresholds are $\pt^{\textrm{MS}}={20, 30, 35, 40, 45, 55, 70}$~GeV. 

In order to compare the simulation to the data distributions, the corresponding templates of $m_{\mu\mu}^{\textrm{ID}}$, $m_{\mu\mu}^{\textrm{MS}}$, and $\rho$ are built using the MC samples of the $\Jpsimm$ and  $\Zmm$ signals.  The background in the $\Zmm$ mass region  is added to the templates using the simulation and corresponds to approximately 0.1\% of the $\Zmm$ candidates. The 
 non-resonant background to $\Jpsimm$, coming from decays of light and  heavy hadrons and from  Drell-Yan production, accounts for about 15\% of the selected $\Jpsimm$ candidates. As it is not possible to accurately simulate it, a data driven approach is used to evaluate it: an analytic model of the background plus the $J/\psi$ signal is fitted to the dimuon mass spectrum of the $\Jpsimm$ candidates in a mass range $2.7 - 4.0$~GeV, then the background model and its normalization are used in the template fit from which the momentum correction are extracted. The analytic fit is performed independently on the ID and MS event candidates. The non-resonant dimuon background is parametrized with an exponential function, while the  $J/\psi$ and $\psi^{2S}$ resonances are parametrized  by a Crystal-Ball function~\cite{crystalball2} in the ID fits, or by a Gaussian distribution convoluted with a Landau in the MS fits, where energy loss effects due to the calorimeter material are larger.

The template fit machinery involves several steps: first a binned likelihood function  $\mathcal{L}$ is built to compare the data to the MC templates of signal plus background. Then modified templates are generated by varying the correction parameters in Eq.~\ref{eq:pt_cor} and applying them to the muon momentum of the simulated signal events. The   $-2\ln\mathcal{L}$ between data and the modified template is then minimized using MINUIT~\cite{James:1975dr}. 
 The procedure is iterated across all the ROFs: the first fit is performed using only events with both muons in the ROF, the following fits allow also one of the muons in a previously analysed ROF and one in the ROF under investigation. After all the detector ROFs have been analysed, the fit procedure is iterated twice in order to improve the stability of the results.  The correction extraction is performed first for the ID and then for the MS, such that the ID transverse momentum present in Eq.~\ref{eq:rho} can be kept constant during the MS correction extraction.

Although the use of $\pt$ bins for the construction of the templates gives a good sensitivity to the $\pt$ dependence of the scale corrections, the fit is not very sensitive to the resolution correction terms $\Delta r^{\textrm{MS}}_0(\eta,\phi)$ and $\Delta r^{\textrm{MS}}_2(\eta,\phi)$ of Eq.~\ref{eq:pt_cor}. The reasons for this are, 
 at low $\pt$, the $\pt>8$~GeV selection cut applied to the $J/\psi$ data sample,  which limits the sensitivity to  $\Delta r^{\textrm{MS}}_0(\eta,\phi)$, and,  at high $\pt$, the limited statistics of the $\Zmm$ data sample with $\pt^{\textrm{MS}}>100$~GeV, which limits the sensitivity to  $\Delta r^{\textrm{MS}}_2(\eta,\phi)$.
As the energy loss fluctuations do not show significant disagreement between data and MC for $|\eta|>0.8$, the parameter $\Delta r^{\textrm{MS}}_0(\eta,\phi)$ has been fixed to zero in this region. The effect of the misalignment of MS chambers in real data, which is expected to be the largest contribution to  $\Delta r^{\textrm{MS}}_2(\eta,\phi)$, is already taken into account in the simulation as described in Sect.~\ref{sec:mc_samples}. Therefore the $\Delta r^{\textrm{MS}}_2(\eta,\phi)$ term is also fixed to zero in the MS correction extraction. Two of the systematic uncertainties described in Sect.~\ref{sec:fit_sys} are used to cover possible deviations from zero of these two  terms.

\subsubsection{Systematic uncertainties}\label{sec:fit_sys}

Systematic uncertainties  cover imperfections in the model used for the muon momentum correction and in the fit procedure used for the extraction of the correction terms. In particular the correction extraction procedure has been repeated using the following different configurations:
\begin{itemize}
\item variation of $\pm 5$~GeV in the dimuon mass window used for the $\Zmm$ event selection. This is intended to cover resolution differences between data and MC that are beyond a simple Gaussian smearing. This results in one of the largest systematic uncertainties on the resolution corrections, with an average effect of $\approx$ 10\% on the  $\Delta r_{1}^{\textrm{ID}}$, $\Delta r_{2}^{\textrm{ID}}$, and $\Delta r_{1}^{\textrm{MS}}$ parameters.
\item Two variations of the $J/\psi$ templates used in the fit. The first concerns the  $J/\psi$ background parametrization: new $m_{\mu\mu}^{\textrm{MS}}$ and $m_{\mu\mu}^{\textrm{ID}}$ background templates are generated using a linear model, for the MS fits, and a linear-times-exponential model, for the ID fits. The second variation concerns the  $J/\psi$ event selection: the minimum muon $\pt^{MS,ID}$ cut is raised from 8 to 10 GeV, thus reducing the weight of low-$\pt$ muons on the corrections. The resulting variations on the resolution correction parameters are $\approx 10$\% of  $\Delta r_{1}^{\textrm{ID}}$ and $\Delta r_{1}^{\textrm{MS}}$. The effect is also relevant for the MS scale corrections with a variation of $\approx 0.01$~GeV on $s_0^{\textrm{MS}}$ and of $\approx 4\times 10^{-4}$ on $s_1^{\textrm{MS}}$.
\item The ID correction extraction is repeated using \linebreak $\Jpsimm$ events only or $\Zmm$ events only.  Since such configurations have a reduced statistical power, only the $s_1^{\textrm{ID}}$ correction parameter is left free in the fit, while the resolution correction terms are fixed to nominal values. The resulting uncertainty on  $s_1^{\textrm{ID}}$, ranging from 0.01\% to 0.05\% from the central to the forward region of the ID, accounts for non-linear effects on the ID scale.
\item The parameter  $\Delta r^{\textrm{MS}}_0$ of Eq.~\ref{eq:pt_cor} is left free in all the regions, instead of fixing it to zero for $|\eta|>0.8$.  The largest variation of 0.08~GeV is applied as an additional systematic uncertainty on the parameter.
\item The MS correction is extracted using a special $\Zmm$ MC sample with ideal geometry, i.e. where no simulation of the misalignment of the MS chambers is applied.
This is needed because the standard simulation has a too pessimistic resolution in the  \mbox{$|\eta|<1.25$} region, forcing the $\Delta r^{\textrm{MS}}_1$ parameter to values compatible with zero. The template fit performed with the ideal-geometry $\Zmm$ MC sample gives $\Delta r^{\textrm{MS}}_1> 0$ in the region \mbox{$0.4<|\eta|<1.25$}. The largest variation of $\Delta r^{\textrm{MS}}_1$, corresponding to $0.012$, is applied as an additional systematic uncertainty for this region.
\item  Variation of the normalization of the MC samples used in  $\Zmm$ background estimate by factors of two and one half. The resulting systematic uncertainty is small except for the detector regions with $|\eta|>2.0$, where the effect is comparable to the other uncertainties.
\end{itemize}
Independently from the fit procedure, the following studies are used to derive additional systematic uncertainties:
\begin{itemize}
\item The simulation of the ID includes an excess of material for $|\eta|>2.3$  resulting in a muon momentum resolution with is too pessimistic. Such imperfection is covered by adding a systematic uncertainties of $2\times 10^{-3}$ on the $s^{\rm ID}_{1}$ parameter, and of 0.01 on the $\Delta r^{\rm ID}_{1}$  parameter, both for $|\eta|>2.3$. These are the largest systematic uncertainties on the ID correction parameters.
\item  The position of the mass peak in the $\Zmm$ sample is studied in finer $\eta$ bins than those used to extract the corrections, using the fit that will be discussed in Sect.~\ref{sec::validation} as an alternative to the template fitting method.  An additional uncertainty of $2\times 10^{-4}$ on  the $s_1^{\textrm{ID}}(\eta)$ parameter is found to cover all the observed deviations between data and corrected MC. 
\item The effect of the measurement of the angle of the muon tracks has been checked by using the $J/\psi$ MC and conservatively increasing the track angular resolution by $\approx 40\%$. The maximum effect is an increase of the resolution correction $\Delta r^{\textrm{ID}}_1$ of $0.001$, which is added to the systematic uncertainties.
\item Special runs with the  toroidal magnetic field off have been  used to evaluate the quality of the MS chamber alignment.  These results are compared to the chamber misalignments in the simulation to define the  systematic uncertainty on  the $\Delta r^{\textrm{MS}}_2(\eta,\phi)$ resolution correction parameter.
\end{itemize}

 The final uncertainty on each of the eight muon momentum correction parameters is derived from the sum in quadrature of all the listed uncertainty sources. This is simplified for use in standard physics analyses, for which only four systematic variations are provided: global upper and lower scale variations and independent resolution variations for the ID and the MS. The upper and lower scale variations are obtained by a simultaneous variation of all the ID and MS scale correction parameters by $1\sigma$. The resolution variation for ID (MS) is obtained by the simultaneous  variation of all the ID (MS)  correction parameters.

The MC-smearing approach of Eq.~\ref{eq:pt_cor} cannot be used to correct the MC when the resolution in real data is better than in the simulation.
To deal with these cases, the amount of resolution that should be subtracted in quadrature from the simulation to reproduce the data is included in the positive ID and MS resolution variations. Then the prescription for physics analysis is to symmetrize the effect of the positive variation of resolution parameters around the nominal value of the physical observables under study. 

\begin{table*}[!thb]
  \centering
\begin{tabular}{cccc}
\toprule
Region & $\Delta r_1^{\textrm{ID}}$ & $\Delta r_2^{\textrm{ID}}$ [TeV$^{-1}$] & $s_1^{\textrm{ID}}$\\
\midrule
$|\eta|<1.05$         & $0.0068  ^{+0.0010}$     & $ 0.146 ^{+0.039}$ &  $-0.92^{+0.26}_{-0.22}\times 10^{-3}$ \\
$1.05\le |\eta|<2.0$  & $0.0105  ^{+0.0018}$ & $ 0.302 ^{+0.046}$ &  $-0.86^{+0.30}_{-0.35}\times 10^{-3}$ \\
$|\eta|\ge 2.0$       & $0.0069  ^{+0.0121}$    & $ 0.088 ^{+0.084}$ &  $-0.49^{+1.17}_{-1.63}\times 10^{-3}$ \\
\bottomrule
\end{tabular}
  \caption{Summary of ID muon momentum resolution and scale corrections used in Eq.~\ref{eq:pt_cor}, averaged over three main detector regions.
The corrections are derived in 18 $\eta$ detector regions, as described in Sect.~\ref{sec:corr_extract}, and averaged according to the $\eta$ width of each region. The uncertainties are the result of the sum in quadrature of the statistical and systematic uncertainties.
Only upper uncertainties are reported for the $\Delta r$ parameters; lower uncertainties are evaluated by symmetrization, as described in Sect.~\ref{sec:fit_sys}.}
  \label{tab:id_cor}
\end{table*}
\begin{table*}[!thb]
  \centering
\begin{tabular}{cccccc}
\toprule
Region & $\Delta r_0^{\textrm{MS}}$ [GeV] & $\Delta r_1^{\textrm{MS}}$ & $\Delta r_2^{\textrm{MS}}$ [TeV$^{-1}$]  & $s_0^{\textrm{MS}}$ [GeV] & $s_1^{\textrm{MS}}$\\
\midrule
$|\eta|<1.05$ (small)        & $0.115  ^{+ 0.083}$ & $ 0.0030  ^{+ 0.0079}$& $0 ^{+ 0.21}$   & $-0.035^{+0.017}_{-0.011}$ & $+3.57^{+0.38}_{-0.60}\times 10^{-3}$ \\
$|\eta|<1.05$ (large)        & $0.101   ^{+ 0.090}$ & $ 0.0034  ^{+ 0.0081}$& $0 ^{+ 0.11}$ & $-0.022^{+0.007}_{-0.014}$ & $-0.22^{+0.37}_{-0.24}\times 10^{-3}$ \\
$1.05\le |\eta|<2.0$  (small)& $0      ^{+ 0.080}$ & $ 0.0171  ^{+ 0.0059}$& $0 ^{+ 0.22}$& $-0.032^{+0.017}_{-0.016}$ & $-1.07^{+0.77}_{-0.93}\times 10^{-3}$ \\
$1.05\le |\eta|<2.0$  (large)& $0      ^{+ 0.080}$ & $ 0.0190  ^{+ 0.0047}$& $0 ^{+ 0.17}$ & $-0.026^{+0.009}_{-0.017}$ & $-1.46^{+0.45}_{-0.57}\times 10^{-3}$ \\
$|\eta|\ge 2.0$ (small)      & $0         ^{+ 0.080}$ & $ 0.0022  ^{+ 0.0075}$& $0  ^{+ 0.06}$ & $-0.031^{+0.029}_{-0.031}$ & $-0.91^{+1.63}_{-0.91}\times 10^{-3}$ \\
$|\eta|\ge 2.0$ (large)      & $0          ^{+ 0.080}$ & $ 0.0171  ^{+ 0.0052}$& $0 ^{+ 0.29}$ & $-0.057^{+0.019}_{-0.021}$ & $+0.40^{+1.22}_{-0.50}\times 10^{-3}$ \\
\bottomrule
\end{tabular}
\caption{Summary of MS momentum resolution and scale corrections for small and large MS sectors, averaged over three main detector regions.
 The corrections for large and small MS sectors are derived in 18 $\eta$ detector regions, as described in Sect.~\ref{sec:corr_extract}, and averaged according to the $\eta$ width of each region.  The parameters $\Delta r_0^{\textrm{MS}}$, for $|\eta|>1.05$, and $\Delta r_2^{\textrm{MS}}$, for the full $\eta$ range, are fixed to zero.
The uncertainties are the result of the sum in quadrature of the statistical and systematic uncertainties.
Only upper uncertainties are reported for the $\Delta r$ parameters; lower uncertainties are evaluated by symmetrization, as described in Sect.~\ref{sec:fit_sys}.}
  \label{tab:ms_cor}
\end{table*}

\subsubsection{Result of the muon momentum scale and resolution corrections}
The ID and MS  correction parameters used in Eq.~\ref{eq:pt_cor} are shown in Tab.~\ref{tab:id_cor} and~\ref{tab:ms_cor}, averaged over three $\eta$ regions. The scale correction to the simulated ID track reconstruction is always below 0.1\% with an uncertainty ranging from $\approx 0.02$\%, for $|\eta|<1.0$, to 0.2\%, for $|\eta|>2.3$. The correction to the MS scale is $\lesssim 0.1$\% except for the large MS sectors in the barrel region of the detector, where a correction of $\approx$0.3\% is needed, and for specific MS regions with \mbox{$1.25<|\eta|<1.5$} where a correction of about $-0.4$\% is needed. An energy loss correction of approximately $30$~MeV is visible for low values of $\pt$ in the MS reconstruction. This correction corresponds to about $1\%$ of the total energy loss in the calorimeter and in the dead material in front of the spectrometer and is compatible with the accuracy of the material budget used in the simulation. Depending on the considered $\pt$ range,  total resolution smearing corrections below 10\% and below 15\%  are  needed for the simulated ID and MS track reconstructions.

 \subsection{Measurement of the dimuon mass scale and resolution}\label{sec::validation}

The collected samples of  $\Jpsimm$, $\Upsilonmm$ and $\Zmm$ decays have been used to study the muon momentum resolution and to validate the momentum corrections obtained with the template fit method described in the previous section with a different methodology. In addition the $\Upsilon$ sample, not used in the extraction of the corrections, provides an independent validation.

Neglecting angular effects, the invariant mass resolution $\sigma(m_{\mu\mu})$ is related to the momentum
resolution by
\begin{equation}
\frac{\sigma(m_{\mu\mu})}{m_{\mu\mu}} \, = \, \frac{1}{2}\frac{\sigma(p_1)}{p_1}  \oplus  \frac{1}{2}\frac{\sigma(p_2)}{p_2}\textrm{,}
\label{eq:dmom0}
\end{equation}
where $p_1$ and $p_2$ are the momenta of the two muons. If the momentum resolution is similar for the two muons then the relative mass resolution is proportional to the relative momentum resolution:
\begin{equation}
  \frac{\sigma(m_{\mu\mu})}{m_{\mu\mu}} \, = \,  \frac{1}{\sqrt{2}} \frac{\sigma(p)}{p}.
\label{eq:dmom}
\end{equation}

The mass resolution has been obtained by fitting the  width of the invariant mass peaks. In
the  $\Jpsimm$ and $\Upsilonmm$ decays, the intrinsic width of the resonance is negligible with respect to the experimental resolution. In the $\Zmm$ case the fits have been performed using a convolution of the true line-shape obtained from the MC simulation with an experimental resolution function.
The momentum scale was obtained by comparing the mass peak position in data and in MC.  Details of the event selection and of the invariant mass fits are given below.

\begin{figure*}[th]
  \centering
  \includegraphics[width=0.32\linewidth]{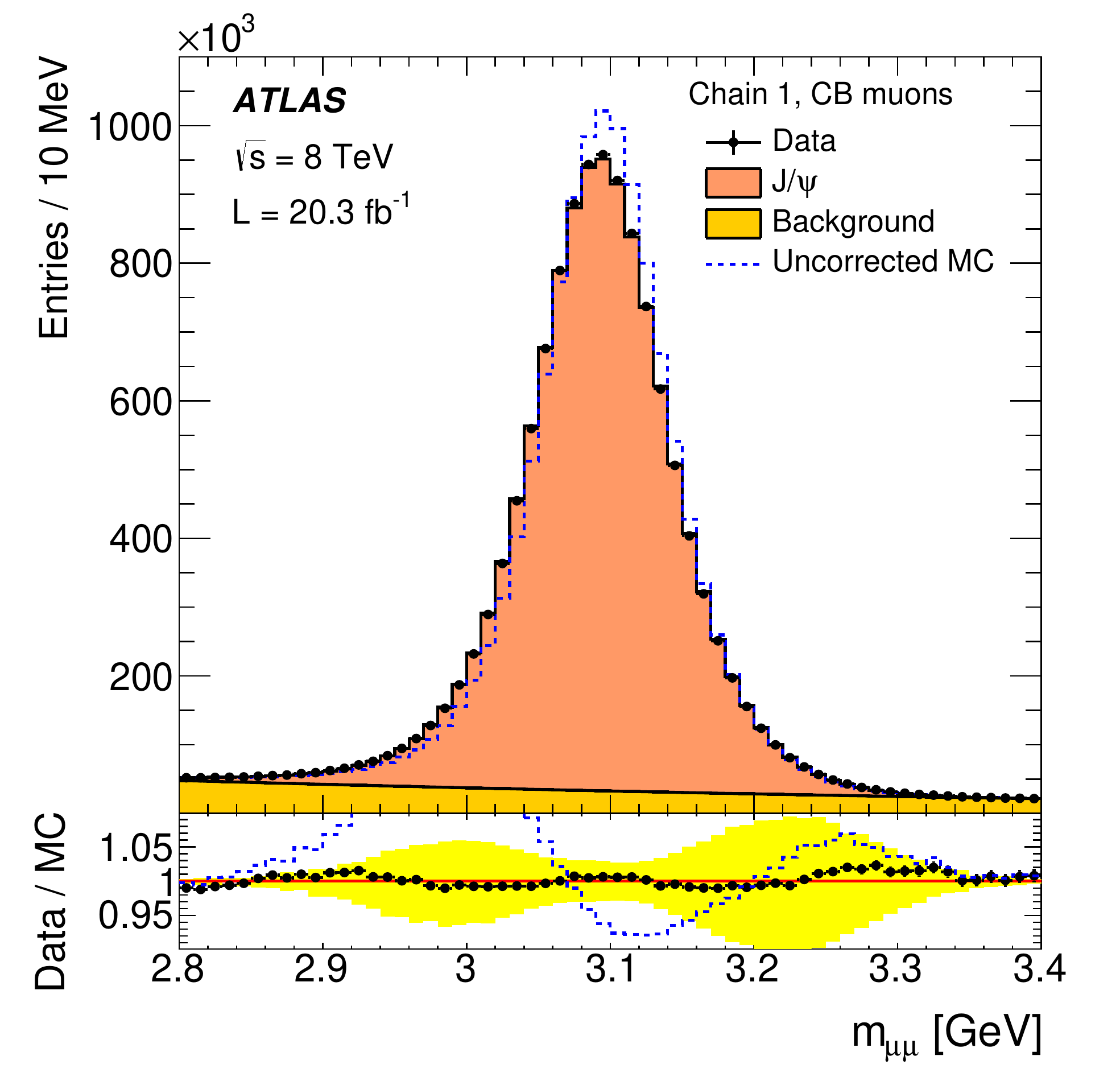}
  \includegraphics[width=0.32\linewidth]{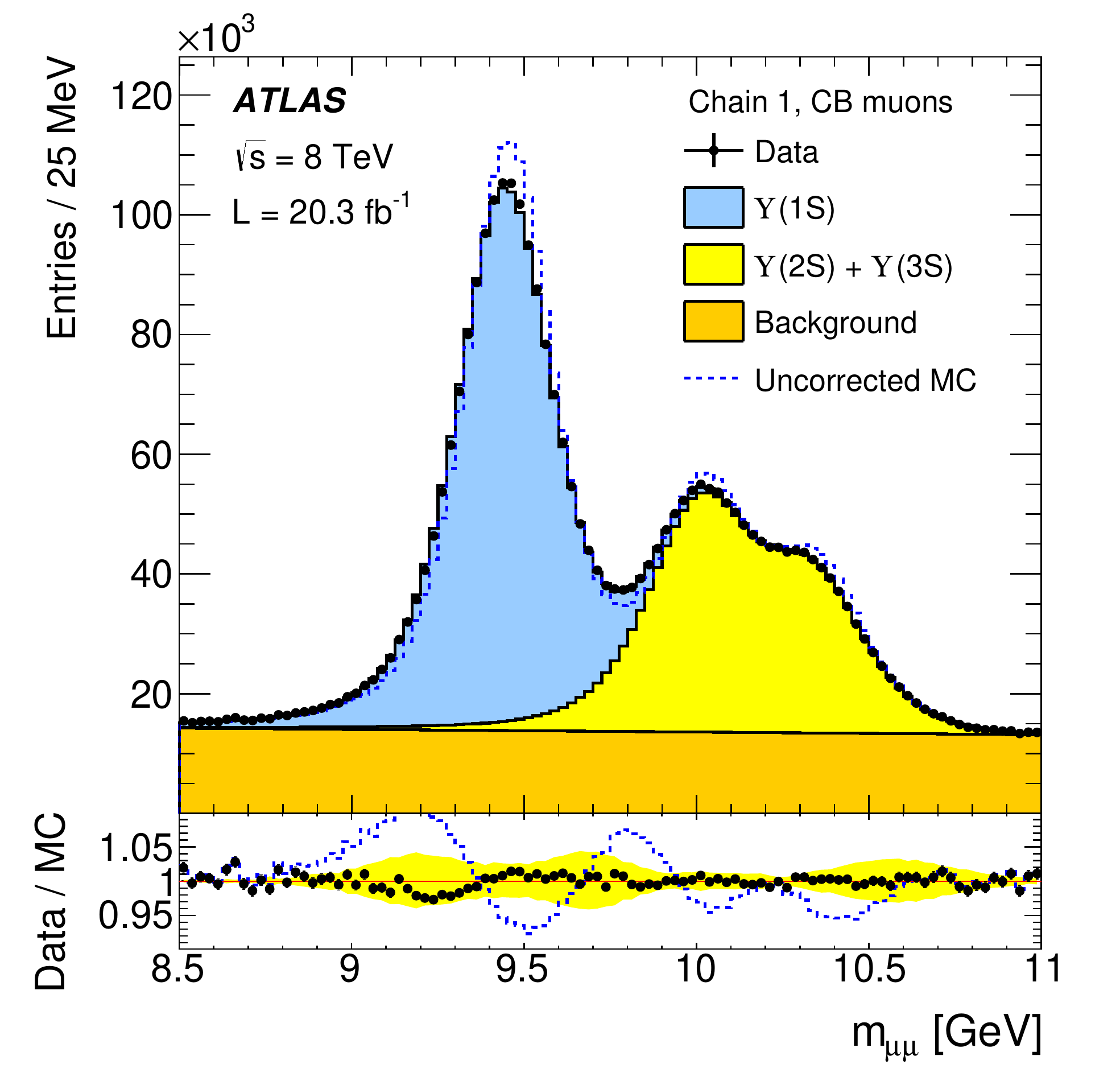}
  \includegraphics[width=0.32\linewidth]{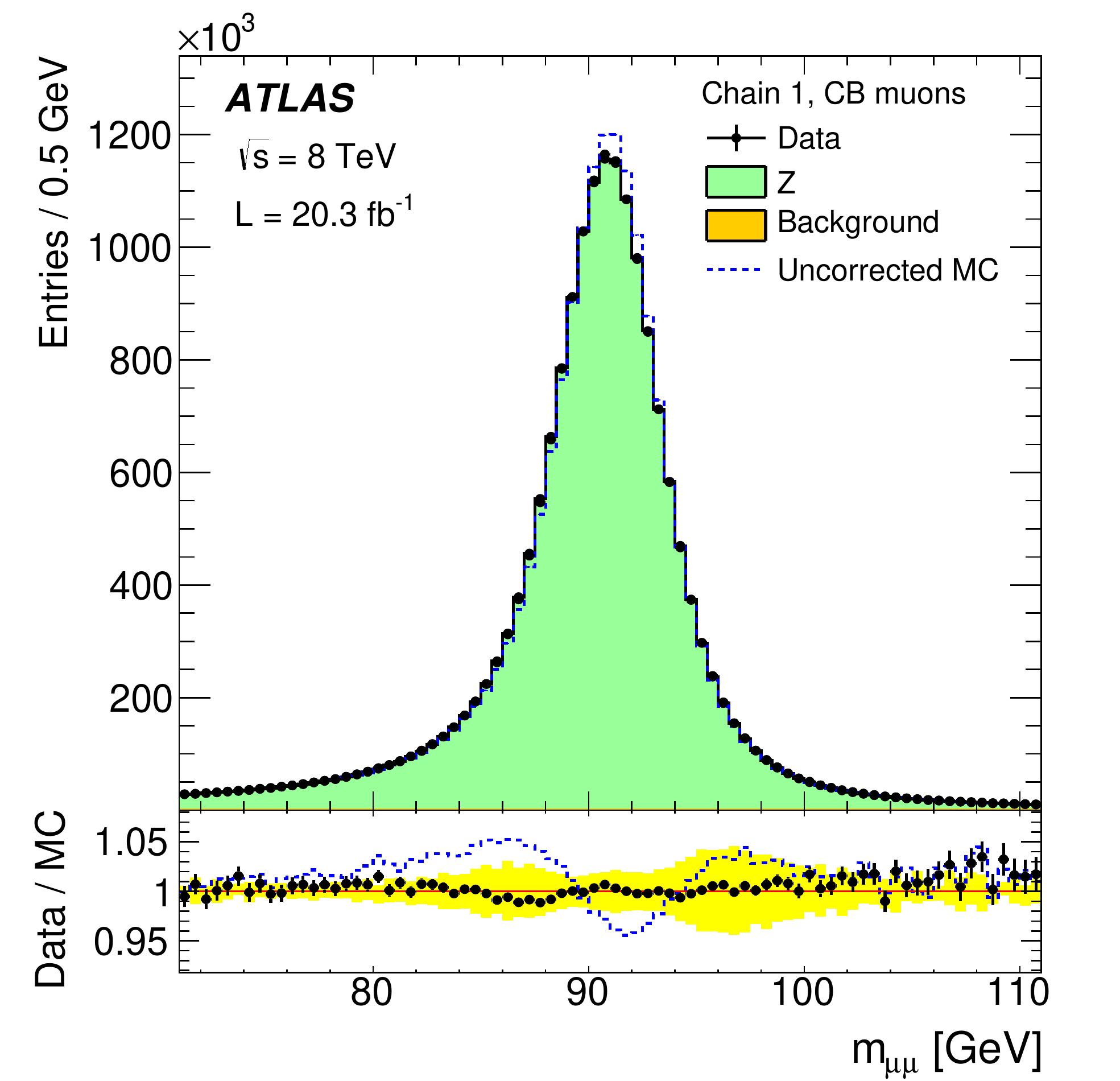}
  \caption{Dimuon invariant mass distribution of 
$\Jpsimm$ (left), $\Upsilonmm$ (center) and  $\Zmm$ (right) candidate events reconstructed with CB muons.
The upper panels show the invariant mass distribution for data and for the signal MC simulation plus the background estimate. The points show the data, the filled histograms show the simulation with the MC momentum corrections applied and the dashed histogram shows
the simulation when no correction is applied. Background estimates are added to the signal simulation. The lower panels show the Data/MC ratios. The band represents the effect of the systematic uncertainties on the MC momentum corrections. In the $J/\psi$ case the background was fitted in a sideband region as described in the text. In the $\Upsilon$ case a simultaneous fit of the normalization of the three simulated $\Upsilonmm$ distributions and of a linear background was performed. In the $Z$ case, the MC background samples are added to the signal sample according to their expected cross sections. The sum of background and signal MC is normalized to the  data.
}
\label{fig:CB_corr_lineshapes}
\end{figure*}

\subsubsection{Event selection and mass fitting}
The $J/\psi$ and $\Upsilon$  events are selected online by the dedicated dimuon triggers described in Sect.~\ref{sec:data_samples}.
The offline event selection requires in addition that both muons are reconstructed as CB muons and have $\pt>7$~GeV. The trigger acceptance limits the muons to the region $|\eta|<2.4$. The resulting data samples consist of $17$M and $4.7$M  candidates for $J/\psi$ and $\Upsilon$, respectively. The $\Zmm$ sample was selected online with the single-muon trigger described in Sect.~\ref{sec::muon_reco_central}. One of the two muons can be outside the trigger acceptance, allowing coverage of the full range $|\eta|<2.7$. The offline selection requires two opposite-charge  muons, one with $\pt>25$~GeV and one with  $\pt>20$~GeV. The two  muons are required to be isolated, to have opposite charges and to be compatible with the primary interaction vertex. 

The invariant mass distribution of the $\Jpsimm$, \linebreak
$\Upsilonmm$ and $\Zmm$ samples are shown in Fig.~\ref{fig:CB_corr_lineshapes} and compared with uncorrected and corrected MC. With the uncorrected MC the signal peaks have smaller width and are slightly shifted with respect to data. After correction, the lineshapes of the three resonances agree very well with the data.
For a detailed study, the position $\langle m_{\mu\mu} \rangle$ and the width $\sigma(m_{\mu\mu})$ of the mass peaks are extracted in bins of $\eta$ and $\pt$ from fits of the invariant mass distributions of the three resonances. 

In the $J/\psi$ case, for each bin, the background is obtained from a fit of two sideband regions outside the $J/\psi$ mass peak (\mbox{$2.55<m_{\mu\mu}<2.9$} and \mbox{$3.3<m_{\mu\mu}<4.0$}~GeV) using a second order polynomial. The background is then subtracted from the signal mass window. The parameters $\langle m_{\mu\mu} \rangle$ and $\sigma(m_{\mu\mu})$ of the background subtracted signal distribution are obtained with a Gaussian fit in the range $\langle m_{\mu\mu} \rangle \pm 1.5 \sigma(m_{\mu\mu})$, obtained using an iterative procedure. Systematic uncertainties associated to the fit are evaluated by repeating the fit using a third order polynomial as the background model and by varying the fit range to $\pm 1 \times$ and $\pm 2 \times \sigma(m_{\mu\mu})$.

As shown in Fig.~\ref{fig:CB_corr_lineshapes}, the three $\Upsilon$ resonances (1S, 2S, 3S) partially overlap. Moreover in the $\Upsilon$ case the mass window imposed by the trigger limits considerably the size of the sidebands available for fixing the background level. Therefore a different fit strategy is adopted in this case. For each  bin, the whole invariant mass distribution in the range $8.5<m_{\mu\mu}<11.5$~GeV is fitted with a linear background plus three Crystal-Ball functions representing the three resonances. The $\alpha$ and $n$ parameters that fix the tail of the Crystal-Ball function are fixed to the values obtained from a fit of the signal MC mass distribution. The relative mass shifts of the three signal peaks are fixed using the PDG masses of the three resonances, while the widths of the three peaks, divided by the corresponding PDG masses, are constrained to be equal. The remaining free parameters in the fit are the mass scale, the width $\sigma(m_{\mu\mu})$ of the $\Upsilon(1S)$, the relative normalizations of the  $\Upsilon(2S)$ and  $\Upsilon(3S)$ distributions with respect to  $\Upsilon(1S)$ and two parameters for the linear background. A similar fit is performed on the MC simulation of the invariant mass distribution obtained by adding  the three signal peaks and a flat background distribution.
The fit systematic uncertainties have been evaluated by chaining the fit range to $8.25< m_{\mu\mu}< 11.75$ and $8.75< m_{\mu\mu}< 11.0$ GeV and by varying the $\alpha$ and $n$ parameters in the range allowed by fits to the simulation.

In the $\Zmm$ case,  for each bin, the true  lineshape predicted by the MC simulation is parametrized with a Breit-Wigner function. The measured dimuon mass spectrum is fitted with a Crystal-Ball function, representing the experimental resolution effects, convoluted with the Breit-Wigner parametrization of the true lineshape. The fit is repeated in different ranges around the mass peak (corresponding approximately to one to two standard deviations) and the spread of the results is used to evaluate the systematic uncertainty of the fit.

\subsubsection{Mass scale results}

\begin{figure}[!ht]
  \centering
  \includegraphics[width=0.95\linewidth]{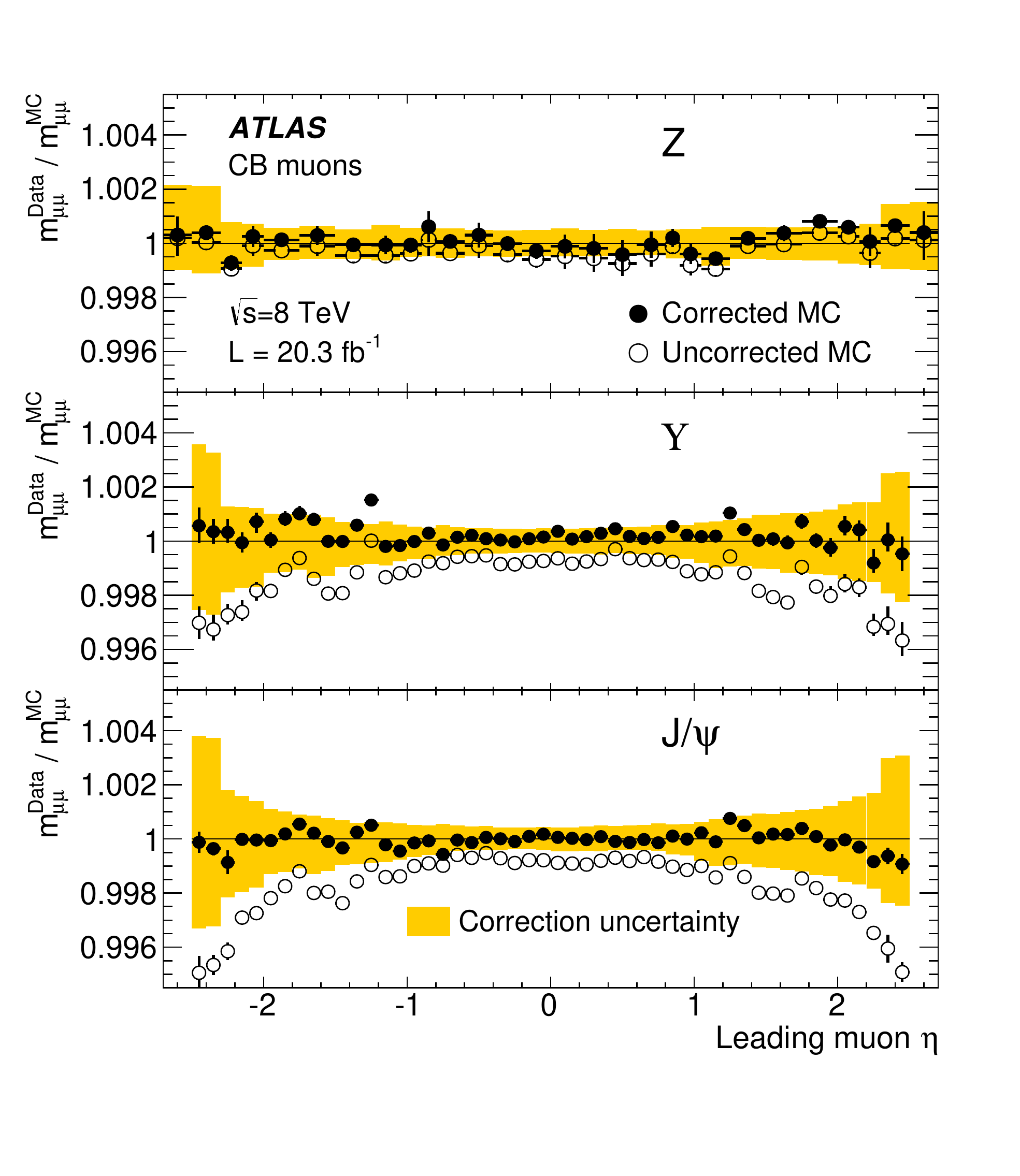}
  \caption{Ratio of the fitted mean mass,  $\langle m_{\mu\mu} \rangle$, for data and corrected MC
    from $Z$ (top), $\Upsilon$ (middle), and $J/\psi$ (bottom) events as a  function of the pseudorapidity of the highest-$\pt$ muon.
    The ratio is shown for corrected MC (filled symbols) and uncorrected MC (empty symbols).
    The error bars represent the statistical and the systematic uncertainty on the mass fits added in quadrature.
    The bands show the uncertainty on the MC corrections calculated separately for the three samples.
}\label{fig:mass_eta}
\end{figure}

Figure~\ref{fig:mass_eta} shows the Data/MC ratio of the mean mass $\langle m_{\mu\mu} \rangle$ obtained from the fits to the $Z$, $J/\psi$, $\Upsilon$ samples 
described above, as a function of the pseudorapidity of the highest-$\pt$ muon for pairs of CB muons. For the uncorrected MC, the ratio deviates from unity in the large $|\eta|$ region of the $J/\psi$ and $\Upsilon$ cases by up to $5\%$. This is mainly due to imperfections in the simulation of the muon energy loss that have a larger effect at low $\pt$ and in the forward $\eta$ region where the MS measurement has a larger weight in the MS-ID combination. The corrected MC is in very good agreement with the data, well within the scale systematics that are $\approx 0.035\%$ in the barrel region and increase with $|\eta|$ to reach $\sim 0.2\%$ in the region $|\eta|>2$ for the $\Zmm$ case.

\begin{figure*}[!htb]
\begin{center}
  \includegraphics[width=0.32\linewidth]{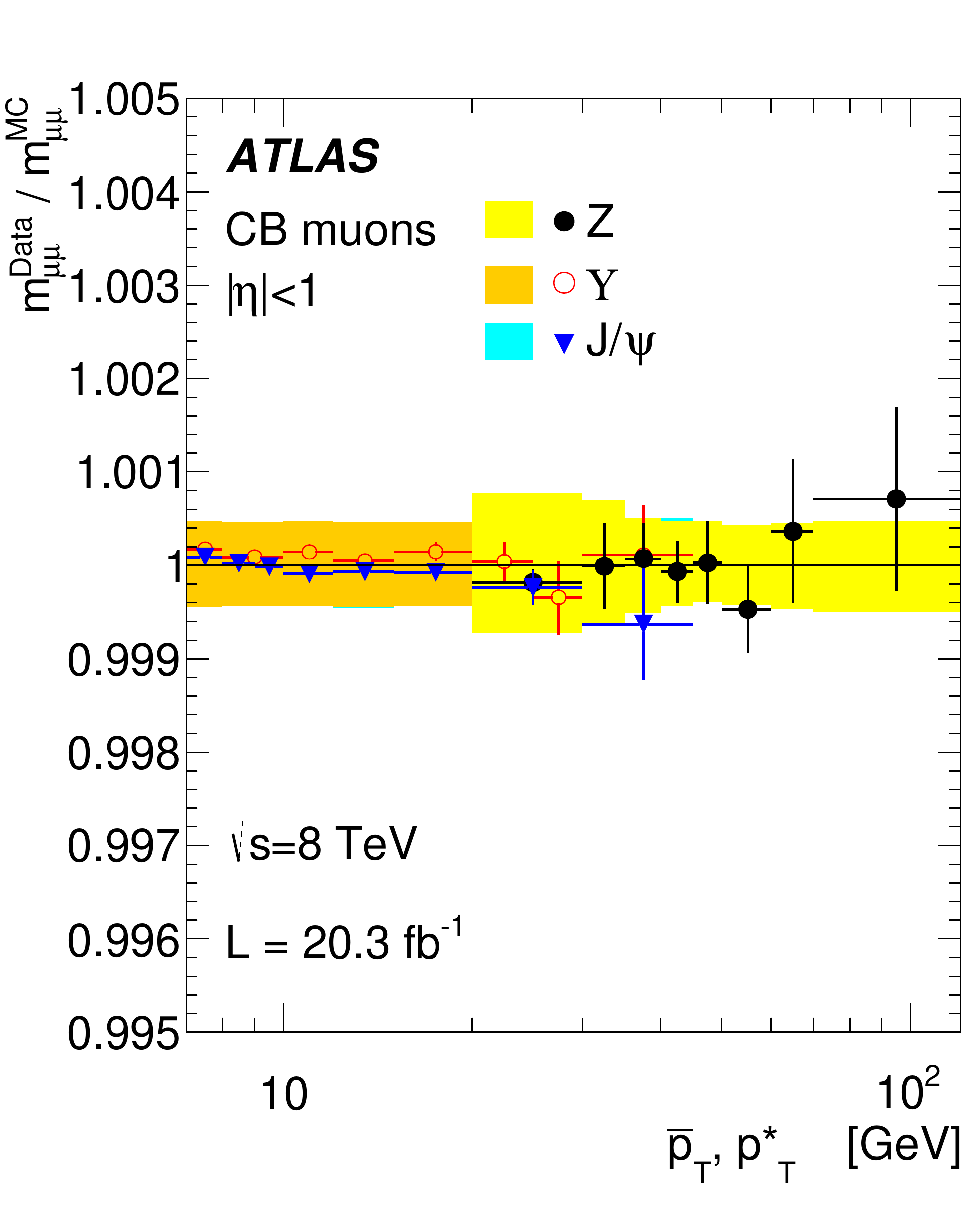}
  \includegraphics[width=0.32\linewidth]{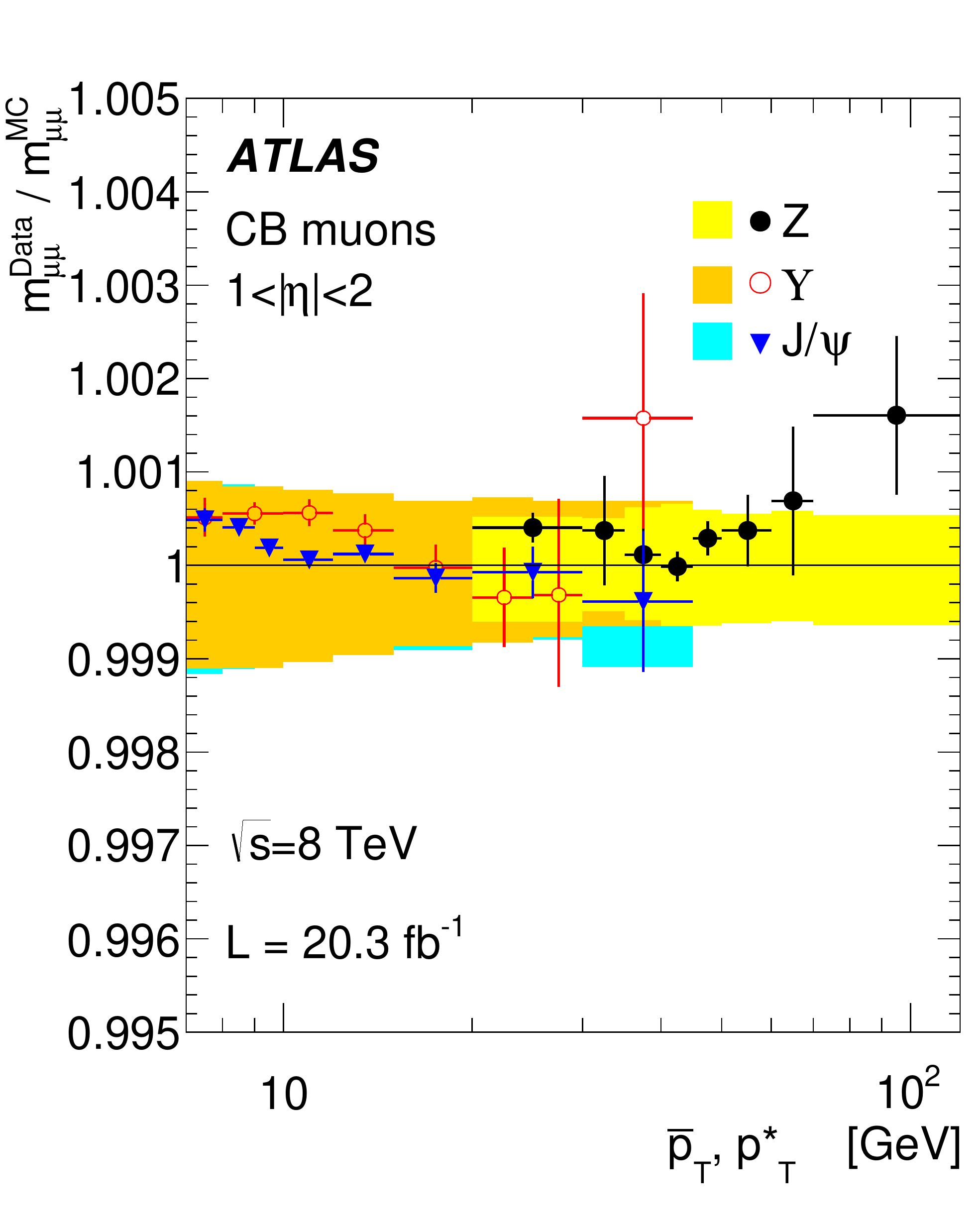}
  \includegraphics[width=0.32\linewidth]{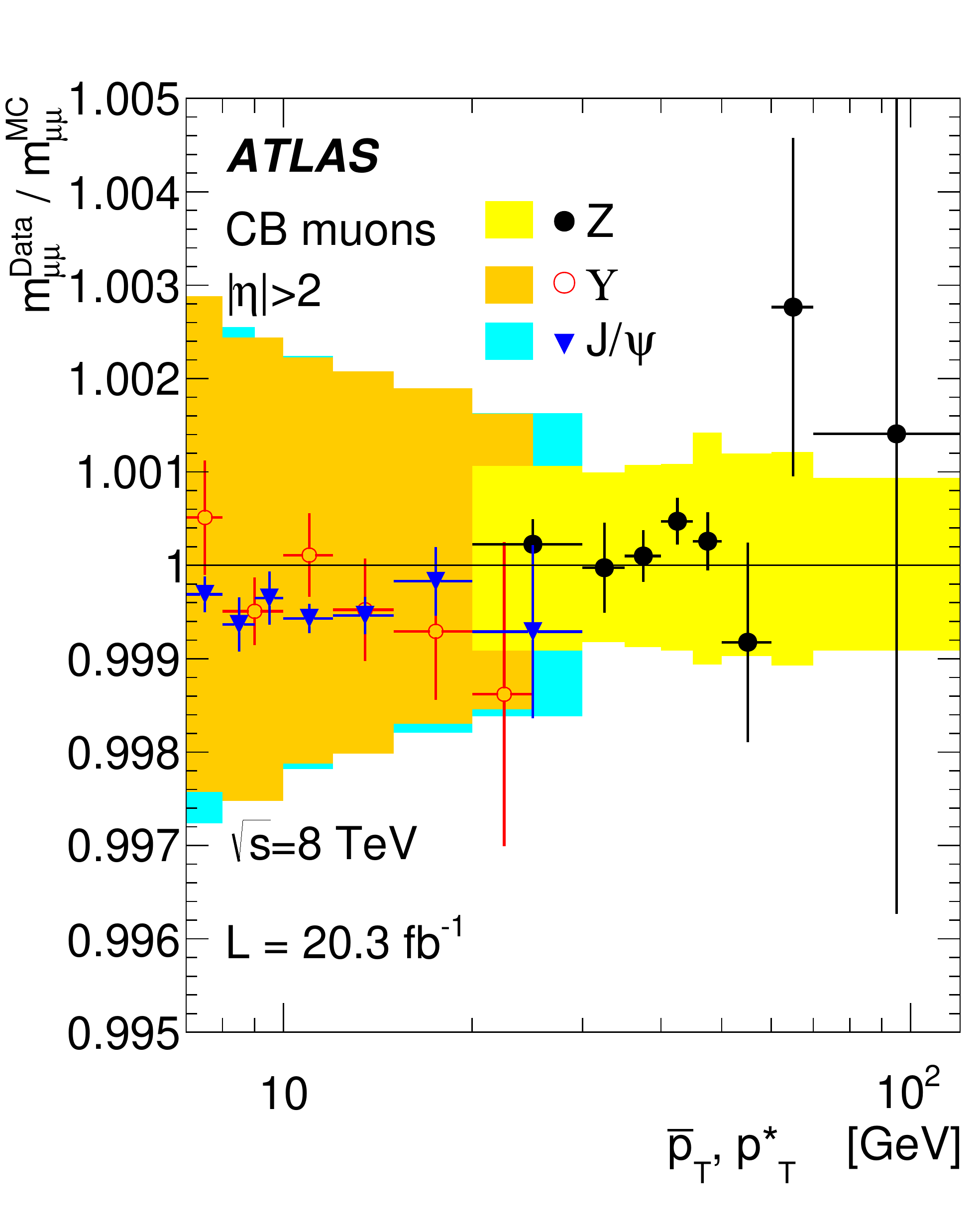}
\end{center}
  \caption{Ratio of the fitted mean mass,  $\langle m_{\mu\mu} \rangle$, for data and corrected MC
    from  $J/\psi$, $\Upsilon$ and  $Z$ events as a  function of the 
    average transverse momentum in three $|\eta|$ ranges.  Both muons are required to be in the same $|\eta|$ range.  The $J/\psi$ and $\Upsilon$ data are shown as a function
    of the  $\bar{p}_{\rm T} = \frac{1}{2}(p_{\mathrm{T},1}+p_{\mathrm{T},2})$ while
    for $Z$ data are plotted as a function of $\pt^*$
    as defined in  Eq.~\ref{eq:angles}.   The error bars represent the statistical uncertainty and the systematic uncertainty on the fit added in quadrature.
The bands show the uncertainty on the MC corrections calculated separately for the three samples.
  \label{f:valid_summary_scale_pt}}
\end{figure*}

Figure~\ref{f:valid_summary_scale_pt} shows the data/MC ratio for   $\langle m_{\mu\mu} \rangle$  as a function of the  transverse momentum $\langle \pt \rangle$ for muons in three different pseudorapidity regions.

For the $J/\psi$ and $\Upsilon$ cases, $\langle \pt \rangle$ is defined as the average momentum $\bar{p}_T = \frac{1}{2}(p_{\mathrm{T},1}+p_{\mathrm{T},2})$ 
while in the $Z$ case it is defined as 
\begin{equation}
\pt^{*} = m_{Z} \sqrt{ \frac{\sin{\theta_1}\sin{\theta_2}}{2 (1-\cos \alpha_{12})}},
\label{eq:angles}
\end{equation}
where $m_{Z}$ is the $Z$ pole mass~\cite{pdg}, $\theta_1$, $\theta_2$ are the polar angles of the two muons and $\alpha_{12}$ is the opening angle of the muon pair. This definition, based on angular variables only, removes the correlation between the measurement of the dimuon mass and of the average $\pt$ that is particularly relevant around the Jacobian peak at $\pt=m_Z/2$ in the distribution of muons from $Z$ decays.

The data from the three resonances span from $\langle \pt  \rangle =7$~GeV to $\langle \pt \rangle =120$~GeV and show that the momentum scale is well known and within the assigned systematic uncertainties in the whole $\pt$ range.

\subsubsection{Resolution results}
\begin{figure*}[!hbt]
  \begin{center}
  \subfigure[]{\includegraphics[width=0.32\linewidth]{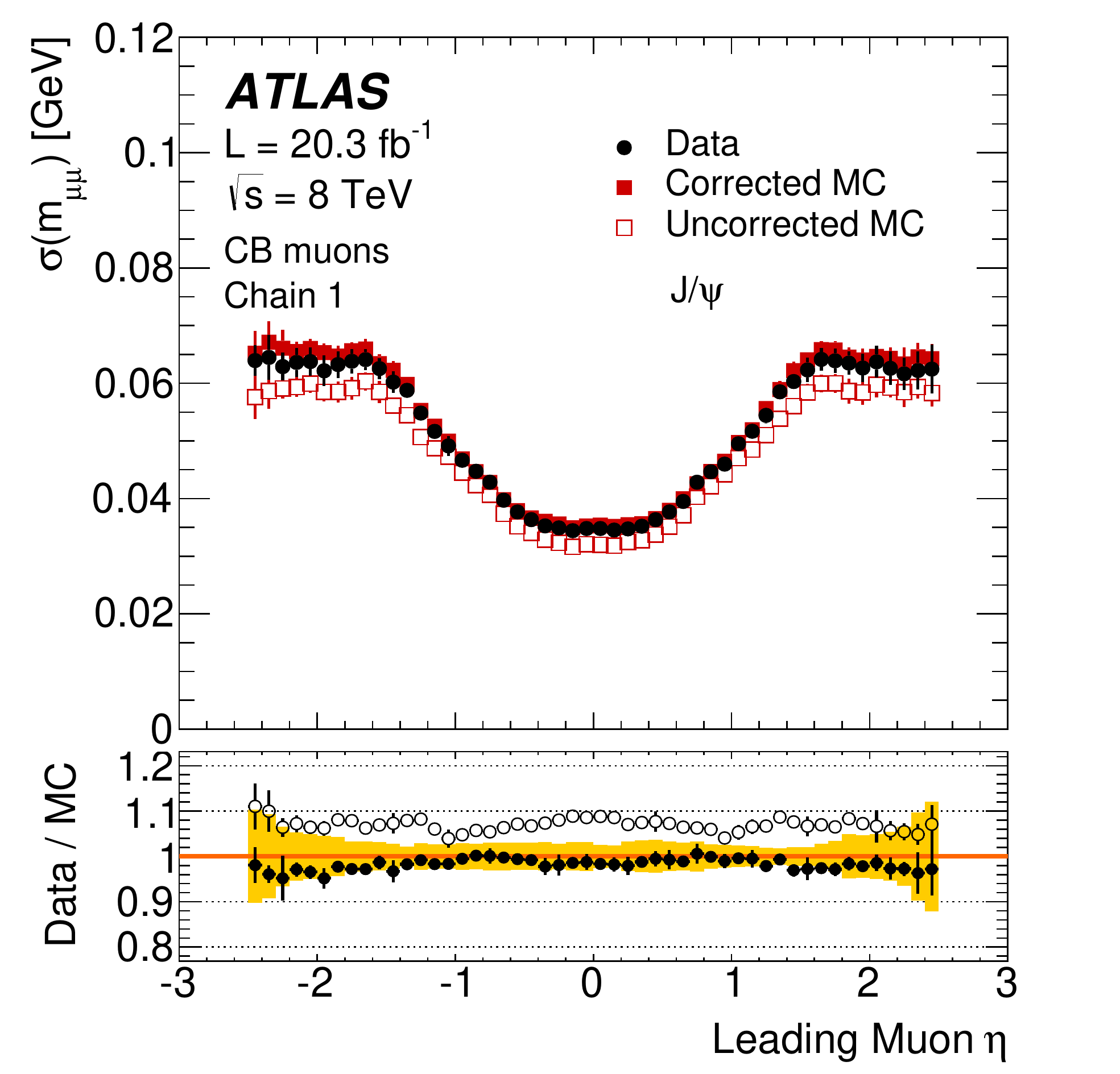}}
  \subfigure[]{\includegraphics[width=0.32\linewidth]{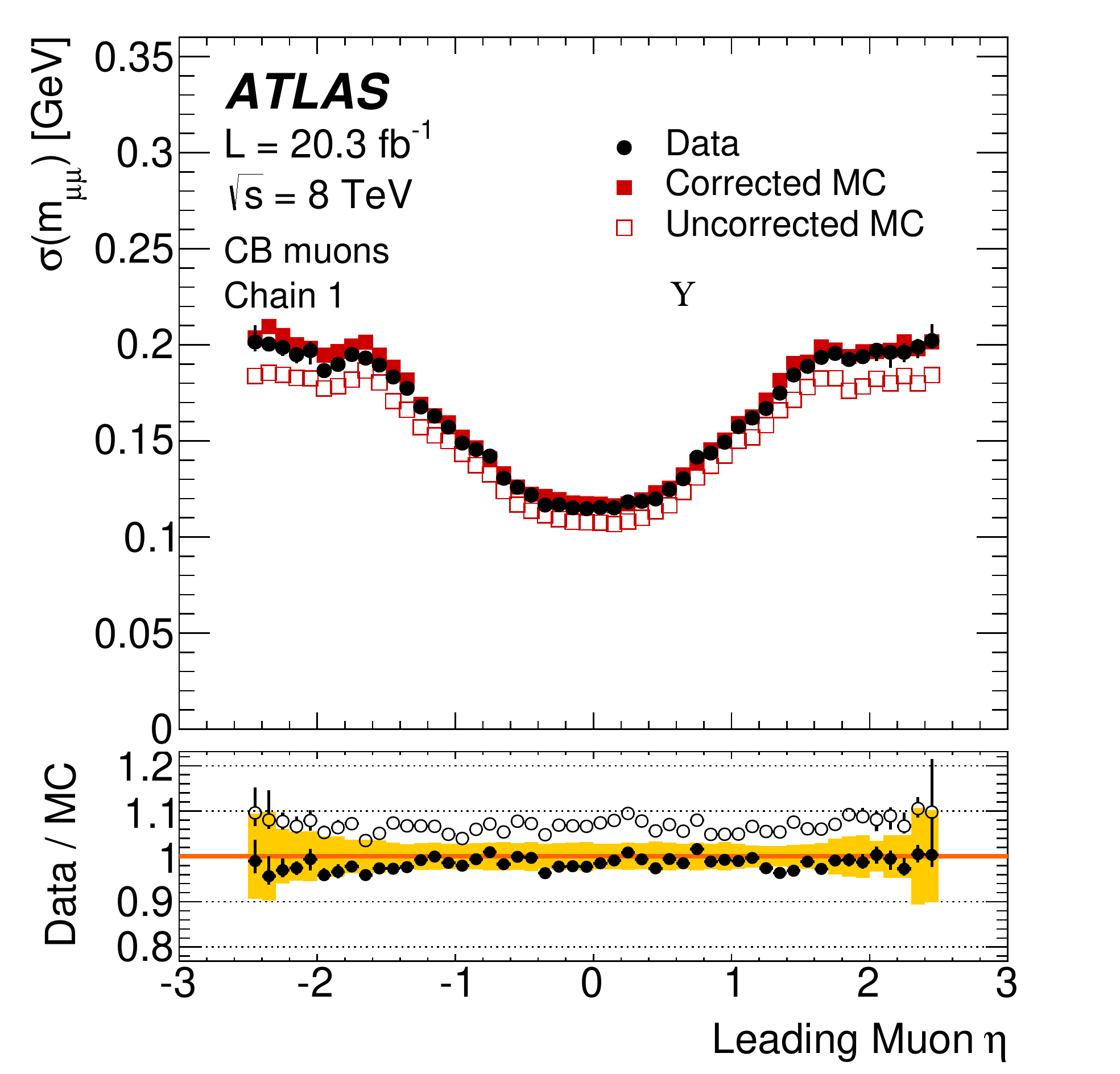}}
  \subfigure[]{\includegraphics[width=0.32\linewidth]{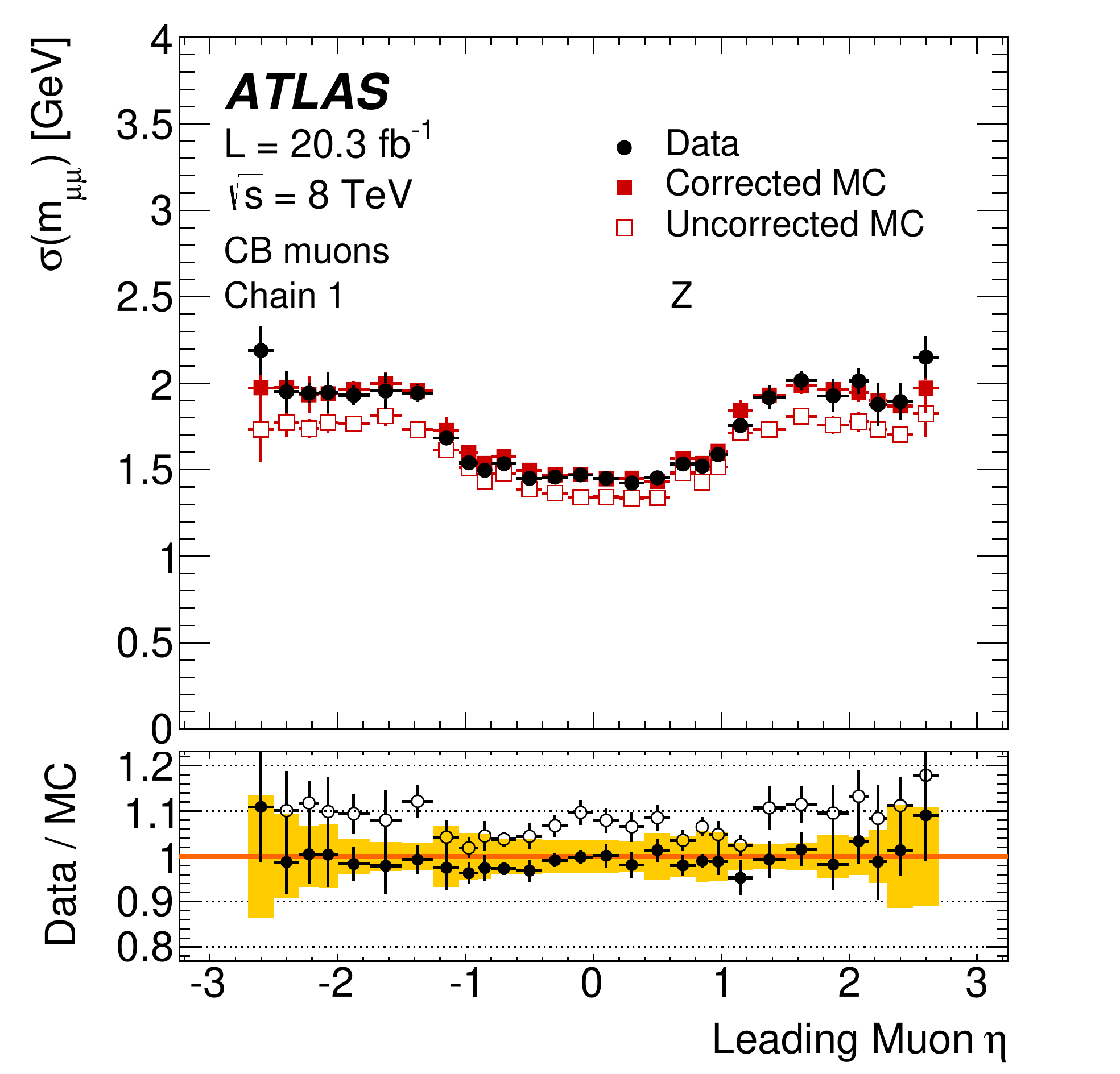}}
    \caption{Dimuon invariant mass resolution for CB muons   for $\Jpsimm$ (a), $\Upsilonmm$ (b) and $\Zmm$ (c) events for
 data and for uncorrected and corrected MC  as a function of the pseudorapidity of the highest-$\pt$ muon.
      The upper plots show the fitted resolution parameter for data, uncorrected MC
      and corrected MC. The lower panels show the data/MC ratio, using uncorrected  and
      corrected MC.  The error bars represent the statistical uncertainty and the systematic uncertainty on the fit added in quadrature. The bands in the lower panels represent the systematic uncertainty on the correction.
}\label{fig:width_eta}
\end{center}
\end{figure*}

The dimuon mass width $\sigma(m_{\mu\mu})$ for CB muons is shown as a function of the leading-muon $\eta$ in Fig.~\ref{fig:width_eta}  for the three resonances. 
The width of the uncorrected MC is 5-10\% smaller than that of the data. After correction the MC reproduces the width of the data well within the correction uncertainties. 

\begin{figure*}[!p]
  \centering
  \includegraphics[width=0.3\linewidth]{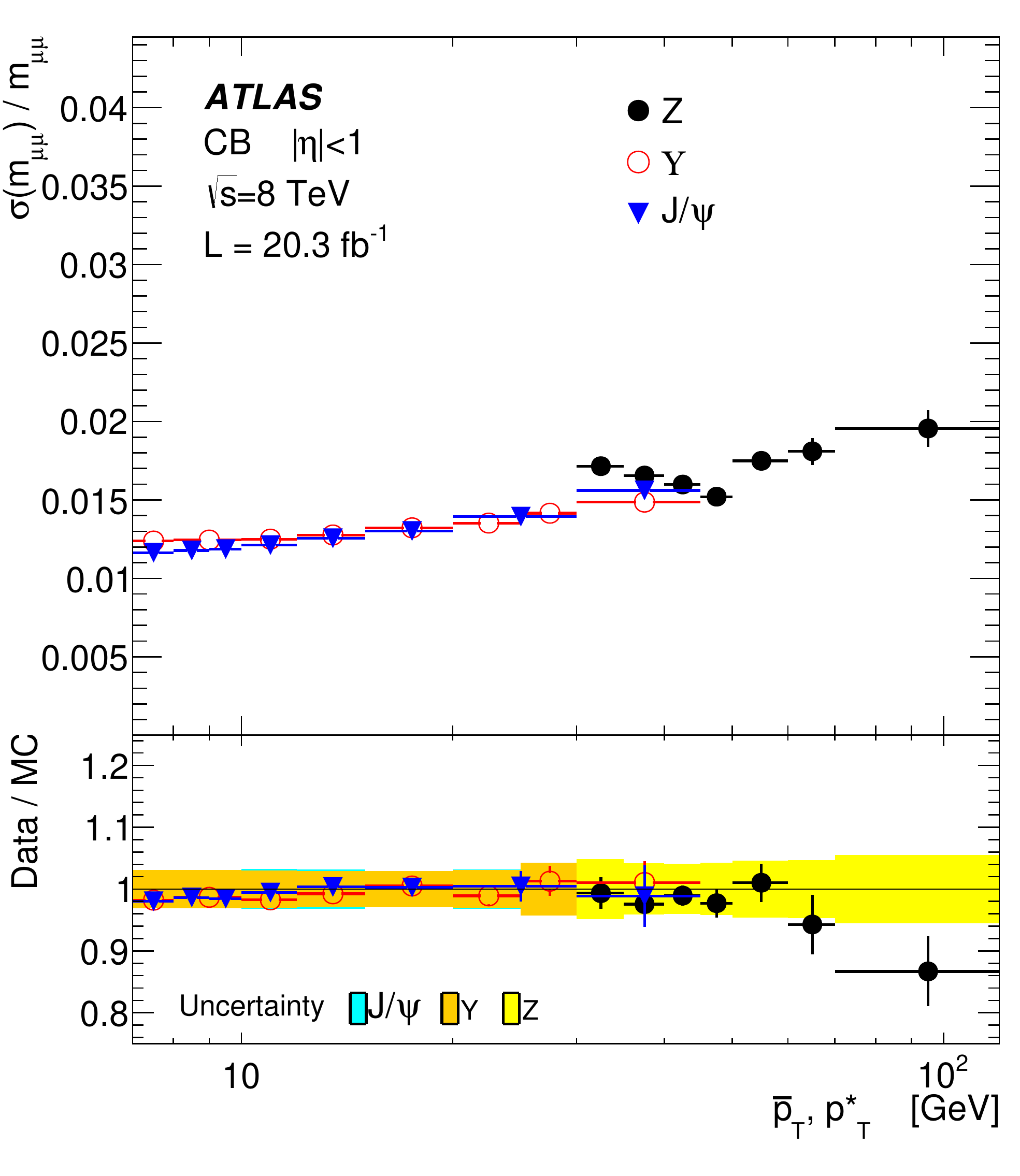}
  \includegraphics[width=0.3\linewidth]{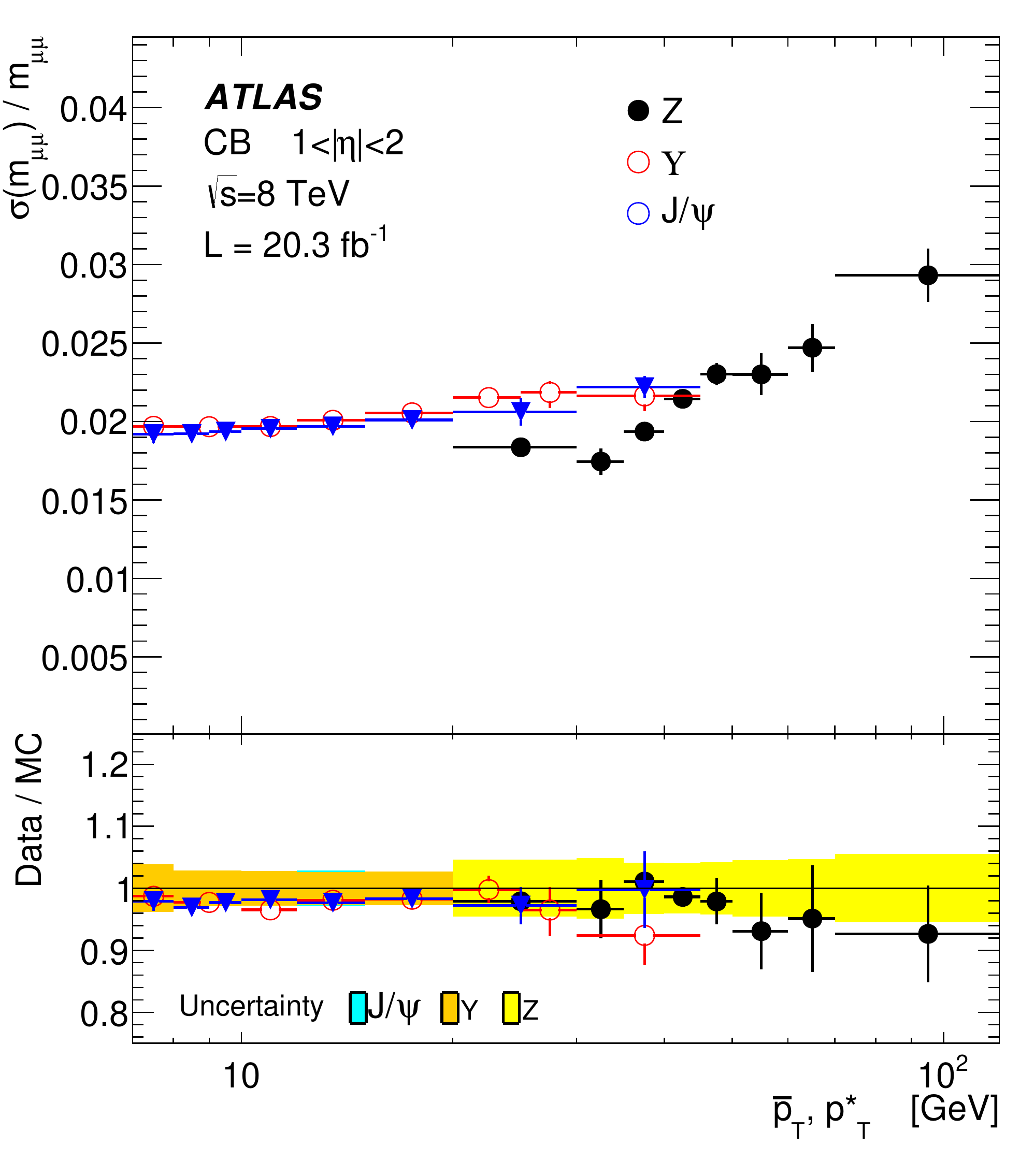}
  \includegraphics[width=0.3\linewidth]{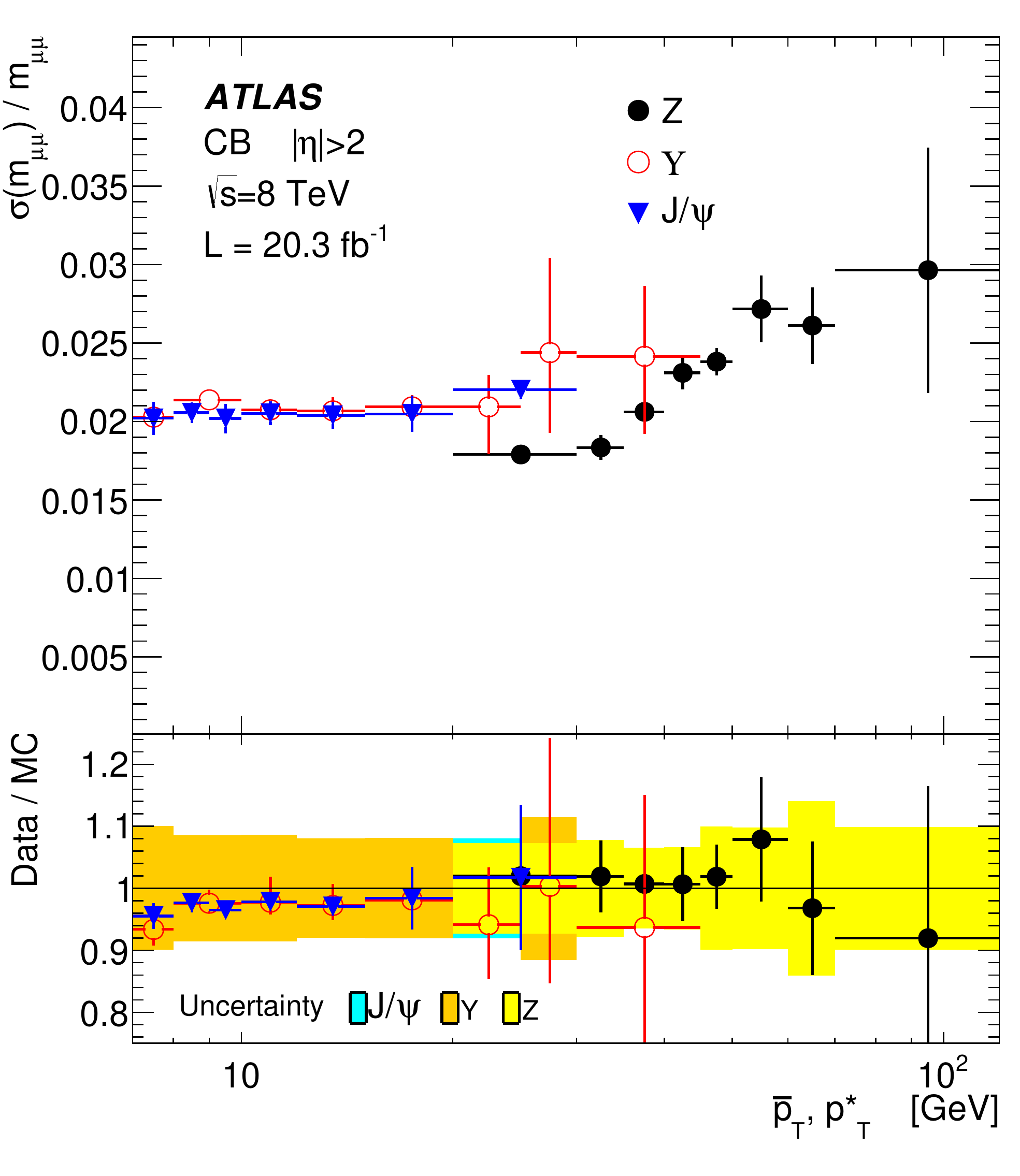}
  \caption{Dimuon invariant mass resolution for CB muons measured  from  $J/\psi$, $\Upsilon$ and  $Z$ events as a  function of the
    average transverse momentum in three $|\eta|$
    ranges.  Both muons are required to be in the same $|\eta|$ range.  The $J/\psi$ and $\Upsilon$ data are plotted as a function
    of  $\bar{p}_{\rm T} = \frac{1}{2}(p_{\mathrm{T},1}+p_{\mathrm{T},2})$ while
    for $Z$ data are plotted as a function of $\pt^*$ as defined in  Eq.~\ref{eq:angles}. The error bars represent statistical and systematic errors added in quadrature.
    The lower panel shows the ratio between data and the corrected MC, with bands representing 
    the uncertainty on the MC corrections  for the three calibration samples.
  \label{f:valid_summary_resolution_pt_cb}}
\end{figure*}

At a given $\eta$, the relative dimuon mass resolution \linebreak 
$\sigma(m_{\mu\mu})/m_{\mu\mu}$ depends approximately on  $\langle \pt \rangle$ (Eq.~\ref{eq:dmom}). This allows a direct comparison of the momentum resolution using different resonances. This is shown in Fig.~\ref{f:valid_summary_resolution_pt_cb}, where the relative mass resolution from  $\Jpsimm$, \linebreak $\Upsilonmm$  and $\Zmm$  events is compared in three regions of $|\eta|$. The $\Jpsimm$ and $\Upsilonmm$ resolutions are in good agreement.

In the $\Zmm$ sample, due to the decay kinematics, below  $\langle \pt \rangle = m_Z /2$ there is a strong correlation between $\langle \pt \rangle$ and the pseudorapidity of the muons, in such a way that the lower is the   $\langle \pt \rangle$, the larger is the $|\eta|$ of the muons. 
Above $\langle \pt \rangle = m_Z /2$, the correlation effect is strongly reduced and the $Z$ measurements are well aligned with those from the lighter resonances. In the barrel region, $|\eta|<1$, the mass resolution increases from $\sigma(m_{\mu\mu})/m_{\mu\mu} \approx 1.2\%$ at $\pt<10$~GeV to $\sigma(m_{\mu\mu})/m_{\mu\mu}  \approx 2\%$ at  $\pt=100$~GeV.
For  $|\eta|>1$ it goes from  $\sigma(m_{\mu\mu})/m_{\mu\mu}  \approx 2\%$ to  $\approx 3\%$ in the same $\pt$ range. This behavior is very well reproduced by the corrected MC.
Following Eq.~\ref{eq:dmom}, it is possible to scale $\sigma(m_{\mu\mu})/m_{\mu\mu}$ by $\sqrt{2}$ to
extract a measurement of the relative momentum resolution $\sigma(p)/p$, which ranges from $\approx 1.7\%$ in the central region and at low $\pt$ to $\approx 4\%$ at large $\eta$ and  $\pt = 100$~GeV.  

\begin{figure*}[!p]
  \centering
 \includegraphics[width=0.3\linewidth]{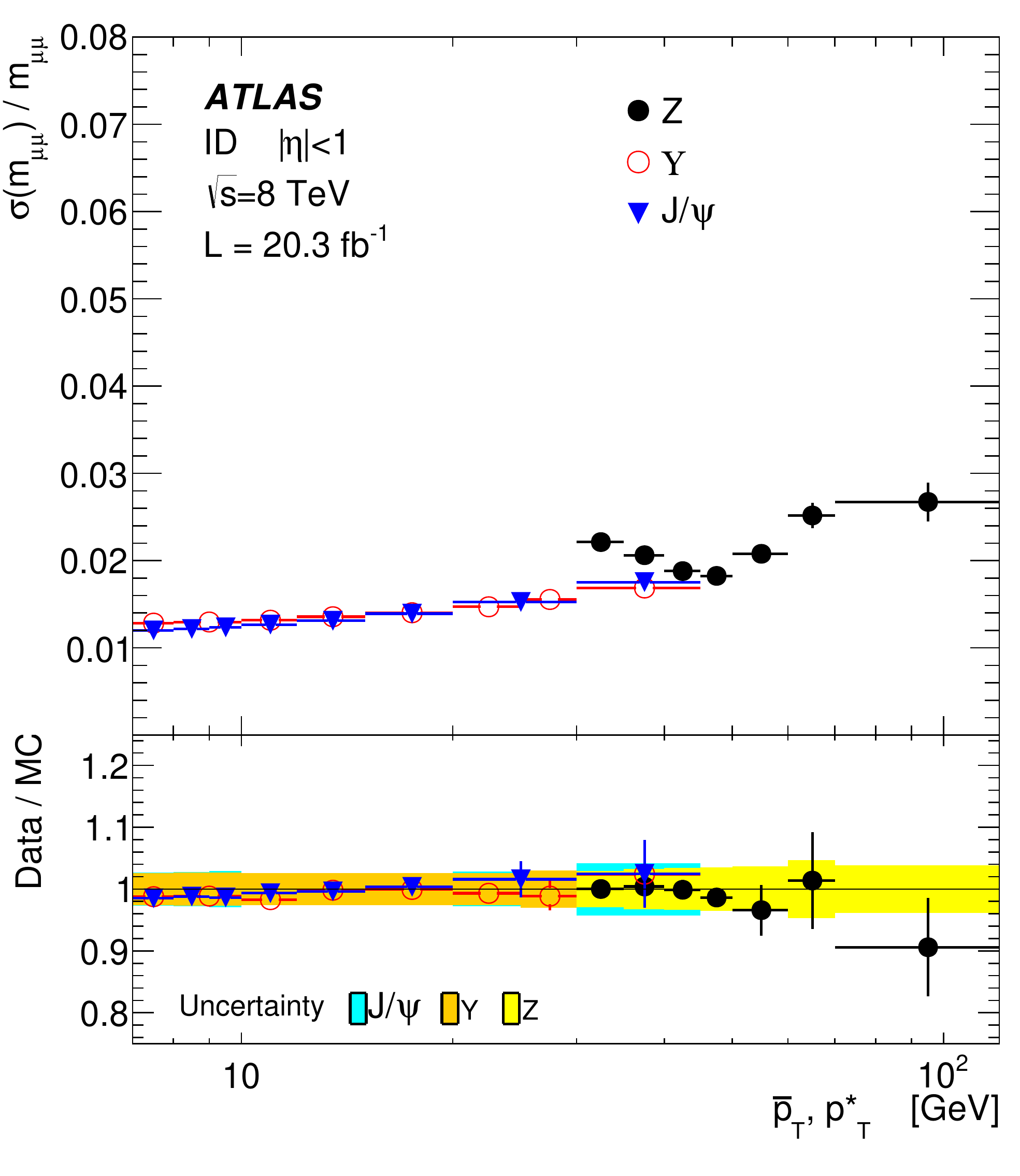}
 \includegraphics[width=0.3\linewidth]{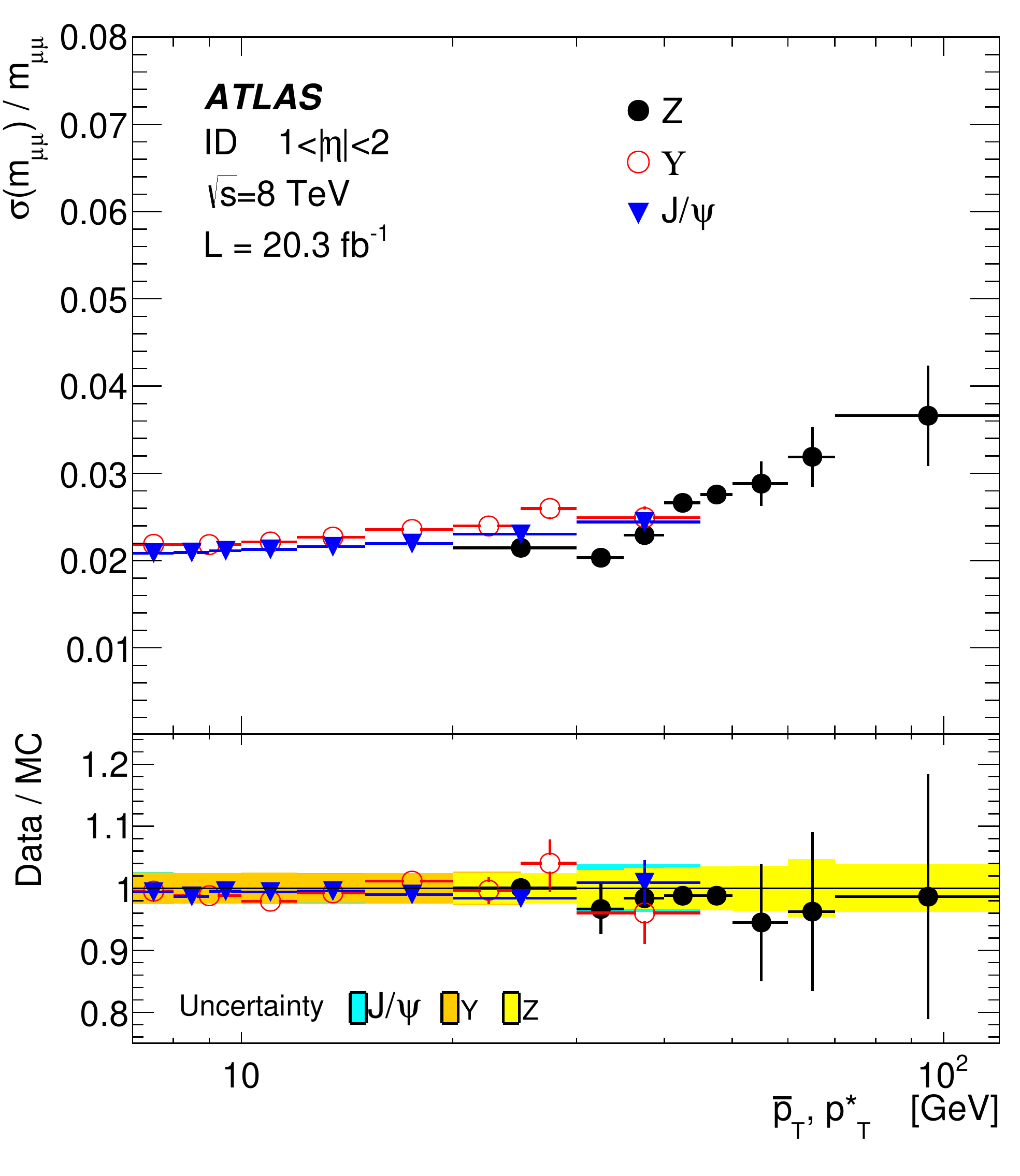}
 \includegraphics[width=0.3\linewidth]{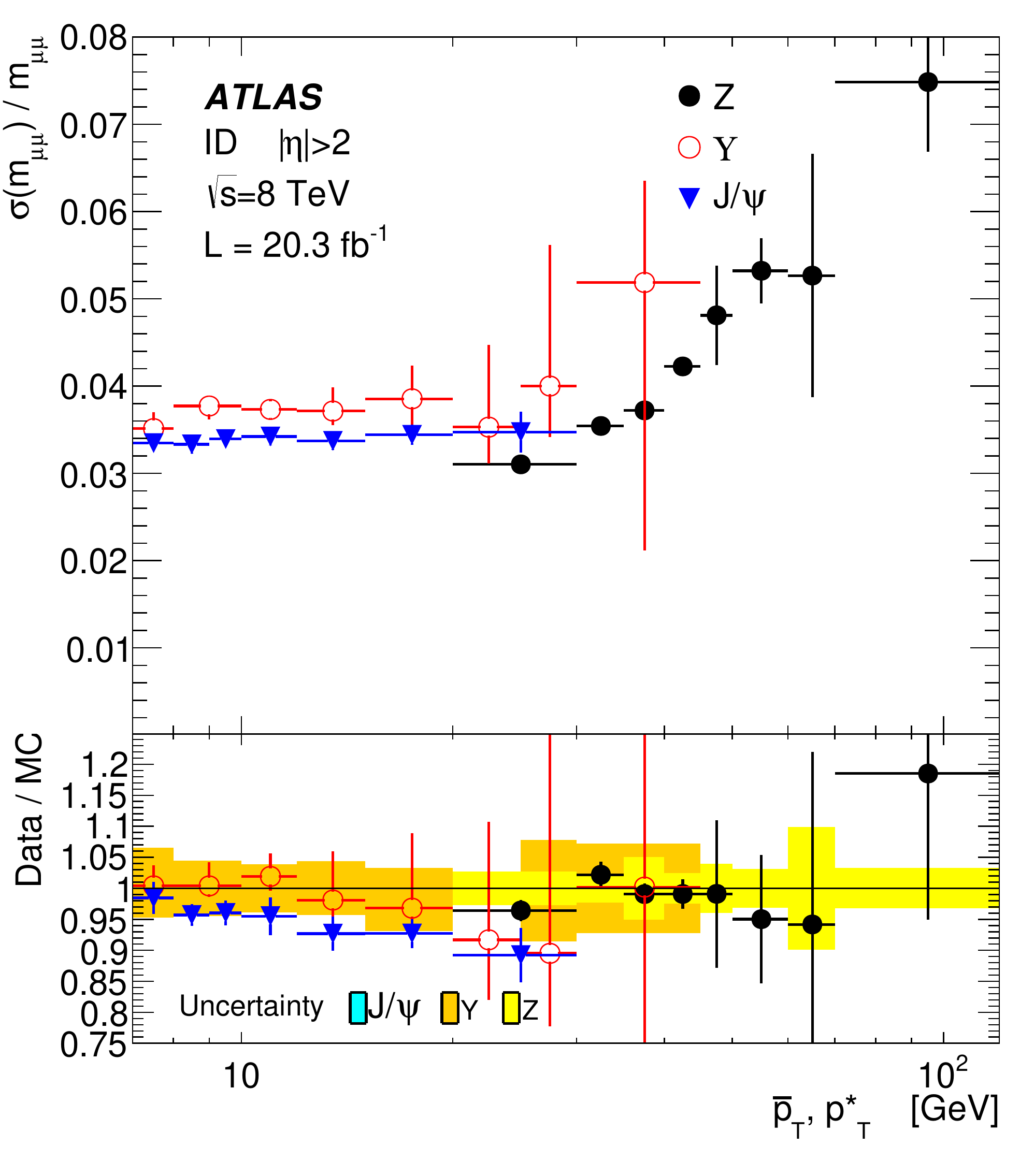}
  \caption{Dimuon invariant mass resolution for muons reconstructed with the ID only, measured  from  $J/\psi$, $\Upsilon$ and  $Z$ events as a  function of the
    average transverse momentum in three $|\eta|$ ranges. Other details as in Fig.~\ref{f:valid_summary_resolution_pt_cb}.
  \label{f:valid_summary_resolution_pt_id}}
\end{figure*}

\begin{figure*}[!p]
  \centering
  \includegraphics[width=0.3\linewidth]{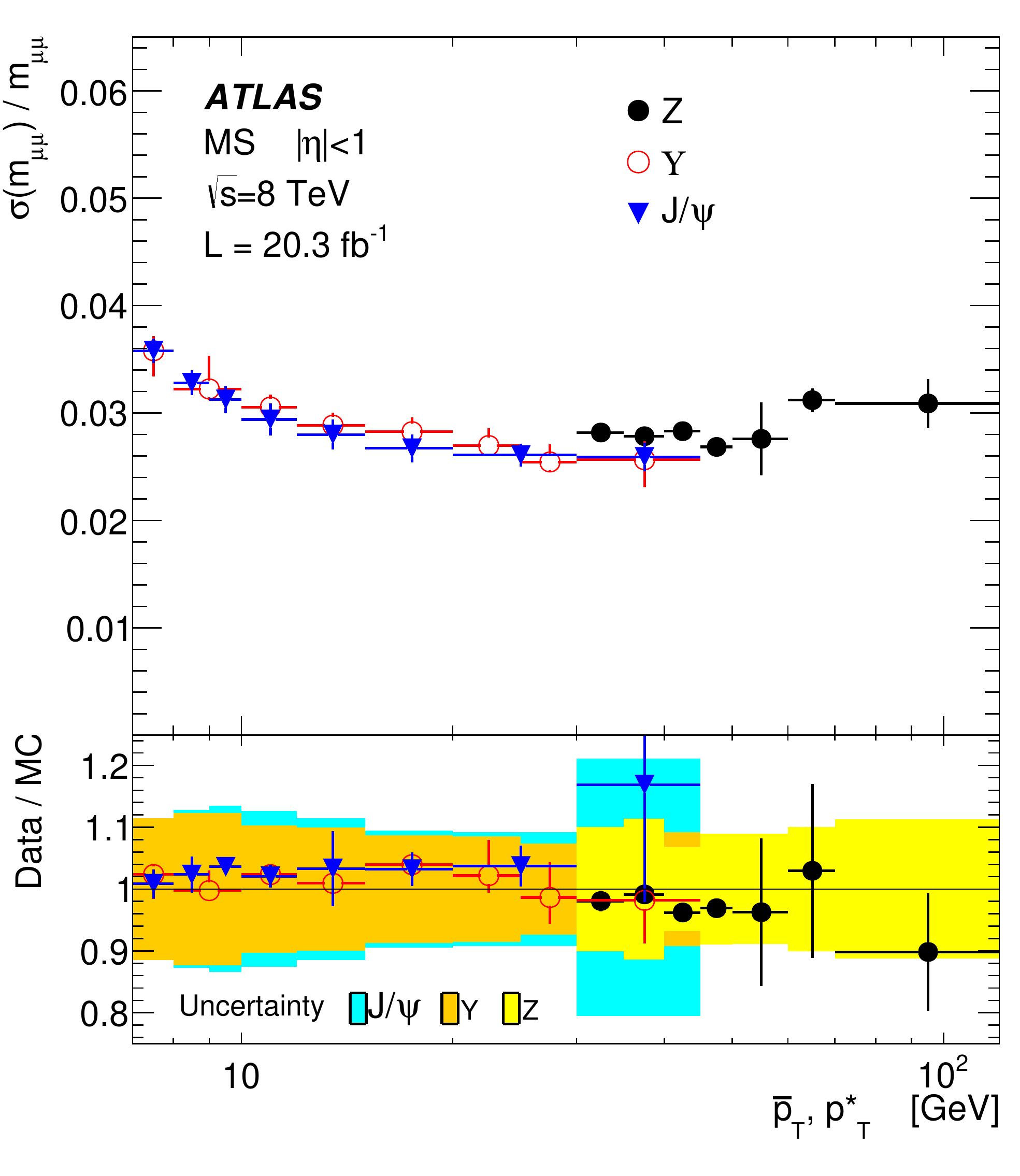}
  \includegraphics[width=0.3\linewidth]{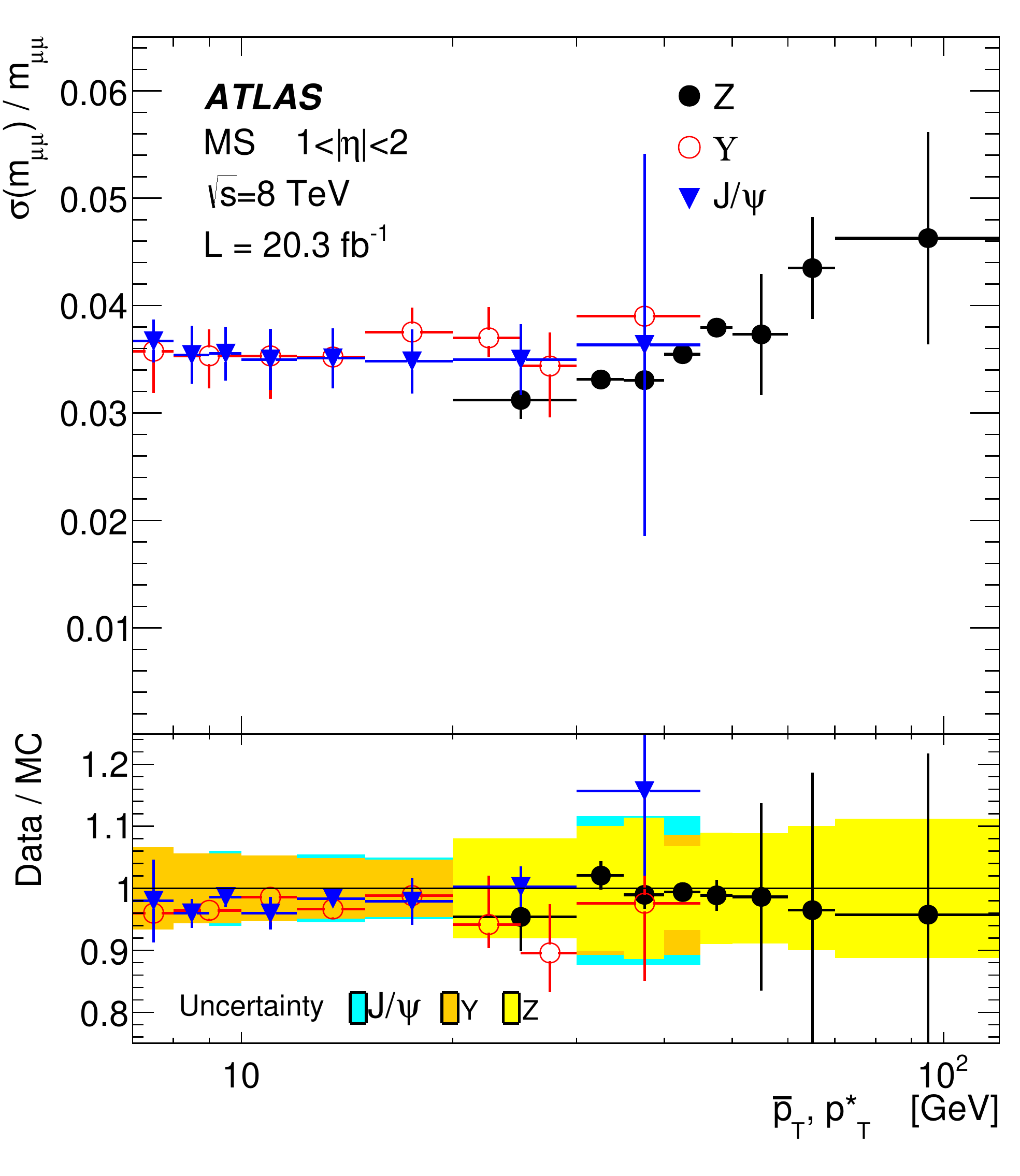}
  \includegraphics[width=0.3\linewidth]{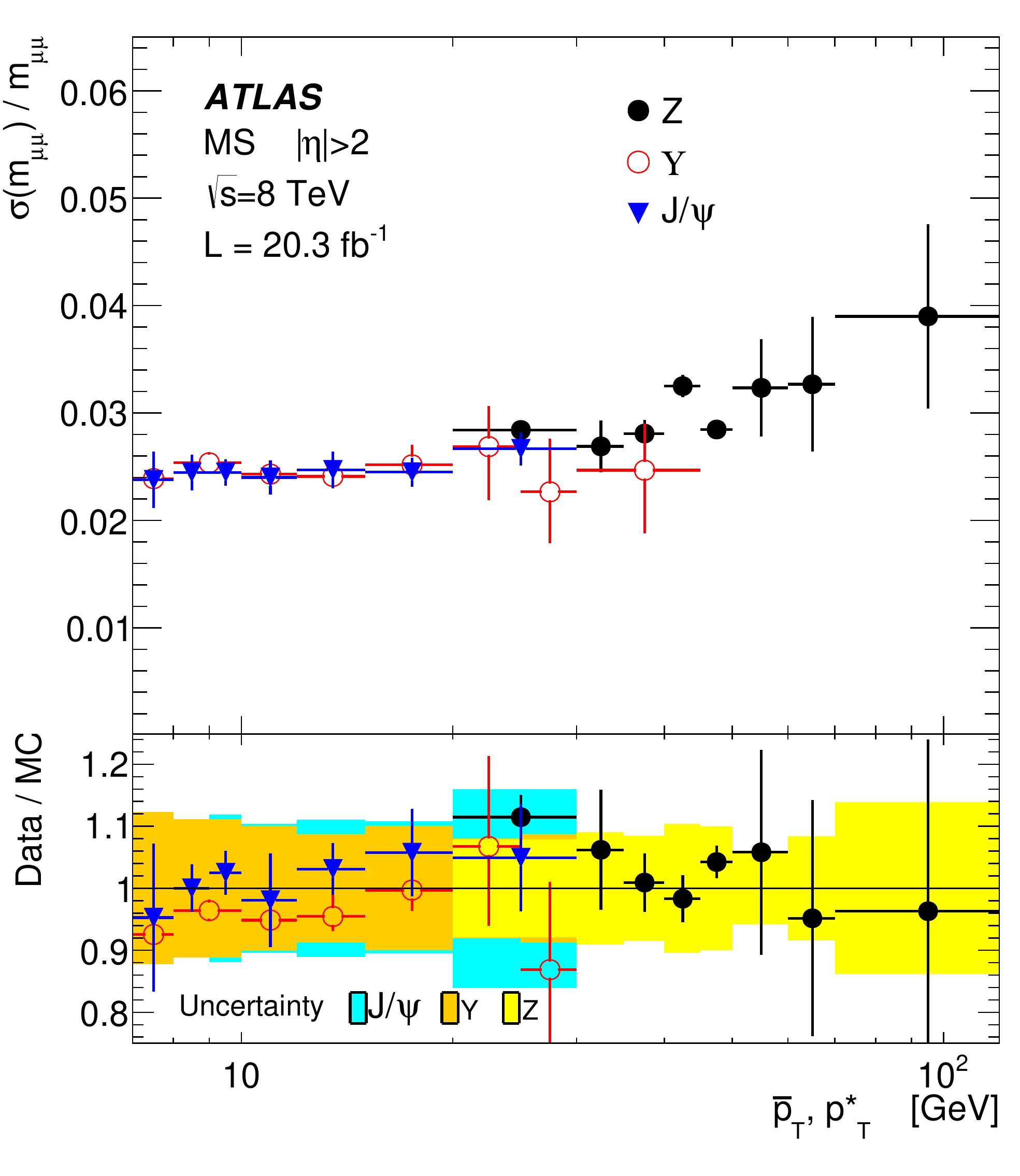}
  \caption{Dimuon invariant mass resolution for muons reconstructed with the MS only, measured  from  $J/\psi$, $\Upsilon$ and  $Z$ events as a  function of the
    average transverse momentum in three $|\eta|$  ranges. Other details as in Fig.~\ref{f:valid_summary_resolution_pt_cb}.
  \label{f:valid_summary_resolution_pt_ms}}
\end{figure*}

To understand better the $\pt$ dependence of the momentum resolution of CB muons, it is useful to study separately the resolution of the ID and of the MS measurements, as shown in Fig.~\ref{f:valid_summary_resolution_pt_id} and~\ref{f:valid_summary_resolution_pt_ms}.   The ID measurement has a better resolution than the MS in the $\pt$ range under study for $|\eta|<2$ while the MS has a better resolution at larger $|\eta|$. The resolution of the CB muons is significantly
better than the ID or the MS measurements taken separately in the whole $|\eta|$  range. The ID resolution has an approximately linear increases with $\pt$, corresponding to a non-zero $r_2$ term in Eq.~\ref{eq:qoverpt_reso}.  The MS resolution is largest in the region $1<|\eta|<2$ which contains the areas with the lowest magnetic field integral. In the region $|\eta|<1$ there is a visible increase at low $\pt$ that corresponds to the presence of a non-zero $r_0$ term in Eq.~\ref{eq:qoverpt_reso}.  The $\pt$
dependence of the resolutions for both the ID and the MS measurements is well reproduced by the corrected MC.
According to studies based on MC, the MS measurement is expected to dominate over the ID in the whole $|\eta|$ range for sufficiently large $\pt$.

\section{Final State Radiation recovery}
\label{Sec:FSR}
The invariant mass distributions of resonances that decay into
muons, such as $\Zmm$ and $H\rightarrow
ZZ\rightarrow 4\ell$, is affected by QED final state
radiation of photons,  causing the mass reconstructed using muons
to be shifted to lower values. 
In this section, a dedicated method to include FSR photons in
the reconstruction of resonances decaying into muons is introduced and 
tested with $\Zmm$ data. This method has been used in several ATLAS publications~\cite{Higgs-mass,Aad:2014fia}.

Final state radiation photons emitted collinearly to muons
can be reconstructed with the LAr calorimeter:
electromagnetic clusters are searched for within a narrow cone around the
axis defined by the muon momentum direction at the interaction point
(i.e. the direction which would be followed by an uncharged particle). The longitudinal segmentation of the LAr
calorimeter is exploited to reduce fake photon clusters produced
by muon energy losses in the calorimeter. This is achieved by using as
a discriminant the fraction $\mathrm{f_1}$ of the cluster energy
deposited in the first segment of the calorimeter divided by
the total cluster energy. 
Collinear FSR photon candidates are required to have 
$E_{T}>1.5$~GeV, $\Delta R_{\mathrm{cluster},\mu}<0.15$ and $\mathrm{f_1}>0.1$.
In addition, non-collinear FSR photons are recovered using the standard ATLAS photon reconstruction,
selecting isolated photons emitted with  $\Delta R_{\mathrm{cluster},\mu}>0.15$ and 
with $E_{T}>10$~GeV~\cite{photonCx}.

The effect of adding a collinear or non-collinear  FSR photon to the
$\Zmm$ invariant mass in data is studied in a sample obtained with a
dedicated selection of  $\Zmm$ candidates plus at least one radiated photon candidate.

The correction for collinear FSR is applied for events in the mass
window $66$~GeV$ < m_{\mu\mu} <89$~GeV
while the correction for non-collinear FSR photons is applied only if the collinear search has failed and the dimuon mass
satisfies $m_{\mu\mu} < 81$~GeV. 

\begin{figure}[!htb]
\centering
\includegraphics[width=0.8\linewidth]{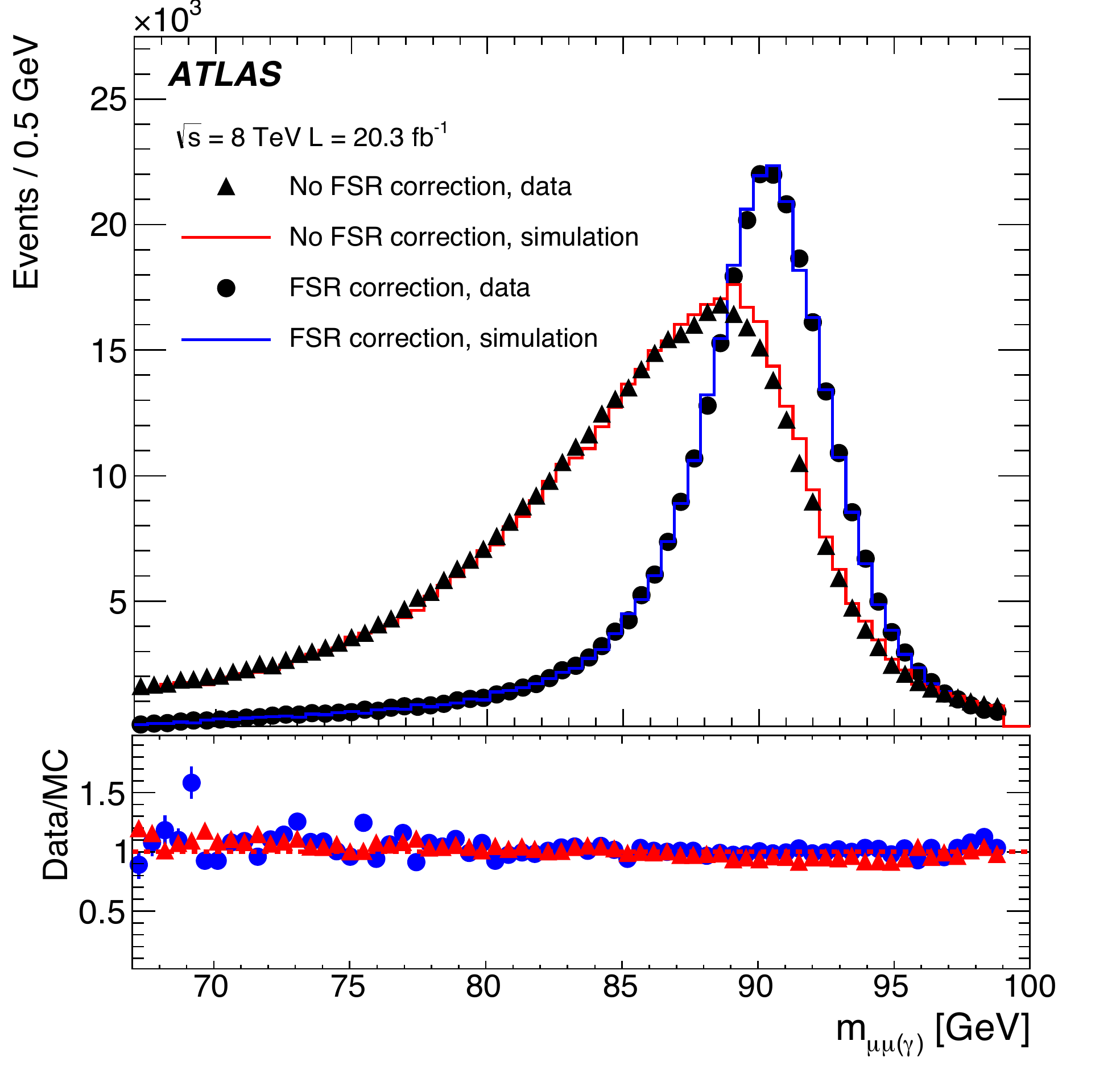}
\includegraphics[width=0.8\linewidth]{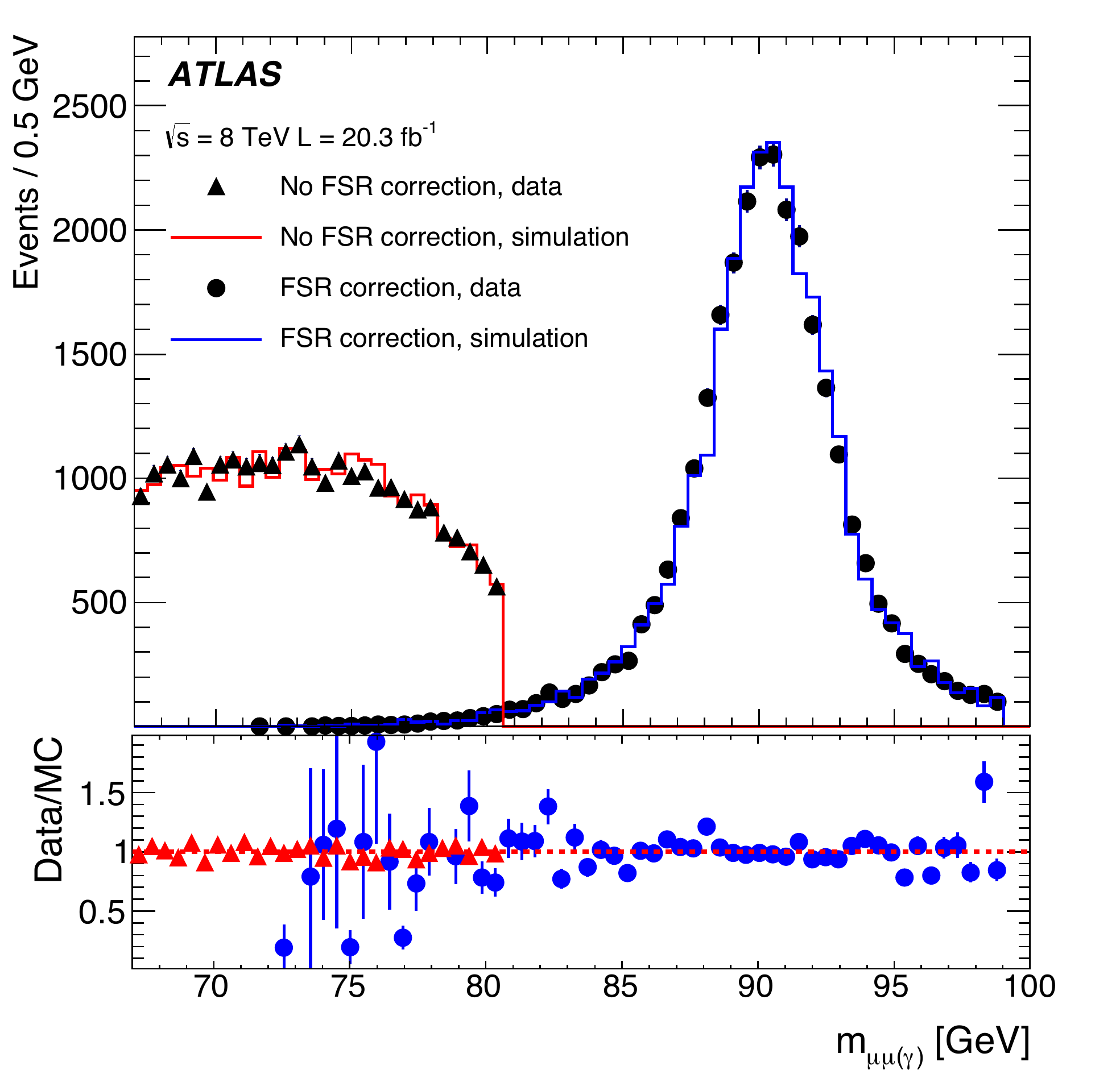}
\caption{Invariant mass distribution of $\Zmm$ events with identified FSR in data
  before (filled triangles) and after (filled circles) FSR correction, for collinear (top) and non-collinear (bottom) FSR.
  The MC prediction is shown before correction (red histogram) and after
  correction (blue histogram). 
}
\label{fig:FSRZMassCorrection}
\end{figure}

In Fig.~\ref{fig:FSRZMassCorrection} the invariant mass
distributions for the sample of $\Zmm$ events with a FSR photon
candidate are shown  before and after the addition of
collinear and non-collinear FSR photons.
A good agreement between data and MC is observed for the corrected $\Zmm$ events.
According to MC studies, the collinear FSR selection has an efficiency
of $70 \pm 4$\% for FSR photons emitted with $E_{T}>1.5$~GeV and $\Delta
R_{\gamma,\mu}<0.15$ in the fiducial region defined requiring $|\eta|<2.37$ and excluding the calorimeter crack region $1.37< |\eta| < 1.52$.
About $85\%$ of the corrected events have genuine FSR photons, with the remaining 
photons coming from muon bremsstrahlung or ionization or from random
matching with energy depositions from other sources. The fraction of
all $\Zmm$ events corrected with a collinear FSR photon is $\simeq$ 4\%. 
The non-collinear FSR selection has an efficiency of $60\pm 3$\% in
the fiducial region and a purity of $\geq 95$\%. The fraction of  $\Zmm$ events corrected with a 
non-collinear FSR photon is $\simeq 1$\%.

The FSR correction may introduce systematic variations in the
invariant mass scale and resolution. To study these effects,
a Gaussian fit of the $\Zmm$ distribution has been performed
in the mass range $91.18 \pm 3.00$~GeV. 
The FSR correction induces a mass shift of $+40\pm3$~MeV and an improvement of the resolution of $3 \pm 1$\% in
the full $\Zmm$ sample. The effects observed in the data are well reproduced by the MC.
The systematic uncertainty introduced by the FSR recovery on the inclusive $Z$ mass scale can be understood by considering a 0.5\% photon energy scale uncertainty, the fact that
only 5\% of the $Z$ events are corrected, and that the fraction of
energy carried by the photons is a few \%. This leads to a systematic uncertainty smaller than 2~MeV.

The effect of pile up on the FSR correction has been estimated by 
dividing the data and the MC into three categories based on the average
number of interactions per bunch crossing: $\langle \mu \rangle=$0-17, 17-23, 23-40. 
A comparison of the fitted $Z$ mass between data and MC has been 
performed in the three categories and no dependence on  $\langle \mu \rangle$ was observed.
Good agreement between data and MC within the statistical uncertainties was found.

\section{Conclusions}
\label{sec:conclusions}

The performance of the ATLAS muon reconstruction has been measured using data from LHC $pp$ collisions at $\sqrt{s} = 7-8$~TeV.
The muon reconstruction efficiency is close to $99\%$ over most of the pseudorapidity range of $|\eta|<2.5$ and for $\pt>10$~GeV.
The large collected sample of 9M  $\Zmm$ decays allows the measurement of the efficiency over the full acceptance of $|\eta|<2.7$, and with a precision at the 1 per-mille level for $|\eta|<2.5$.
By including $\Jpsimm$ decays, the efficiency  measurement  has been extended over the transverse momentum range from $\pt \approx 4$~GeV to $\pt \approx 100$~GeV.

The muon momentum scale and resolution has been studied in detail using large calibration samples of \linebreak
 $\Jpsimm$, $\Upsilonmm$ and  $\Zmm$ decays.  These studies have been used to correct the MC simulation to improve the data-MC agreement and to minimize the uncertainties in physics analyses.
The momentum scale for combined muons is known with an uncertainty of $\pm 0.05\%$
for \mbox{$|\eta| <1$},  which  increases to \mbox{$\lesssim 0.2\%$} for  \mbox{$|\eta|>2.3$} for $\Zmm$ events. 
The dimuon mass resolution is $\approx 1.2\%$ ($2\%$)  at low-$\pt$ increasing to $\approx 2\%$ ($3\%$) at $\pt \approx 100$~GeV for $|\eta|<1$ ($|\eta|>1$). The resolution is reproduced by the corrected simulation within relative uncertainties  of $3\%$ to $10\%$ depending on $\eta$ and $\pt$.

The mass resolution for the $\Zmm$ resonance was found to improve when photons from QED final state radiation are recovered. The FSR recovery allows to recover $\approx 4\%$ of the events from the low-mass tail to the peak region, improving the dimuon mass resolution by $\approx 3\%$.

\begin{appendix}
\section{Results with different reconstruction ``Chains''}\label{appendix:A}

\begin{figure}[hbt]
\begin{center}
  {\includegraphics[width=0.9\linewidth]{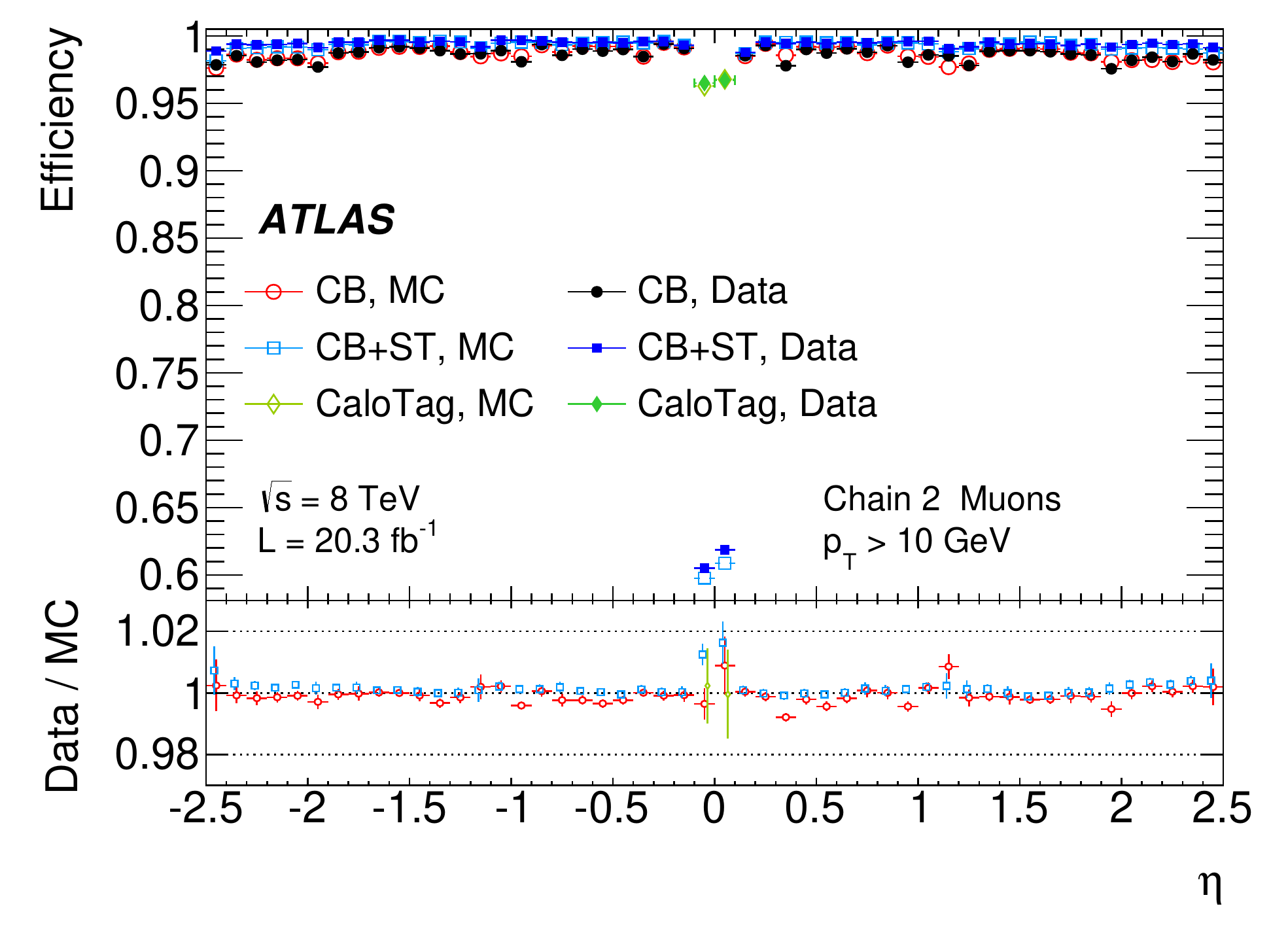}}
  {\includegraphics[width=0.9\linewidth]{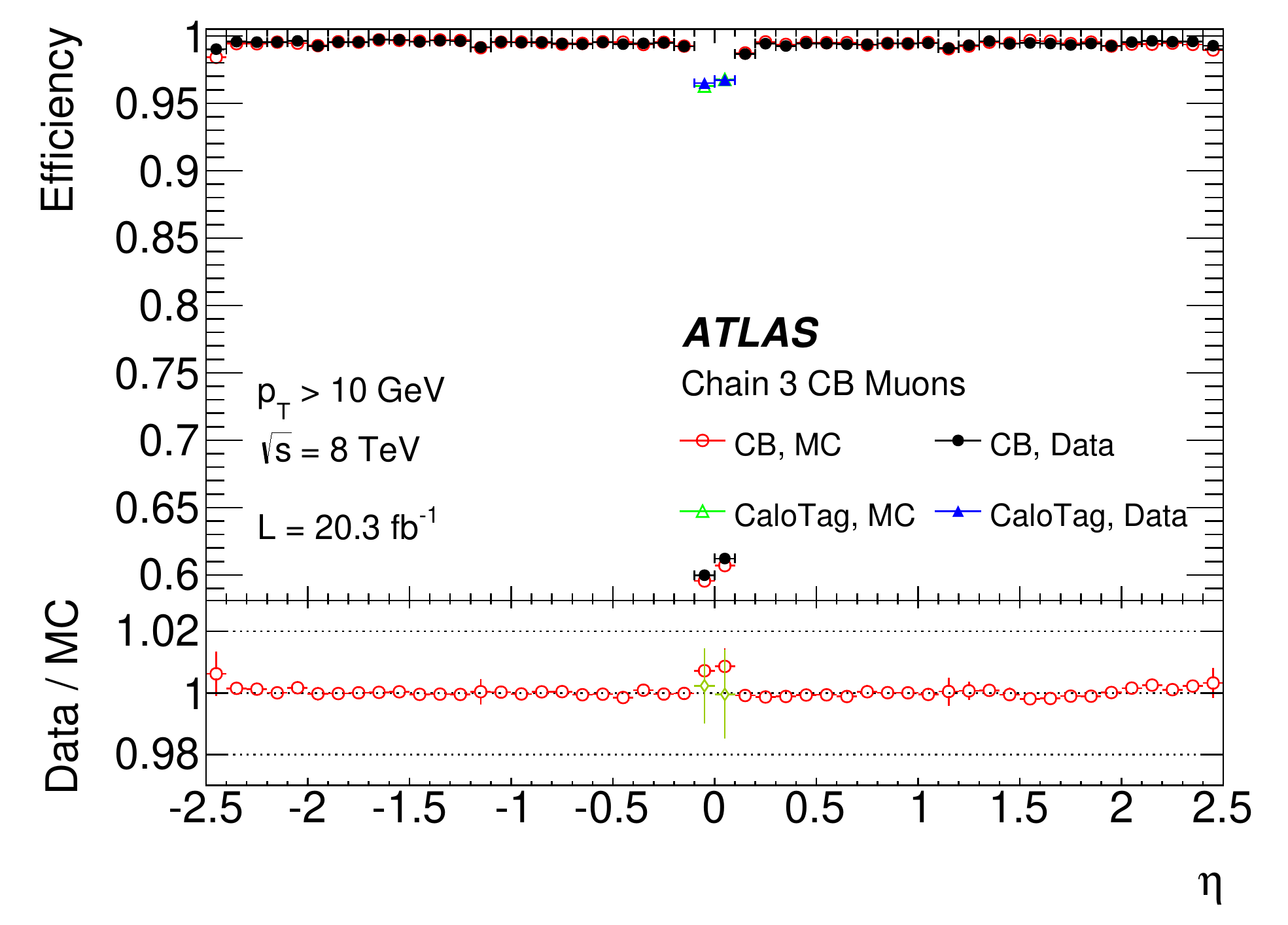}}
    \caption{Muon reconstruction efficiency as a function of $\eta$, measured using $\Zmm$ events, for muons reconstructed with Chain-2 (top) and Chain-3 (bottom), for different muon reconstruction types.    CaloTag muons are shown in the region $|\eta|<0.1$, where they are used in physics analyses. The error bars shown for the efficiencies represent the statistical uncertainty. The panel at the bottom shows the ratio between the measured and predicted efficiencies. The error bars show statistical and systematic uncertainties added in quadrature. }\label{Fig:AllTypeEff_Chain23}
\end{center}
\end{figure}

This appendix reports the main results obtained with the other two muon reconstruction software packages used to process 2012 data, Chain~2 and the  unified  reconstruction programme Chain-3.
Figure~\ref{Fig:AllTypeEff_Chain23}  shows the efficiency as a function of $\eta$ for Chain~2 and Chain~3 and is similar to Fig.~\ref{Fig:AllTypeEff_staco}
for Chain~1. 
The efficiency drop that is observed in Chain~1 for CB muons at $|\eta| \simeq 1.2$ is not present in the other two packages due to the less strict selection on the number of measurements in the MS. These relaxed requirements also improve the data/MC agreement. In Chain~2 the CB+ST efficiency is higher than the CB efficiency alone, similarly to Chain~1. For Chain~3,  the distinction between CB and ST muons is not applicable anymore since a ID-MS combined momentum fit is performed also in the case of muons that traversed only one MS chamber, a category that is assigned to ST muons in Chain~1 and (with some exceptions) in Chain~2. Therefore  only one type of Chain~3 muons is considered, which was tuned to provide a purity similar to that of the CB muons of Chain~1.

The momentum resolution of the three chains is very similar, with Chain~3 having approximately 2\% better resolution than Chain~1.
The data/MC agreement and the amount of correction applied to the simulation is  compatible among the three packages.

\section{Results on 2011 data}
\label{appendix:B}
During the 2011 data taking period, the LHC delivered $pp$ collisions at a center of mass energy of $\sqrt{s}=7$~TeV. A sample corresponding to
an integrated luminosity of  $4.5$~fb$^{-1}$ has been used to measure the muon reconstruction performance with 2011 data.
The ID and MS configurations were the same in 2011 as in 2012, with the exception of additional MDT chambers installed between the two periods to increase the number of MS layers from one to two at $\eta=-1.2$ and in part of the region at $\eta=1.2$. 
The trigger thresholds were in general lower in 2011. The reconstruction programs used for 2011 data were similar to those used in 2012, although several improvements have been introduced between the two periods. Tighter requirements on the ID tracks associated to the muon track were applied in 2011. Similar MC samples as those used for the study of 2012 data have been generated at $\sqrt{s}=7$~TeV for the study of muon performance in 2011, using the same simulation based on GEANT4. The reconstruction of the 2011 simulated data was performed with ideal alignment in the MS. 

\begin{figure}[!b]
\begin{center}
  {\includegraphics[width=0.9\linewidth]{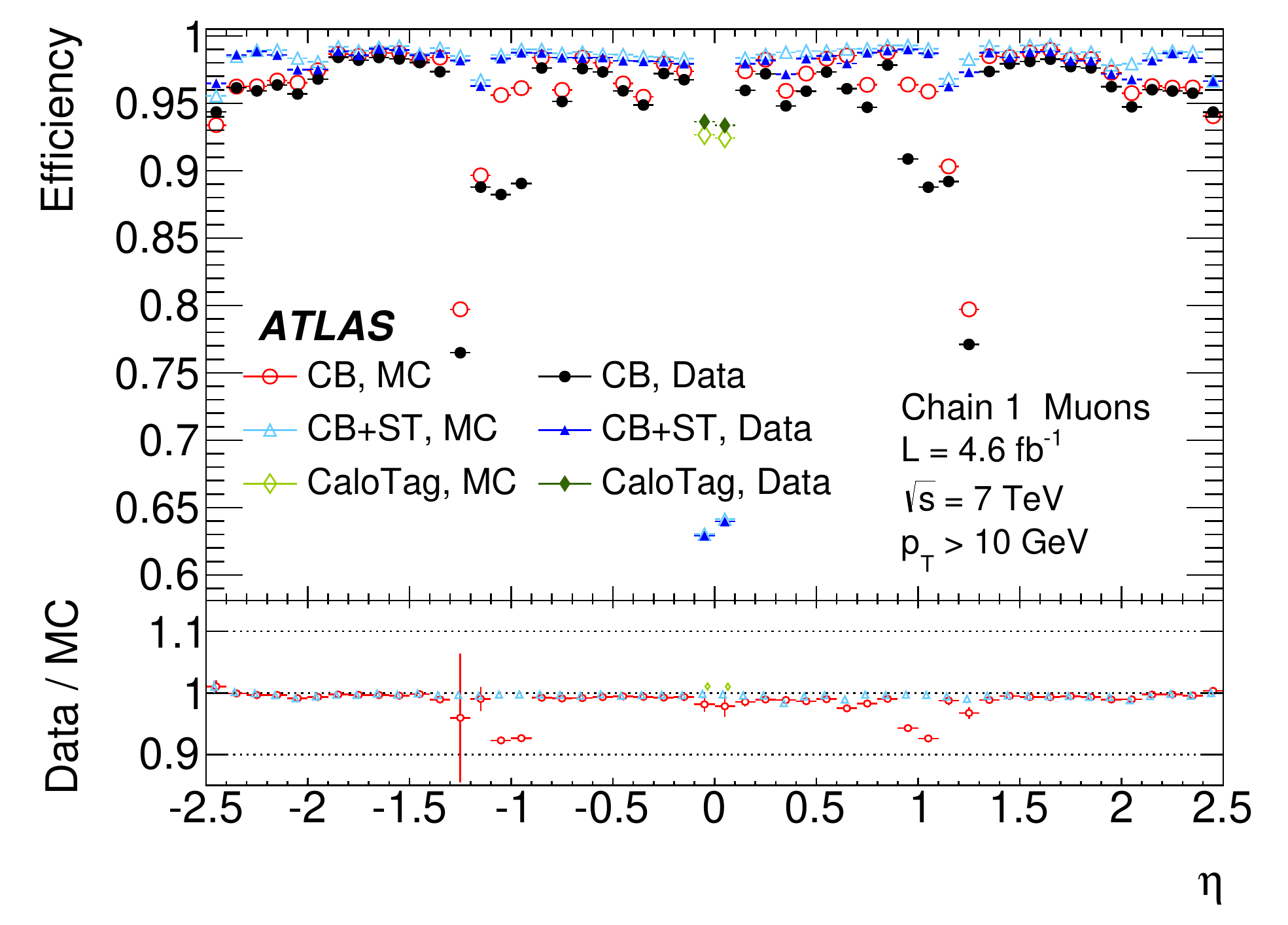}}
    \caption{Muon reconstruction efficiency as a function of $\eta$ measured in $\Zmm$ events in the 2011 data sample for different muon reconstruction types.
    CaloTag muons are only shown in the region $|\eta|<0.1$, where they are used in physics analyses. For the efficiency, the error bars indicate the statistical uncertainty. The panel at the bottom shows the ratio between the measured and MC efficiencies. The error bars on the ratios show the combination of statistical and systematic uncertainties.  The lower efficiency of CB muons at $|\eta|\approx 1.2$ is due to the fact that some of the MS chambers were not yet installed. }\label{fig:eff_2011}
\end{center}
\end{figure}

\begin{figure}[!hb]
  \centering
  \includegraphics[width=0.99\linewidth]{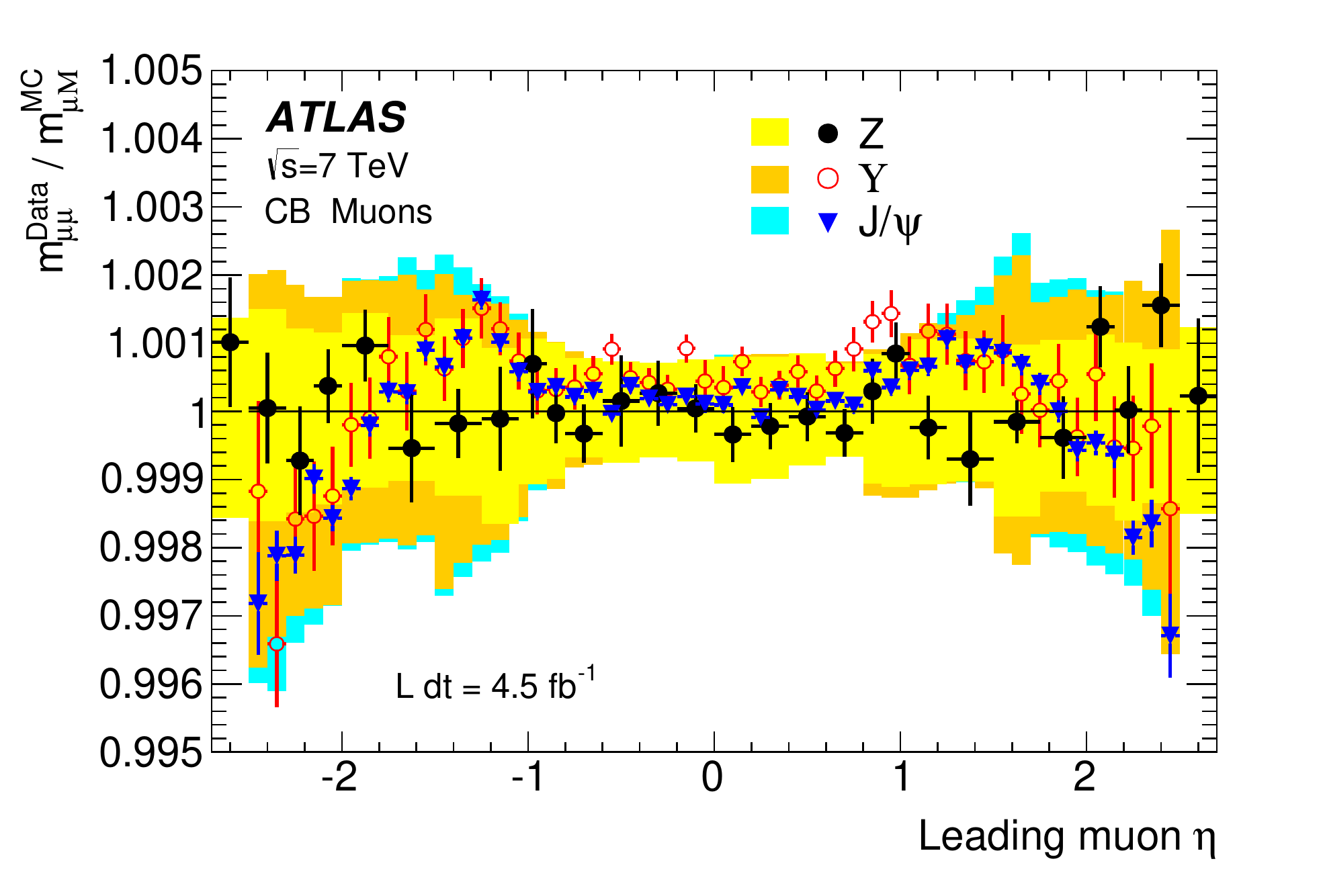}
  \caption{Ratio of the fitted mean mass,  $\langle m_{\mu\mu} \rangle$, for data and corrected MC in the 2011 data samples.  Measurements from $J/\psi$, $\Upsilon$ and  $Z$ events are shown as a  function of $\eta$ of the highest-$\pt$ muon. The bands show the uncertainty on the MC corrections extracted for the three calibration samples.}
  \label{fig:scale_2011}
\end{figure}

\begin{figure}[!hb]
  \begin{center}
   \includegraphics[width=0.85\linewidth]{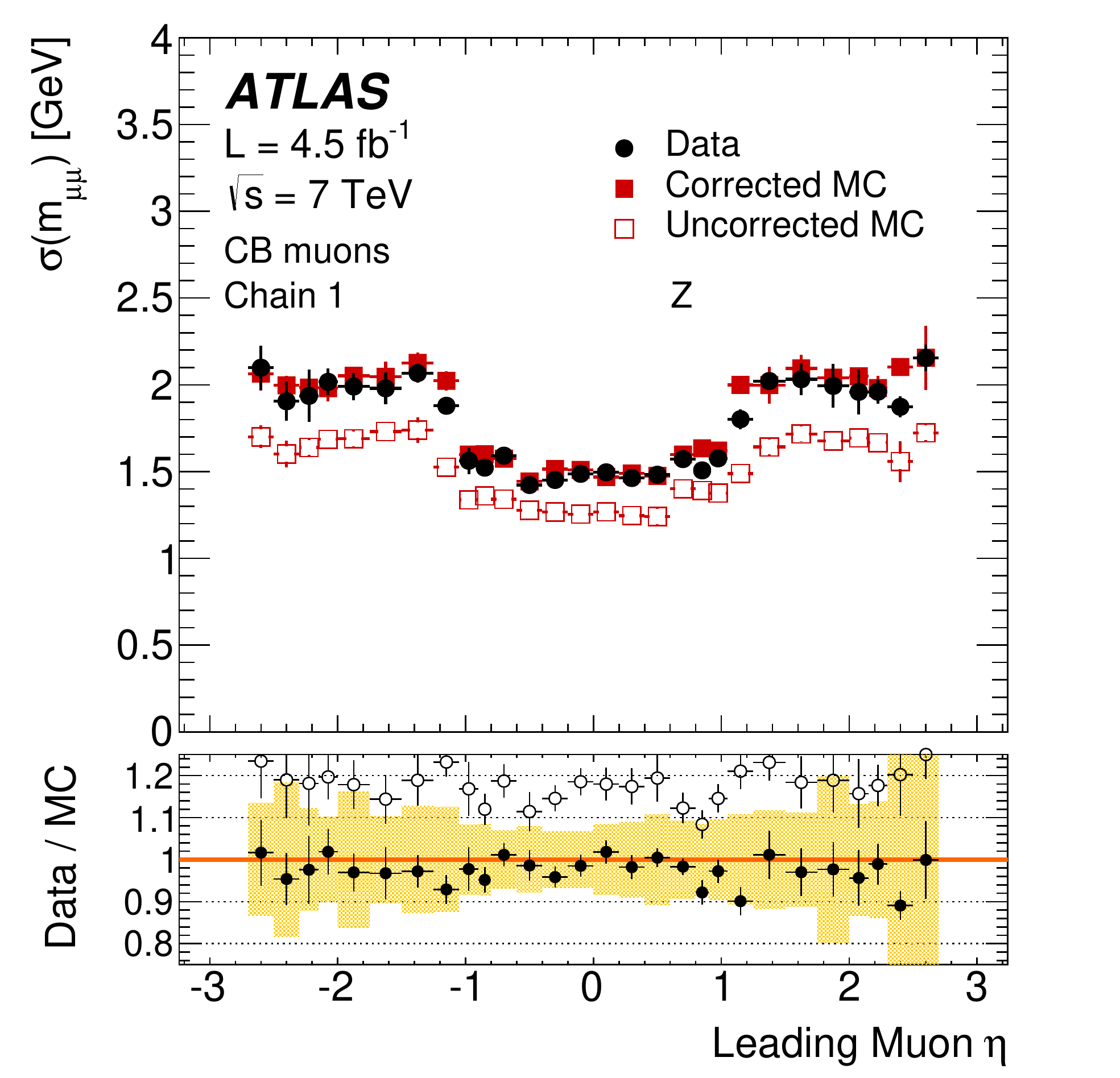}
  \end{center}
  \caption{Dimuon mass resolution $\sigma(m_{\mu\mu})$ reconstructed with Chain~1 CB muons for $\Zmm$ events recorded in 2011 for data and for uncorrected and corrected MC,  as a function of the pseudorapidity of the highest-$\pt$ muon.
  The lower panel shows the data/MC ratio and the band shows the systematic uncertainty from the momentum corrections.}
  \label{fig:resolution_2011}
\end{figure}
The efficiency, calculated with the ``tag and probe'' method as in 2012, is presented in 
Figure~\ref{fig:eff_2011}  for Chain~1 muons. The main difference with respect to 2012 is the lower efficiency of CB muons at  $|\eta|\simeq 1.2$, in which a layer of MDT chambers was missing, and the inefficiency introduced by the tighter ID selection.

The momentum corrections have been derived for the  2011 MC in the same way as for the 2012 MC. After correction, the mass scales of data and MC are in good agreement as shown in Fig.~\ref{fig:scale_2011}.
Due to the smaller data sample, the momentum corrections have larger uncertainties than in 2012. 
The resolution for CB muons obtained with $Z$ events is presented in Fig.~\ref{fig:resolution_2011}. 
The resolution of the uncorrected MC is $\approx 20\%$ smaller than data, significantly worse than in the 2012 case. This is due to the improvements introduced in the reconstruction of 2012 data, including a better knowledge of the ID and MS alignments, and  to the use of the ideal MS alignment in the 2011 simulation.

\end{appendix}



\section{Acknowledgements}

We thank CERN for the very successful operation of the LHC, as well as the
support staff from our institutions without whom ATLAS could not be
operated efficiently.

We acknowledge the support of ANPCyT, Argentina; YerPhI, Armenia; ARC,
Australia; BMWFW and FWF, Austria; ANAS, Azerbaijan; SSTC, Belarus; CNPq and FAPESP,
Brazil; NSERC, NRC and CFI, Canada; CERN; CONICYT, Chile; CAS, MOST and NSFC,
China; COLCIENCIAS, Colombia; MSMT CR, MPO CR and VSC CR, Czech Republic;
DNRF, DNSRC and Lundbeck Foundation, Denmark; EPLANET, ERC and NSRF, European Union;
IN2P3-CNRS, CEA-DSM/IRFU, France; GNSF, Georgia; BMBF, DFG, HGF, MPG and AvH
Foundation, Germany; GSRT and NSRF, Greece; ISF, MINERVA, GIF, I-CORE and Benoziyo Center,
Israel; INFN, Italy; MEXT and JSPS, Japan; CNRST, Morocco; FOM and NWO,
Netherlands; BRF and RCN, Norway; MNiSW and NCN, Poland; GRICES and FCT, Portugal; MNE/IFA, Romania; MES of Russia and ROSATOM, Russian Federation; JINR; MSTD,
Serbia; MSSR, Slovakia; ARRS and MIZ\v{S}, Slovenia; DST/NRF, South Africa;
MINECO, Spain; SRC and Wallenberg Foundation, Sweden; SER, SNSF and Cantons of
Bern and Geneva, Switzerland; NSC, Taiwan; TAEK, Turkey; STFC, the Royal
Society and Leverhulme Trust, United Kingdom; DOE and NSF, United States of
America.

The crucial computing support from all WLCG partners is acknowledged
gratefully, in particular from CERN and the ATLAS Tier-1 facilities at
TRIUMF (Canada), NDGF (Denmark, Norway, Sweden), CC-IN2P3 (France),
KIT/GridKA (Germany), INFN-CNAF (Italy), NL-T1 (Netherlands), PIC (Spain),
ASGC (Taiwan), RAL (UK) and BNL (USA) and in the Tier-2 facilities
worldwide.

\bibliographystyle{atlasBibStyleWoTitle}
\bibliography{references}

\onecolumn
\clearpage
\begin{flushleft}
{\Large The ATLAS Collaboration}

\bigskip

G.~Aad$^{\rm 84}$,
B.~Abbott$^{\rm 112}$,
J.~Abdallah$^{\rm 152}$,
S.~Abdel~Khalek$^{\rm 116}$,
O.~Abdinov$^{\rm 11}$,
R.~Aben$^{\rm 106}$,
B.~Abi$^{\rm 113}$,
M.~Abolins$^{\rm 89}$,
O.S.~AbouZeid$^{\rm 159}$,
H.~Abramowicz$^{\rm 154}$
% ...
\bigskip
\\
% ...
$^{11}$ Institute of Physics, Azerbaijan Academy of Sciences, Baku, Azerbaijan\\
% ...
$^{84}$ CPPM, Aix-Marseille Universit{\'e} and CNRS/IN2P3, Marseille, France\\
% ...
$^{89}$ Department of Physics and Astronomy, Michigan State University, East Lansing MI, United States of America\\
% ...
$^{106}$ Nikhef National Institute for Subatomic Physics and University of Amsterdam, Amsterdam, Netherlands\\
% ...
$^{112}$ Homer L. Dodge Department of Physics and Astronomy, University of Oklahoma, Norman OK, United States of America\\
$^{113}$ Department of Physics, Oklahoma State University, Stillwater OK, United States of America\\
% ...
$^{116}$ LAL, Universit{\'e} Paris-Sud and CNRS/IN2P3, Orsay, France\\
% ...
$^{152}$ Institute of Physics, Academia Sinica, Taipei, Taiwan\\
% ...
$^{154}$ Raymond and Beverly Sackler School of Physics and Astronomy, Tel Aviv University, Tel Aviv, Israel\\
% ...
$^{159}$ Department of Physics, University of Toronto, Toronto ON, Canada\\
% ...
\end{flushleft}

\end{document}